\documentclass[twoside,12pt]{article}
\usepackage{graphicx}
\usepackage{amsmath, amsfonts,amssymb}
\usepackage{bm}

\usepackage[colorlinks,linkcolor=blue,citecolor=blue]{hyperref}
\newcommand{\dd}{\mathrm{d}}

\newcommand\TL{\hfil$\displaystyle{##}$}
\newcommand\TR{$\displaystyle{{}##}$\hfil}

\def\eqalign#1{\vcenter{\openup1\jot
    \halign{\strut\span\TL & \span\TR\cr #1 \cr}}}
\def\eno#1{(\ref{#1})}
\newcommand{\eqn}[2]{\begin{equation}\eqalign{#2}\label{#1}\end{equation}}
\def\diag{\rm diag\,}

\def\SSG#1{}

\begin{document}

\title{ \vspace{1cm} \Huge Heavy ions and string theory}
\author{Oliver DeWolfe,$^1$ Steven S.\ Gubser,$^2$ Christopher Rosen,$^3$\\[4pt]
and Derek Teaney$^4$\\[20pt]
\small\sl$^1$Department of Physics, 390 UCB, University of Colorado, Boulder, CO  80309, USA\\
\small\sl$^2$Joseph Henry Laboratories, Princeton University, Princeton, NJ 08544, USA\\
\small\sl$^3$Crete Center for Theoretical Physics, University of Crete, PO Box 2208, 71003 Heraklion, Greece\\
\small\sl$^4$Department of Physics \& Astronomy, SUNY at Stony Brook, Stony Brook, NY  11794, USA}\maketitle
\begin{abstract}
\begin{picture}(0,0)(0,0)\put(314,500){PUPT-2446}\end{picture}
\begin{picture}(0,0)(0,0)\put(287,513){COLO-HEP-579}\end{picture}
\begin{picture}(0,0)(0,0)\put(291,487){CCTP-2013-06}\end{picture}
We review a selection of recent developments in the application of ideas of string theory to heavy ion physics.  Our topics divide naturally into equilibrium and non-equilibrium phenomena.  On the non-equilibrium side, we discuss generalizations of Bjorken flow, numerical simulations of black hole formation in asymptotically anti-de Sitter geometries, equilibration in the dual field theory, and hard probes.  On the equilibrium side, we summarize improved holographic QCD, extraction of transport coefficients, inclusion of chemical potentials, and approaches to the phase diagram.  We close with some possible directions for future research.
\end{abstract}
\eject
\tableofcontents
\vfill
\eject

\section{Introduction}
\label{INTRO}

\def\st{\begin{equation}}
\def\stp{\end{equation}}
\def\Sect#1{Section~\ref{#1}}
\def\Ref#1{Ref.~\cite{#1}}
\def\p{{\bm p}}
\def\llangle{\left\langle}
\def\rrangle{\right\rangle}
\def\gsim{\mbox{~{\protect\raisebox{0.4ex}{$>$}}\hspace{-1.1em}
	{\protect\raisebox{-0.6ex}{$\sim$}}~}}
\def\lsim{\mbox{~{\protect\raisebox{0.4ex}{$<$}}\hspace{-1.1em}
	{\protect\raisebox{-0.6ex}{$\sim$}}~}}

\subsection{Gauge-gravity duality  and the strong force}

Quantum chromodynamics (QCD) has been understood to be the correct theory of the strong interaction for four decades. However, because the theory is strongly coupled at low energies, many strong interaction phenomena remain difficult to successfully characterize theoretically. Lattice QCD is a powerful method that has had numerous successes, but certain kinds of phenomena, notably real-time and finite density physics, are not so easily accessible using lattice techniques. The physics of heavy ion collisions is both real-time and apparently strongly coupled, and as a result alternate theoretical tools to help in understanding heavy ion physics are welcome.

The idea that QCD might simplify in the limit of a large number of colors $N_c$ is almost as old as the theory itself \cite{tHooft:1973jz}, and has led to significant conceptual progress.  Non-planar diagrams drop out of perturbative calculations when the 't Hooft coupling $\lambda \equiv g^2 N_c$ is kept finite, and meson and glueball states become stable as the large-$N_c$ limit is approached. Calculations of real-time processes remained difficult to address even in the large-$N_c$ context until a fundamental breakthrough took place in 1997, with the formulation of the AdS/CFT, or gauge/gravity, correspondence \cite{Maldacena:1997re, Gubser:1998bc, Witten:1998qj}; for reviews see for example \cite{Aharony:1999ti, Klebanov:2000me, DHoker:2002aw, Maldacena:2003nj}. Motivated from calculations in string theory involving the dynamics of D-branes,
the correspondence states that certain non-Abelian gauge theories can be described in a wholly different way, as theories of quantum gravity living in a higher-dimensional spacetime, in particular a spacetime with the asymptotic behavior of anti-de Sitter space (AdS).  The AdS/CFT correspondence provides a concrete realization of the holographic principle, the idea ---  motivated by the scaling of the entropy of black holes as the surface area rather than the volume --- that quantum gravitating theories are in some sense hugely redundant, and can be described by a non-gravitational theory in fewer dimensions (\cite{tHooft:1993gx, Susskind:1994vu}; for a review see \cite{Bousso:2002ju}).

The AdS/CFT correspondence relies essentially on the asymptotic properties of anti-de Sitter space. In AdS space, massive particles must stay at finite spatial values (the ``bulk") but massless trajectories can reach spatial infinity, called the ``boundary". As a result, describing physics in an asymptotically AdS space requires more than ordinary initial conditions: it requires boundary conditions as well, fixing the behavior of the various dynamical fields at infinity. The geometry of the boundary has one less dimension than the bulk,\footnote{Neglecting additional compact space factors, such as the five-sphere in $AdS_5 \times S^5$.} and is identified with the space on which the dual quantum field theory lives. The precise statement of the AdS/CFT correspondence is then that for every field $\phi(r, \vec{x})$ in the bulk, there is a an associated ``dual" operator ${\cal O}_\phi(\vec{x})$  in the quantum field theory, and that the suitably-defined boundary conditions $\phi_0(\vec{x})$ on the field $\phi$ correspond to sources in the Lagrangian for the dual operator. Schematically, this may be thought of as an equality between path integrals:
\eqn{}{
Z_{\rm grav}[ \phi \to \phi_0] = \langle e^{i \int \phi_0 {\cal O}} \rangle_{\rm QFT} \,.
}
As part of the correspondence, an identification exists between the isometries of the geometry of the gravity theory, and the symmetries of the dual quantum field theory. 
Five-dimensional anti-de Sitter space may be described by the metric
\eqn{AdS}{
ds^2 = {r^2 \over L^2} \left( -dt^2 + d\vec{x}^2 \right) + {L^2 \over r^2} dr^2 \,,
}
where slices of constant radial coordinate $r$ are four-dimensional Minkowski space, with the boundary at $r \to \infty$, and $L$ is the radius of curvature, related to the number of degrees of freedom in the dual field theory.\footnote{When the dual is ${\cal N}=4$ super-Yang-Mills, the radius of curvature is related to the number of colors by $L^3/\kappa^2 = (N_c/2\pi)^2$, where $\kappa^2 = 8\pi G_5$ and $G_5$ is the five-dimensional gravitational coupling.}  The $AdS_5$ geometry has the isometry group $SO(4,2)$, which is also the conformal group in four spacetime dimensions; thus the dual quantum field theory has no scale, and is a conformal field theory (CFT). The most celebrated of the original dualities discovered by Maldacena is the duality between type IIB string theory living in the spacetime $AdS_5 \times S^5$ with $N_c$ units of flux through the five-sphere $S^5$, and the maximally supersymmetric four-dimensional gauge theory ${\cal N}=4$ Super-Yang-Mills with gauge group $SU(N_c)$, which is indeed an exactly conformal theory.  The isometry realizing overall scale transformations in AdS corresponds to a translation in the fifth (radial) direction along with an overall rescaling of the coordinates shared with the field theory. Consequently the fifth, holographic direction can be identified with a change of scale, with the region near the boundary corresponding to the high energy (ultraviolet) limit and the region far from the boundary encoding the low energy (infrared) physics; this is realized by the redshift factor $r^2/L^2$ weighting the Minkowski metric in \eno{AdS}.  Other gauge symmetries of the gravity side are mapped to globally conserved currents in the field theory.

\begin{figure}[tb]
\begin{center}
\includegraphics[width=4.5in]{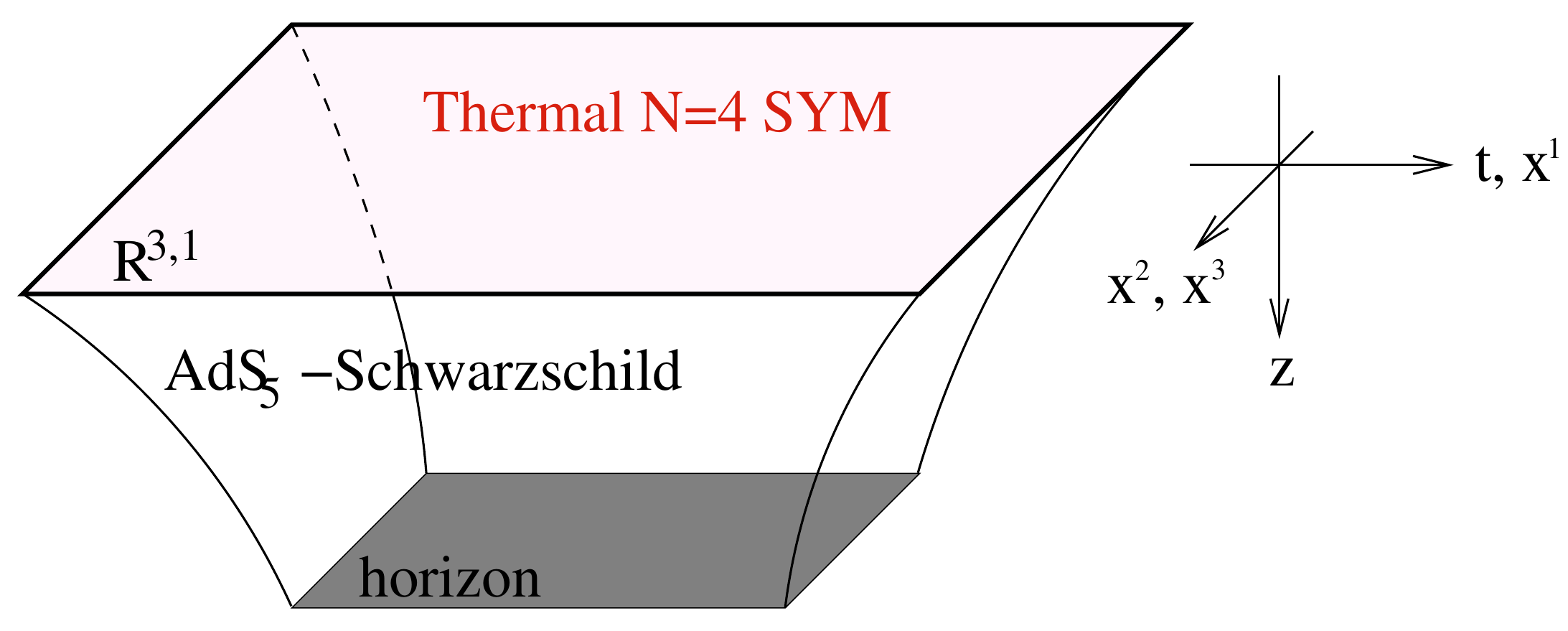}
\begin{minipage}[t]{16.5 cm}
\caption{ The $AdS_5$-Schwarzschild geometry is dual to a thermal state of ${\cal N}=4$ super-Yang-Mills theory, which can be understood as living on the boundary. (Color online.)  \label{Fig:AdSSch}}
\end{minipage}
\end{center}
\end{figure}

 Natural generalizations of the original AdS/CFT relation are possible, notably introducing features that break conformal symmetry, producing a duality with a non-conformal field theory. One notable way to break conformal symmetry and introduce a scale is to generalize to the AdS-Schwarzschild solution, a black hole in anti-de Sitter space:
\eqn{AdSSchw}{
ds^2 = {r^2 \over L^2} \left( - h(r) dt^2 + d\vec{x}^2 \right) + {L^2 \over r^2 h(r)} dr^2 \,, \quad \quad
h(r) \equiv 1  - {r_0^4 \over r^4} \,,
}
interpreted as being dual to a thermal state of the dual field theory \cite{Witten:1998qj, Witten:1998zw}. The surface gravity and surface area of the black hole horizon are identified with the temperature and entropy density of the QFT, providing a concrete realization of the thermodynamic properties of the black hole.

In addition to the gravity-side flux parameter $N_c$ being identified with the number of colors of the gauge theory, the string coupling $g_s$ is identified with the Yang-Mills coupling $g^2_{\rm YM}$. The strongest statement of the duality is that the two systems are equivalent for all values of the two parameters. However, over generic parts of parameter space, the string theory coupling is not small, and when $N_c$ is not large, the characteristic size of string excitations is comparable to the curvature scale; both mean that stringy and quantum gravity corrections, which are in general not well-understood, must be taken into account. However, in the limit of large $N_c$ and large 't Hooft coupling, string theory reduces to smooth, classical gravity, and these corrections become unimportant. Thus precisely for this limit, when the field theory is strongly coupled and at large $N_c$, the gravity description becomes tractable. This is the source of the power of the AdS/CFT correspondence: difficult strongly coupled large-$N_c$ gauge theories are described by relatively simple classical gravity.

The utility of AdS/CFT took another step forward with the study of transport coefficients and other real-time phenomena: see for example \cite{Policastro:2001yc, Son:2002sd}.
It was found that across a wide swath of gravity theories, the shear viscosity over entropy density is constant and small, $\eta/s = 1/4 \pi$ \cite{Policastro:2001yc, Buchel:2003tz,Kovtun:2004de}. When the results from the Relativistic Heavy Ion Collider (RHIC) indicated that the quark-gluon plasma can be understood as a near-perfect fluid with $\eta/s$ numerically close to the AdS/CFT result (see below for a review), the idea that holography could provide a useful, experimentally testable model for heavy ion collisions took root.

We believe that the correspondence between $AdS_5 \times S^5$ and ${\cal N}=4$ Super-Yang-Mills is exact; however, ${\cal N}=4$ Super-Yang-Mills is not QCD. It is at best a cousin, sharing the features of being an $SU(N_c)$ non-Abelian gauge theory coupled to fermions, but adding the properties of adjoint matter, scalars, conformality and supersymmetry not shared by the theory of the strong force. Nonzero temperature breaks conformality and supersymmetry, bringing the theories closer together, but they  remain distinct. The ideal would be to formulate via string theory a gravity dual for QCD itself; even in the large-$N_c$ limit this would be monumentally useful. However, prospects for formulating such a dual seem unlikely.  Nevertheless, since no perfect tractable theoretical framework for studying all aspects of heavy ion physics exists, it is still of natural interest to discover just how useful gauge/gravity calculations can be.

There are several approaches for how to model QCD using AdS/CFT. One is to use ${\cal N}=4$ Super-Yang-Mills itself or one of its related cousins. QCD at higher energies becomes more closely conformal; it is possible that ${\cal N}=4$ Super-Yang-Mills may provide a description of  certain aspects of the quark-gluon plasma that is at least approximately realistic. It is also possible to search for quantities like $\eta/s$ that are universal over a wide class of accessible theories, and attempt to probe whether general lessons for the properties of strongly coupled non-Abelian gauge theories can be derived from these that may also apply to QCD. Lastly, it is possible to ``engineer" gravity duals that do not descend directly from string theory; the duality map is not precisely known in these so-called ``bottom-up" models, but certain aspects such as symmetries and thermodynamics can be built in by the theorist constructing them, in the hopes of matching certain aspects of QCD as closely as possible. All of these techniques will be described in this review.

\subsection{Heavy ion collisions, QCD at finite density,  and the scope of this review}

Before continuing we wish to summarize the dynamics and observables in a heavy
ion collision, which will  motivate the  AdS/CFT computations described in
the rest of this work. For a fairly complete pre-LHC summary of heavy ion
collisions, see
Refs.~\cite{Adcox:2004mh,Adams:2005dq,Teaney:2009qa,Majumder:2010qh}.  For a
post-LHC review see \Ref{Muller:2012zq}, and especially \Ref{Heinz:2013th} for
an up-to-date account of the observed flow that provides the  primary
motivation for this review. To
our knowledge, there is one available review of jet-quenching in the LHC era~\cite{Mehtar-Tani:2013pia}, in addition to conference the summaries~\cite{CasalderreySolana:2012bp,Schukraft:2012cy}.

In a heavy ion collision the initial nuclei pass through each 
other leaving behind an excited  non-equilibrium state of matter. 
There is compelling experimental evidence that this non-equilibrium state
evolves and thermalizes, and is reasonably described by viscous hydrodynamics
for  the bulk of the evolution. 
Indeed, the hydrodynamic modeling of heavy ion collisions has become remarkably sophisticated.  
On an event by event basis the azimuthal distribution of particles around the 
beam pipe can be expanded in a Fourier series
\st
\frac{dN}{d\phi_\p} = \frac{N}{2\pi} \left(1 +  2 \sum_n v_n \cos(n(\phi_\p - \Psi_n)) \right) \, , 
\stp
where the phases and magnitudes, $\Psi_n$ and $v_n$, fluctuate from event to event.
The observed rms values of the harmonic coefficients, $\sqrt{\llangle v_n^2
\rrangle}$, are significantly too large to be explained by $1/N$ statistical
fluctuations. The heavy ion community has largely accepted that these fluctuations
arise from event-by-event spatial inhomogeneities in the initial energy density. The associated pressure gradients then induce collective flow converting
the spatial anisotropy to a momentum space anisotropy, which is ultimately measured.
Hydrodynamics simulations account for the rms magnitudes $\sqrt{\llangle v_n^2 \rrangle}$ \cite{CMS:2012xxa,Luzum:2012wu,Adare:2011tg,Gardim:2012yp},
the fluctuations in  $v_n$ ({\it i.e.} $P(v_n)$) \cite{Bhalerao:2011yg,Gale:2012in,Gale:2012rq}, and the angular correlations  between the harmonics of different orders ({\it i.e.} $\llangle \cos(5\Psi_5 - 2\Psi_2-3\Psi_3) \rrangle$) \cite{Qiu:2012uy,Gardim:2011xv,Teaney:2012ke,Teaney:2012gu}.
Further, the harmonic flows depend on transverse momentum, the centrality of the collision, the particle mass, and the beam energy in characteristic ways \cite{Shen:2012vn}. All
of these features are reproduced qualitatively by viscous
hydrodynamics, and striking quantitative agreement is 
found for most observables.  
This overall pattern of agreement is reached if $\eta/s$ is in the range $0.08\lsim \eta/s \lsim 0.4$ \cite{Heinz:2013th}, though it is significantly more difficult to  place a lower bound~\cite{Luzum:2012wu}.  

Clearly, one of the open problems in heavy ion collisions is to provide a
coherent theoretical explanation for the transition to hydrodynamics.
Gauge-string duality should be a useful starting point for addressing this
transition because it provides a tractable framework which interpolates
smoothly between non-thermal physics in the ultraviolet and hydrodynamics in
the infrared.  Using the correspondence the equilibration and subsequent
hydrodynamic expansion of the system is  studied in various ways in
\Sect{BJORKEN} and \Sect{noneqdynamics}.

In \Sect{BJORKEN} we show that for  boost invariant collisions  of conformal field theories much can be said  about the initial collision and the hydrodynamics 
expansion. The power of conformal symmetry  became clear through  gauge-gravity duality where the
conformal group  is manifest in the isometries of $AdS_5$.  
In \Sect{TRAPPED} we will consider an AdS/CFT model for  colliding  nuclei, based on  a dual collision of point particles in the bulk, which historically elucidated the consequences of conformal symmetry for high energy collisions.  Here we will show how the conformal
symmetry of this model can be used to determine the marginally trapped surface, which provides a lower bound on the entropy produced during the
collision.  With a clear understanding of the roles of conformal
symmetry and boost invariance, an exact hydrodynamic solution for a boost invariant flow with a non-trivial radial profile was found in \Ref{Gubser:2010ze}, which
is very useful to the heavy ion community.  
Perturbations to this flow can be classified according to the symmetry
properties of the background fluid, and this provides  an analytical framework for studying how the inhomogeneities in the initial state are transformed into collective flow, and how viscosity affects the resulting $v_n$  and their correlations.

Gauge-gravity duality can also be used to study the thermalization process 
more directly. In \Sect{noneqdynamics}, we review several calculations which investigate
how specific initial states of varying degrees of complexity thermalize. This 
involves numerically determining the evolution of the gravitational fields
in order to see how the fluid-gravity correspondence emerges at late times.
Since this is a challenging numerical enterprise, we have  provided a short review of numerical relativity in \Sect{RELATIVITY}. The results
of these methods can be used to understand the transition to hydrodynamics from
an arbitrary initial state, and the effect of the pre-equilibrium phase on the subsequent evolution. This work is described in \Sect{Hydroization}.

The equilibration of the system can also be studied by examining the 
two-point functions of the system. Indeed, this is not academic as  two-point functions control the rate of photon emission (current-current correlators) and heavy quark diffusion (force-force correlators). As
both of these quantities have active experimental programs \cite{Adare:2012px,delValle:2012qw,Afanasiev:2012dg,Adare:2011zr}, it is important to
quantify the effect of the initial non-equilibrium  phase on these
observables. This is done in \Sect{twopnt} where the emission rate 
of non-equilibrium photons is studied with gauge-gravity duality.

Most of these studies of non-equilibrium phenomena rely on the original
duality between supergravity on ${\rm AdS}_5$ and  ${\cal N}=4$ Super
Yang-Mills theory. We will describe several efforts to make the underlying
field theory look more like QCD. To introduce such ``bottom-up" models, in \Sect{blackbranes}  we will
first provide a brief technical account of black brane geometries and their
thermodynamics, the calculation of transport coefficients, and the energy loss of
energetic particles.  This section explains how a scalar
coupled
to gravity in anti-de Sitter space breaks conformal invariance and affects the
calculation of transport coefficients.  An example of this type of theory is
improved holographic QCD, which we describe in section~\ref{IHQCD}. The
parameters of the model can be fit to lattice data on the equation of state,
and then many other quantities can be computed such as  glueball masses, the
bulk viscosity, and the drag force of a heavy quark.

A common approximation in studying the physics of the quark-gluon plasma is to
neglect the chemical potential $\mu$ for baryons.  At least at mid-rapidity,
this can be quantitatively justified based on the relative abundance of
particle species, as summarized in \cite{Abelev:2013vea} for recent heavy ion
data from the LHC.  But QCD dynamics at finite $\mu$ is quite interesting: in
particular, it is believed that the cross-over at $\mu=0$ between confined and
deconfined phases in QCD sharpens into a first order phase transition as $\mu$
becomes non-zero, with a critical point at finite $\mu$ and $T$ (for a review,
see for example \cite{Rajagopal:2000wf}).  Beam-energy scans at RHIC
\cite{Aggarwal:2010cw} and the CBM project at FAIR \cite{Staszel:2010zza} aim
to explore as much of the $\mu$-$T$ phase diagram as possible, and hopefully to
create laboratory conditions in which an equilibrated quark-gluon plasma
crosses through or near the critical point on its way to hadronization.  A
significant obstacle to making clear predictions about this type of experiment
is that different theoretical approaches lead to no consensus on the position
of the critical point \cite{Stephanov:2005iu}.  As we review in
section~\ref{PHASE}, AdS/CFT offers some insight here, generating from
zero-density lattice data a realistic phase diagram with a first-order
line ending on a critical point, albeit with mean field critical exponents and
suppressed energy-momentum transport. We also discuss a related top-down
construction of the phase transition, and comment briefly on gauge/gravity
approaches to color superconductivity.

Other reviews of the interplay between heavy ion collisions and the AdS/CFT
correspondence exist, notably \cite{CasalderreySolana:2011us} and
\cite{Adams:2012th}. While there is some overlap in content, the topics
emphasized in the respective reviews are distinct, leaving them complementary.
\cite{CasalderreySolana:2011us} has more discussion of heavy quark and
quarkonia topics, while \cite{Adams:2012th} also devotes substantial attention
to ultracold gases. The current work devotes more attention to non-equilibrium
dynamics as well as to physics at nonzero baryon density. For better or for
worse, this review is also shorter.

\section{Bjorken flow and symmetries}
\label{BJORKEN}

According to modern understanding of relativistic heavy ion collisions, the
quark-gluon plasma is created from low-momentum partons which linger near the
collision plane while the nuclear pancakes recede at nearly the speed of light.
A widely used approximation in treating the mid-rapidity region is the
assumption of boost-invariance, which was justified at the level of an
early-time parton model \cite{Bjorken:1982qr} by arguing that the low-momentum
partons' longitudinal velocity shortly after the collision is proportional to
their distance from the collision plane.  Assuming that these partons locally equilibrate into a fluid, and making the further
idealization of a uniform distribution of matter in the transverse plane, one
can derive the famous Bjorken flow solution, in which the rapidity of each
element of fluid equals its spacetime pseudorapidity, and the energy density in
the local rest frame scales as $\tau^{-4/3}$, where $\tau = \sqrt{t^2-x_3^2}$.

With longitudinal boost invariance as a first simplifying assumption, one is
still left with challenging questions about initial conditions, local
equilibration, transverse or radial flow, and anisotropies.  The gauge-string
duality should be a useful starting point for at least some of these questions,
providing a graceful transition from an arbitrary initial state to hydrodynamics.
It
also makes symmetries more obvious through five-dimensional geometric
constructions.  We begin with an exposition in sections~\ref{TRAPPED} of an
$SO(3)$ symmetry which emerges in AdS/CFT as the symmetry group of the
transverse space in collisions of lightlike particles.  In the bulk gravity
theory, this symmetry facilitates the computation of the entropy of trapped
surfaces formed when lightlike particles collide.  The same $SO(3)$ can be used
to generate new solutions to the Balitsky-Kovchegov equation describing the
behavior of Wilson line correlators in the presence of a high-rapidity hadron,
as we explain in section~\ref{BKsymmetry}.  These correlators are key
ingredients to the color-glass condensate (CGC) picture of initial conditions.
The same $SO(3)$ symmetry allows for an analytic generalization of Bjorken
expansion which includes radial flow, as we explain in
section~\ref{BjorkenVariant}.  A somewhat different deformation of Bjorken
flow, aimed at accounting for rapidity dependence in a symmetry-based context,
is reviewed in section~\ref{COMPLEX}.  We defer to section~\ref{noneqdynamics}
a detailed discussion of equilibration.  It is in the study of equilibration
that AdS/CFT provides particularly interesting dynamical information: by
preparing a variety of initial conditions in the bulk and seeing how they
settle into geometries with regular black hole horizons, one can extract a lot
of detailed information about how non-equilibrium states evolve and thermalize
in the dual field theory.

\subsection{Trapped surface estimates of multiplicity}
\label{TRAPPED}

\begin{figure}
 \begin{center}
  \includegraphics[width=4in]{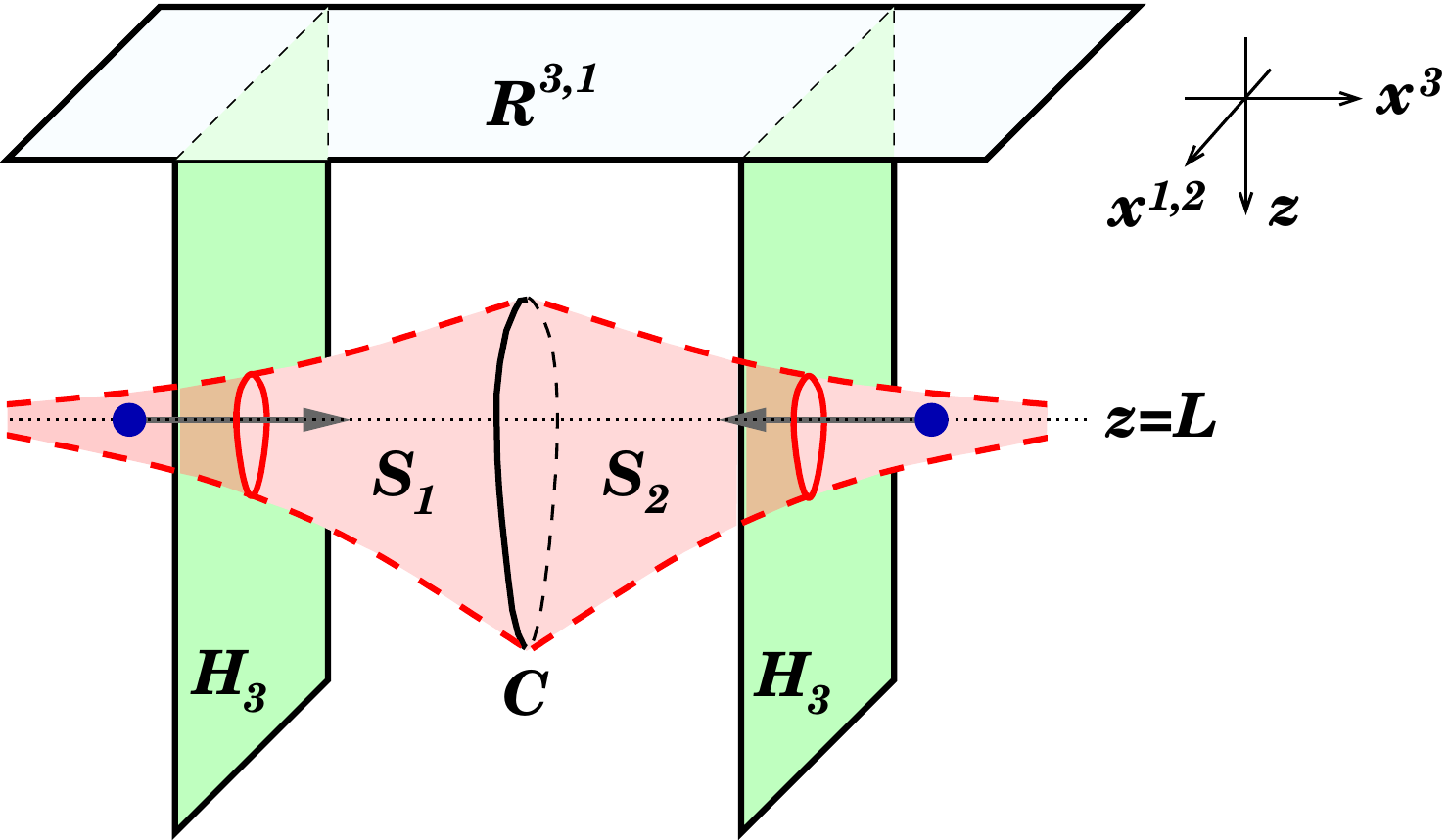}
  \caption{
  A schematic of the head-on collision of two massless point particles  in the 
  bulk. Also shown
  is marginally trapped surface $S_1 \cup S_2$ that forms around the collision.
  The transverse geometry is $H_3$, a three-dimensional analog of the upper half-plane.  Finding the intersection curve $C$ of the two halves of the trapped surface is a key step toward the entropy estimate \eno{Strapped}.  From \cite{Gubser:2008pc}.}\label{3Dtrapped}
 \end{center}
 \end{figure}

In this section we will first consider the head-on  collision of two massless
pointlike particles in ${\rm AdS}_5$ as a model for colliding nuclei.   
As explained in the introduction, this model was historically important in
realizing the powerful constraints of conformal symmetry and boost invariance.
In
Poincar\'e coordinates the incoming trajectories of the colliding point
particles have $x_3 = x_0$ with $x_1=x_2=0$ and $z=L$, {\it i.e.} the incoming
particles are at constant ``depth'' below the boundary.
Figure~\ref{3Dtrapped} presents a schematic of this collision together with
the marginally trapped surface, which will be described below. 

Gauge-string duality makes the conformal symmetries of the problem manifest: the conformal group $SO(4,2)$ of ${\bf R}^{3,1}$ is the group of isometries of $AdS_5$.  This is seen most directly by embedding global $AdS_5$ into ${\bf R}^{4,2}$ as the solution of the equation
 \eqn{GlobalEmbedding}{
  X_{-1}^2 + X_0^2 - X_1^2 - X_2^2 - X_3^2 - X_4^2 = L^2 \,.
 }
These coordinates are related to the usual $(t,\vec{x},z)$ coordinates of the Poincar\'e patch by
 \eqn{CoordinateRelations}{
  X_{-1} &= {z \over 2} \left( 1 + {L^2 + \vec{x}^2 - t^2 \over z^2} \right) \qquad
  X_0 = L {t \over z}  \cr
  X_i &= L {x_i \over z} \qquad
  X_4 = {z \over 2} \left( -1 + {L^2 - \vec{x}^2 + t^2 \over z^2} \right) \,.
 }
Thus, the lightlike trajectories of the incoming particles in $AdS_5$ are given by the global coordinates $X_3 = X_0$ with
$X_1=X_2=X_4=0$.  
Ordinarily, a lightlike particle in four dimensions has a little
group $SO(2)$, corresponding in our case to rotations of the $x_1$-$x_2$ plane.
In $AdS_5$, the little group becomes the $SO(3)$ that rotates $X_1$, $X_2$, and
$X_4$ among themselves.  The transverse ``plane'' in the bulk is obtained by
setting $X_3 = X_0$ in \eno{GlobalEmbedding}: then one finds $X_{-1}^2 - X_1^2
- X_2^2 - X_4^2 = L^2$, which is the equation for the Euclidean manifold ${\bf
H}^3$, also called the three-dimensional hyperbolic plane.  The $SO(3)$ little
group acts as the group of isometries of ${\bf H}^3$ that preserve the point
$X_1=X_2=X_4 = 0$.

An obvious question is how the $SO(3)$ symmetry, which is evident in global
$AdS_5$,  acts in the dual field theory.  As a first step in this direction, we
note (following for instance \cite{Gubser:2008pc}) that the stress tensor
induced in the boundary by  the right-moving lightlike particle at depth $z=L$
takes the form
 \eqn{STdepth}{
  T_{--} = {2L^4 E \over \pi (L^2 + x_1^2 + x_2^2)^3} \delta(x^-) \,,
 }
where we define $x^\pm = t \pm x_3$.  Other components of $T_{\mu\nu}$ in the
$(x^+,x^-,x_1,x_2)$ coordinate system vanish.  
Thus, the collision of two point particles in the bulk  corresponds 
to the collision of two shock waves in the boundary field theory
with a definite transverse profile given by \eno{STdepth}.
Evidently, the profile of
\eno{STdepth} is normalizable in the transverse directions, like a boosted
nucleus. However, it has only a few finite transverse moments, which is quite unlike
the Wood-Saxon profile for a nucleus, whose exponential tails make all
transverse moments finite.  It can be argued \cite{Gubser:2008pc} that the form
\eno{STdepth} of the stress tensor follows from $SO(3)$ symmetry  alone,
without reference to a holographic dual theory.

Although the full dynamics of the collision cannot be determined analytically, the $SO(3)$ symmetry was employed in \cite{Gubser:2008pc,Lin:2009pn,Gubser:2009sx} to obtain an analytical lower bound on the black hole entropy which is
produced during the point-particle collision in $AdS_5$. 
The dual bound on entropy production from the collision 
of shock waves in the boundary theory yields estimates for
the total charged particle production in a model nucleus-nucleus collision.
This section is devoted to summarizing some of the main features of these results, namely: 1) Good agreement with measured multiplicities at top RHIC energies, $\sqrt{s_{NN}} = 200\,{\rm GeV}$; 2) a rapid growth of the multiplicity, $N_{\rm charged} \sim E_{\rm beam}^{2/3}$, which is in conflict with data; 3) A slower asymptotic scaling, $N_{\rm charged} \sim E_{\rm beam}^{1/3}$, when an ultraviolet cutoff is imposed; and 4) Impact parameter dependence which predicts too slow a fall-off of rapidity as centrality decreases.

The key calculation in \cite{Gubser:2008pc} is to locate a trapped surface in the collision of two energetic pointlike particles in $AdS_5$.  These pointlike particles are dual to distributions of stress-energy as indicated in \eno{STdepth}: that is, profiles which are normalizable in the transverse plane and localized in the longitudinal direction.  Such a distribution is a tolerably good approximation of a relativistic nucleus, except that the transverse profile of the energy density of a nucleus falls off exponentially at large transverse radii, whereas distribution indicated in \eno{STdepth}---sometimes called a conformal soliton---falls off as a power of transverse radius.  The advantage of considering collisions of conformal solitons first is that head-on collisions preserve an $SO(3)$ symmetry of the type described in previous sections.  This makes it relatively easy to find the shape of a trapped surface that forms around the collision point.  A cartoon of the trapped surface is shown in Figure~\ref{3Dtrapped}.

Trapped surfaces are a standard means of estimating the position of a black hole horizon.  Intuitively, a trapped surface in a spacetime of $D$ dimensions is a closed, spatial $D-2$-dimensional hypersurface such that any light-ray starting at the surface falls toward its interior rather than out toward infinity.  A marginally trapped surface has the property that of the two light rays starting at a point along the surface and directed normal to the surface, one falls inward and one propagates forward in time without moving outward or inward---in a sense made mathematically precise in terms of the covariant derivative of the light rays along the surface.  A snapshot at fixed Killing time of the surface of a Schwarzschild black hole is a trivial example of a marginally trapped surface, because of the two light rays normal to the horizon, one propagates inward while the other propagates exactly along the horizon.  Spherical surfaces inside the horizon are examples of trapped surfaces, whereas spherical surfaces outside the horizon are not trapped.  Thus in the Schwarzschild geometry, the outermost trapped surface is the horizon.  In more general settings, it is generally believed that trapped and marginally trapped surfaces must reside at or inside the event horizon.

Because the black hole entropy is calculated as the area of the horizon divided by $4G_5$, one can set a lower bound on the entropy by computing the area of a marginally trapped surface and dividing by $4G_5$.  There is a heuristic argument that this lower bound should be close to the true entropy, since after the horizon is formed the subsequent time dependence of the black hole is mostly dual to the hydrodynamical expansion of a thermal plasma with small viscosity.
 This expansion generates little additional entropy.  The entropy of the marginally trapped surface for head-on collisions of conformal solitons as estimated in \cite{Gubser:2008pc} is
 \eqn{Strapped}{
  S_{\rm trapped} \approx \pi \left( {L^3 \over G_5} \right)^{1/3} (2EL)^{2/3} \,.
 }
In order to link this estimate of entropy production in the collision of conformal solitons to the total multiplicity in heavy ion collisions, three numerical estimates are required:
 \begin{itemize}
  \item Lattice results show that $\epsilon/T^4 \approx 11$ for $1.2 T_c \lesssim T \lesssim 2T_c$, which is the approximate temperature range for top-energy RHIC collisions.  This ratio of $\epsilon/T^4$ is recovered for black holes in $AdS_5$ provided $L^3/G_5 \approx 1.9$.
  \item The factor of $L$ in $(2EL)^{2/3}$ is to be understood as the transverse extent of the conformal soliton, dual to the depth in $AdS_5$ of the dual pointlike particle.  If the latter depth is modified to $z=z_*$, then one replaces $(2EL)^{2/3} \to (2Ez_*)^{2/3}$ in \eno{Strapped}, without altering the $(L^3/G_5)^{1/3}$ factor.  $E$ is the beam energy, and $L \approx 4.3\,{\rm fm}$ is the root-mean-square transverse radius of a gold nucleus.
  \item The total multiplicity of charged particles is related to the entropy by $S \approx 7.5 N_{\rm charged}$.  One way of reaching this estimate is to employ free field estimates of both entropy density and number densities of hadron species, summing over known hadron resonances and setting $T = 170\,{\rm MeV}$, close to the transition temperature of QCD.
 \end{itemize}
The result of plugging these numerical estimates into \eno{Strapped} is $N_{\rm charged} \geq 4700$.  This is satisfyingly close to observed values $N_{\rm charged} \approx 5060$ \cite{Back:2002wb}.  However, at LHC energies, one obtains from the same formula $N_{\rm charged} \geq 27000$, whereas data indicates $N_{\rm charged} \approx 17000$.

In \cite{Gubser:2009sx}, a modification of the trapped surface calculation was considered which reduces the asymptotic growth from $E^{2/3}$ to $E^{1/3}$ at high energies.  The modification uses only the part of trapped surface below a depth in $AdS_5$ corresponding to some energy scale $\Lambda_{\rm UV}$.  The rationale is that at sufficiently high energy scales, the dynamics of QCD is weakly coupled, and little entropy is produced.  An additional infrared cutoff was also considered, whereby the part of the trapped surface below a depth corresponding to some energy scale $\Lambda_{\rm IR}$ is discarded.  This additional infrared cutoff does not further alter the $E^{1/3}$ asymptotic scaling, and for reasonable values of $\Lambda_{\rm IR}$ (namely, close to the QCD scale), the infrared cutoff has much less effect on the total trapped surface entropy than the ultraviolet cutoff.

\begin{figure}[tb]
 \begin{center}
  \includegraphics[width=5in]{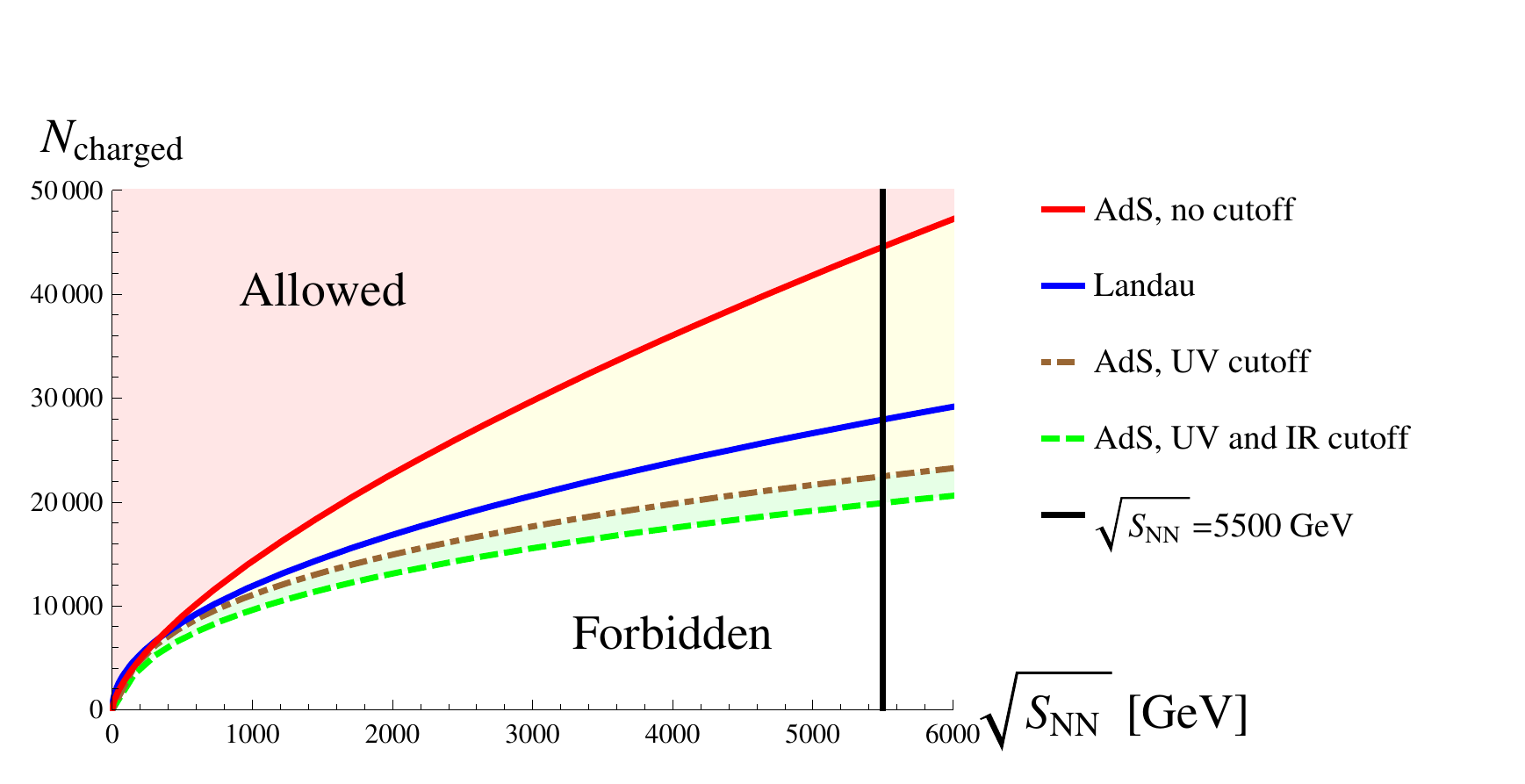}
  \caption{(Color online.)  Lower bounds on the total number of charged particles $N_{\rm charged}$ produced in head-on lead-lead collisions with center of mass energy per nucleon pair $\sqrt{s_{\rm NN}}$.  From \cite{Gubser:2009sx}.}\label{TotalMultiplicityPredictions}
 \end{center}
 \end{figure}

The trapped surface model of \cite{Gubser:2009sx} incorporating an ultraviolet cutoff $\Lambda_{\rm UV} = 2\,{\rm GeV}$ is quite successful: it has only a slightly lower prediction for total multiplicity at top RHIC energies, and at $\sqrt{s_{\rm NN}} = 2.76\,{\rm TeV}$ it predicts $N_{\rm charged} \gtrsim 16800$.  One may object that the parameter $\Lambda_{\rm UV}$ can be adjusted to get any desired $N_{\rm charged}$ (within limits); while this is true, the choice $\Lambda_{\rm UV} = 2\,{\rm GeV}$ was made in 2009, well before LHC heavy-ion data was available.\footnote{The model with both ultraviolet and infrared cutoff model predicts $N_{\rm charged} \gtrsim 3820$ at $\sqrt{s_{\rm NN}} = 200\,{\rm GeV}$ and $N_{\rm charged} \gtrsim 15000$ at $\sqrt{s_{\rm NN}} = 2.76\,{\rm TeV}$.}  The predictions of the UV cutoff model of \cite{Gubser:2009sx} for central lead-lead collisions at higher LHC energies are available from Figure~\ref{TotalMultiplicityPredictions}.

Extensions of the above results have been developed in \cite{Kiritsis:2011yn},
together with some new ideas.  Notably, it was proposed that the UV cutoff in
$AdS_5$ should be chosen to correspond to the saturation scale in the dual
theory, which runs slowly with energy as $Q_s \sim E^\lambda$ where $\lambda
\approx 0.15$.  Adopting such a prescription leads to modestly faster growth of
$N_{\rm charged}$ with energy. Indeed, the particular version treated in detail
in \cite{Kiritsis:2011yn} gives $N_{\rm charged} \sim E^{0.483}$ at large
energies, which can be compared to the Landau expectation $N_{\rm charged} \sim
E^{1/2}$. The model finds good agreement at the LHC ($\sqrt{s_{\rm NN}} = 2.76\,{\rm
TeV}$) as well as at RHIC ($\sqrt{s_{\rm NN}}=0.2\,{\rm TeV}$) due to
the subleading energy dependence in the model.  Similarly good fits to data were achieved in
\cite{Kiritsis:2011yn} in an Improved Holographic QCD (IHQCD) framework 
where the multiplicity increases  as $N_{\rm charged} \sim E^{0.451}$ up to logarithmic enhancements.

\subsection{$SO(3)$ conformal symmetry in initial conditions}
\label{BKsymmetry}

As remarked in the previous section, $SO(3)$ symmetry tightly constrains the
stress tensor $T_{--}$ of a right-moving shock wave, which is dual to a
lightlike particle in $AdS_5$.  In this section we will describe how the same $SO(3)$ symmetry  can be used  in a rather different context to classify solutions to the Balitsky-Kovchegov equation.

In QCD, a useful quantity in describing initial conditions is the correlator $S(r_1,r_2;Y)$ of lightlike Wilson lines at transverse positions $r_1$ and $r_2$, evaluated in the presence of a hadronic target at rapidity $Y$.  It obeys the Balitsky-Kovchegov (BK) equation \cite{Balitsky:1995ub,Kovchegov:1999yj,Kovchegov:1999ua}, which at leading order reads
 \eqn{BKleading}{
  {\partial S(r_1,r_2;Y) \over \partial Y} = 
    {\bar\alpha_s \over 2\pi} \int d^2 z \, {|r_1-r_2|^2 \over |r_1-z|^2 |r_2-z|^2} \left[ 
      S(r_1,z;Y) S(z,r_2;Y) - S(r_1,r_2;Y) \right] \,,
 }
where 
 \eqn{alphabarDef}{
  \bar\alpha_s \equiv {\alpha_s N_c \over \pi} \,.
 }
A standard simplification is to require $S$ to depend on the transverse positions $r_1$ and $r_2$ only through the separation $|r_1-r_2|$.  This amounts to requiring planar symmetry in the transverse plane of the collision.  Because \eno{BKleading} possesses conformal invariance (a feature of which is modified by subleading corrections in QCD), one can simplify in a different way, demanding $SO(3)$ symmetry in the transverse plane rather than planar symmetry.  Precisely what this $SO(3)$ symmetry is can be best understood from the explicit form of the solution: It is
 \eqn{ConformalForm}{
  S(r_1,r_2;Y) = S_q(d_q(r_1,r_2);Y)
 }
where
 \eqn{ChordalDistance}{
  d_q(r_1,r_2) \equiv {|r_1-r_2| \over \sqrt{(1+q^2 |r_1|^2) (1+q^2 |r_2|^2)}} \,,
 }
and $q$ is a parameter with dimensions of inverse length.  If the transverse plane is mapped stereographically to the sphere, as is common in complex analysis, then $d_q$ is the chordal distance from one point on the sphere to another.\footnote{To make the stereographic map precise, we must specify how radius $|r|$ maps to polar angle $\theta$.  The relation needed is $|qr| = \tan(\theta/2)$.}

It is consistent to require that $S(r_1,r_2;Y)$ takes the form indicated in \eno{ConformalForm} because the BK equation respects conformal symmetry: thus initial conditions of the form \eno{ConformalForm} will lead to solutions which have the same form.  A more ambitious claim, conjectured but not proven in \cite{Gubser:2011qva}, is that arbitrary initial conditions with finite extent in the transverse plane lead to solutions that tend toward the form \eno{ConformalForm} at very large $Y$, for some $q$ whose value depends on the initial conditions and characterizes the transverse width of the hadron.  It was also argued in \cite{Gubser:2011qva} that, as a consequence of the form \eno{ConformalForm}, the saturation scale as a function of distance $b$ from the center of the hadron takes the form
 \eqn{sigmaApprox}{
  Q_s(b;Y) = {Q_s^{\rm max}(Y) \over 1 + q^2 b^2} \,.
 }
This form was previously considered \cite{Iancu:2007st} on more phenomenological grounds, and the large $b$ behavior was understood earlier, for example in \cite{Bondarenko:2003ym}.

Formally, the ansatz \eno{ConformalForm} is a natural starting point for describing the gauge theory dual of a lightlike point particle in $AdS_5$.  Recalling that the little group of massless particles is $SO(3)$ in five dimensions, we can think of the symmetry arguments leading to \eno{STdepth} and \eno{ConformalForm} as being consequences of enhancements of the little group from $SO(2)$ to $SO(3)$.  It would be interesting to develop field theory calculations to higher orders in ${\cal N}=4$ super-Yang-Mills theory, where the analog of the BK equation must respect conformal symmetry to all orders.  Possibly some more direct comparison with gravity duals will emerge.

Phenomenologically, a deficiency of the forms \eno{ConformalForm} and \eno{sigmaApprox} for describing individual hadrons (for instance protons) is that confinement strongly modifies the power-law fall-off at large $r_i$ or $b$.  In particular, in order to obtain reasonable agreement with the phenomenon of geometric scaling in deep inelastic scattering, one needs some form of infrared cutoff: for example, $Q_s$ may follow the form \eno{sigmaApprox} out to $b = b_* \approx 0.7\,{\rm fm}$, and then fall immediately to $0$.  Agreement with data is mostly a matter of choosing the correct value of $b_*$.  Values of $q$ between $0.1\,{\rm GeV}$ and $0.5\,{\rm GeV}$, or perhaps even a broader range, can be accommodated once the infrared cutoff is in place.  One can see the need for an infrared cutoff already in the stress tensor \eno{STdepth}: the profile of \eno{STdepth} is normalizable in the transverse directions, like a boosted nucleus; but it has only a few finite transverse moments, which is quite unlike the Wood-Saxon profile for a nucleus, whose exponential tails make all transverse moments finite.

\subsection{Variants of Bjorken flow}
\label{BjorkenVariant}

Having seen in section~\ref{BKsymmetry} that an $SO(3)$ subgroup of $SO(4,2)$ has some interesting consequences for initial states, let's ask how it might constrain the final state.  The easiest context in which to address this is boost-invariant hydrodynamics, where one additionally assumes invariance under boost symmetry in the beamline direction.  Investigations of $SO(3)$-invariant, boost-invariant hydrodynamics, and its perturbations, have been the aim of several works \cite{Gubser:2010ze,Gubser:2010ui,Staig:2010pn,Staig:2011wj}.

Our presentation is organized as follows.  In section~\ref{FORMAL}, we will explain the action of the conformal group on ${\bf R}^{3,1}$ and then outline a systematic procedure in which we start with a specific type of subgroup of the conformal group and develop from it a solution to the Navier-Stokes equations.  In section~\ref{RadialResults}, we will summarize the main features of $SO(3)$-invariant, boost-invariant hydrodynamics, which include radial flow and analytical expressions for energy density only somewhat more complicated than for Bjorken flow.  In section~\ref{ADAPTED}, we explain a choice of coordinates on ${\bf R}^{3,1}$ which amounts to mapping the $SO(3)$-invariant flow into the geometry $dS_3 \times {\bf R}$, where $dS_3$ is three-dimensional de Sitter space.  In section~\ref{ThreePERTURB} we summarize phenomenological studies of perturbations made possible by this mapping.  Finally, in section~\ref{COMPLEX} we detour to a deformation of Bjorken flow \cite{Gubser:2012gy} which retains its usual symmetries in the transverse plane but modifies the beamline boost invariance so as to achieve an interpolation between Bjorken flow at central rapidities and a glasma-like regime at forward rapidities.

\subsubsection{Formal features}
\label{FORMAL}

The conformal group $SO(4,2)$ has an obvious action on $AdS_5$, inherited from its action on ${\bf R}^{4,2}$.  In order to discuss the consequences of conformal symmetry more quantitatively in the dual field theory, we need to explain the action of $SO(4,2)$ on ${\bf R}^{3,1}$, as well as the action of the $SO(3)$ subgroup that we previously identified as the enhanced little group.  To this end, we first enumerate the $15$ continuous symmetries in $SO(4,2)$ in terms of differential operators $\xi = \xi^\mu \partial_\mu$ acting on ${\bf R}^{3,1}$, as follows:
 \begin{itemize}
  \item Translations, $T_{(\mu)} = \partial_\mu$.  Thus, for example, $T_{(1)}$ is translation in the $x_1$ direction.
  \item Spatial rotations, $R_{(ij)} = x_i \partial_j - x_j \partial_i$.
  \item Boosts, $B_{(i)} = t \partial_i + x_i \partial_t$.
  \item Dilations: $D = x^\mu \partial_\mu$.
  \item Special conformal transformations: $K_{(\mu)} = x^\nu x_\nu \partial_\mu - 2 x_\mu x^\nu \partial_\nu$.
 \end{itemize}
Then a general element of the Lie algebra of $SO(4,2)$ may be expressed as
 \eqn{GeneralElement}{
  \ell = \mathfrak{t}^\mu T_{(\mu)} + \mathfrak{r}_{ij} R_{(ij)} + \mathfrak{b}_i B_{(i)} + 
    \mathfrak{d} D + \mathfrak{k}^\mu K_{(\mu)} \,,
 }
where $\mathfrak{t}^\mu$, $\mathfrak{r}_{ij}$, $\mathfrak{b}_i$, $\mathfrak{d}$, and $\mathfrak{k}^\mu$ are real numbers.  The standard isometries of the transverse plane form a group $ISO(2)$ generated by $R_{(12)}$, $T_{(1)}$, and $T_{(2)}$, while the beamline boost-invariance group $SO(1,1)$ is generated by $B_{(3)}$.  Note that all elements of $ISO(2)$ commute with all boosts in $SO(1,1)$.

The $SO(3)$ symmetry of interest to us is based on leaving the $SO(1,1)$ as it is, and also leaving in place the rotational symmetry $R_{(12)}$ of $SO(2) \subset ISO(2)$, but making the replacement
 \eqn{ToSOthree}{
  T_{(i)} \to W_{(i)} \equiv T_{(i)} - q^2 K_{(i)} \qquad\hbox{for}\quad i = 1,2 \,,
 }
where $q$ is a parameter whose dimensions are inverse length.  One may straightforwardly check that the $W_{(i)}$ commute with $B_{(3)}$, and that along with $R_{(12)}$ they form the $SO(3)$ algebra.  It is a little more involved to show that this particular $SO(3)$ algebra is dual to the $SO(3)$ which preserves a pointlike null trajectory in $AdS_5$ at a constant depth $z=1/q$.\footnote{The attentive reader will notice that we had the $AdS_5$ radius $L$ playing the role of a transverse length scale in \eno{STdepth} (alternatively, depth in $AdS_5$), whereas now $1/q$ plays a comparable role.  The explanation for this transition is that we previously chose $q=1/L$ purely for notational convenience, but there is no reason to stick with this choice since we can change $q$ by applying an overall spacetime dilation.}

Consider a spacetime symmetry group of the form $A \times B$ where $B$ is a one-parameter abelian group, while $A$ is a three-parameter non-abelian group.  Describing this group as a spacetime symmetry group means that it must be a subgroup of $SO(4,2)$.  Let ${\cal L}_\xi$ denote the Lie derivative with respect to a vector field $\xi^\mu$: in particular, if $f = f(x^\mu)$ is a function of the spacetime coordinates, then ${\cal L}_\xi = \xi^\mu \partial_\mu f$.  There is an essentially unique combination $g$ of the $x^\mu$ satisfying
 \eqn{gForm}{
  {\cal L}_\xi g = 0 \qquad\hbox{for}\quad \xi \in A \times B \,.
 }
There are four such equations (because there are four generators of $A \times B$), but only three are independent, because one element of $A$ can be generated from the other two.  Each of the three independent equations can be written in the form $\xi^\mu v_\mu = 0$, where
 \eqn{vForm}{
  v_\mu = \partial_\mu g \,.
 }
Three constraints of the form $\xi^\mu v_\mu = 0$ leave only one possible direction for $v_\mu$ at each point in spacetime.  That is why there is essentially only one solution to \eno{gForm}, where by ``essentially'' we mean that given one solution, any other can be written as a function of it.

Although it appears we have started very abstractly, the quantities introduced so far are physically interesting.  For example, when $A = ISO(2)$ and $B = SO(1,1)$ as for Bjorken flow, then $g = t^2 - x_3^2$ is essentially Bjorken time, and we know that the main result of a hydrodynamic analysis will be a functional form $\epsilon = \epsilon(g)$ for the energy density.  For inviscid, conformal hydrodynamics, $\epsilon = \epsilon_0/g^{2/3}$ where $\epsilon_0$ is a constant of integration.  The following paragraphs essentially explain how to generalize the hydrodynamic analysis to more a wider choice of symmetries.

The example of Bjorken flow highlights another constraint that must be imposed: $v_\mu$ must be timelike in the region of physical interest---which for Bjorken flow is the causal future of the collision plane, i.e.~$t>|x_3|$.  Intuitively, then, the strategy is to invent some interesting variant of Bjorken time and then require the fluid velocity to be the gradient of this new ``time'' coordinate.

With a function $g$ in hand, we next seek a solution to the equation
 \eqn{hForm}{
  {\cal L}_\xi h = -{1 \over 4} (\nabla_\lambda \xi^\lambda) h 
    \qquad\hbox{for}\quad \xi \in A \times B \,.
 }
As before, it is possible to find a function $h(x^\mu)$ satisfying \eno{hForm} because only three of the equations are independent.  With $g$ and $h$ constructed, one may straightforwardly show that, for any constant $\alpha$, the general function satisfying the equations
 \eqn{fForm}{
  {\cal L}_\xi f = -{\alpha \over 4} (\nabla_\lambda \xi^\lambda) f
    \qquad\hbox{for}\quad \xi \in A \times B
 }
takes the form $f(x^\mu) = h^\alpha \hat{f}(g)$.  In particular, this shows the sense in which $h$ is essentially unique.  When $A \times B$ is composed entirely of isometries of ${\bf R}^{3,1}$ (as in Bjorken flow), we may choose $h=1$ because $\nabla_\lambda \xi^\lambda = 0$ for all isometries (i.e.~translations, rotations, and boosts).

The construction of a hydrodynamic flow that is invariant under $A \times B$ now hinges on tensors with definite conformal weights, by which we mean tensors $Q_{\mu_1\mu_2\cdots}^{\nu_1\nu_2\cdots}$ satisfying
 \eqn{ZetaWeightDef}{
  {\cal L}_\xi Q_{\mu_1\mu_2\cdots}^{\nu_1\nu_2\cdots} = 
    -{\alpha \over 4} (\nabla_\lambda \zeta^\lambda) Q_{\mu_1\mu_2\cdots}^{\nu_1\nu_2\cdots}
     \qquad\hbox{for}\quad \xi \in A \times B \,,
 }
where $\alpha$ is the conformal weight.  In particular, the metric tensor $g_{\mu\nu}$ automatically has weight $-2$, the invariant scalar $g$ has weight $0$, and the scalar $h$ has weight $1$.  Moreover, the four-velocity profile
 \eqn{umuChoice}{
  u_\mu = \pm {v_\mu \over \sqrt{-g^{\alpha\beta} v_\alpha v_\beta}}
 }
is (up to the sign ambiguity) the only unit vector field with conformal weight $-1$.  The sign ambiguity is fixed by requiring $u_t > 0$.  The projection tensor $P_{\mu\nu} = g_{\mu\nu} + u_\mu u_\nu$ has weight $-2$.  The stress tensor $T_{\mu\nu}$ needs to have weight $2$ in any conformal field theory: see e.g.~\cite{Gubser:2010ui} for a general argument to this effect.  Already from conformal inviscid hydrodynamics, where
 \eqn{TmunuForm}{
  T_{\mu\nu} = \epsilon u_\mu u_\nu + {\epsilon \over 3} P_{\mu\nu} \,,
 }
we see that $\epsilon$ must have weight $4$; this weight assignment doesn't change as one adds viscous corrections.  Recall that $u_\mu$ has already been completely fixed in \eno{umuChoice}, and that the energy density must take the form
 \eqn{epsilonForm}{
  \epsilon = h^4 \hat\epsilon(g) \,.
 }
The hydrodynamic equations $\nabla^\mu T_{\mu\nu} = 0$ must boil down to an ordinary differential equation for $\hat\epsilon(g)$.  Intuitively this is because the hydrodynamic equations are compatible with conformal invariance.

\subsubsection{Bjorken flow generalized to include transverse expansion}
\label{RadialResults}

As soon as we specify the four-parameter symmetry group $A \times B$, the machinery of the previous section can be brought to bear to reduce hydrodynamics to an ordinary differential equation.  Let's choose the $SO(3)$ group indicated in \eno{ToSOthree} for $A$, and of course beamline boost symmetry for $B$.  After a little experimentation, it is not hard to see that
 \eqn{gAndh}{
  g = {1 - q^2 \tau^2 + q^2 x_\perp^2 \over 2q\tau} \qquad\qquad 
    h = {1 \over \tau}
 }
where $\tau = \sqrt{t^2-x_3^2}$ is the usual Bjorken time and $x_\perp = \sqrt{x_1^2+x_2^2}$ is the transverse radius.  If one expresses $u^\mu$ in $(\tau,\eta,x_\perp,\phi)$ coordinates, where $\eta$ is spacetime rapidity and $\phi$ is the azimuthal angle around the beamline, then $u^\tau = \gamma_\perp$ and $u^\perp = v_\perp \gamma_\perp$ where $\gamma_\perp = 1/\sqrt{1-v_\perp^2}$ and the transverse radial velocity is
 \eqn{umu}{
  v_\perp = {2q^2 \tau x_\perp \over 1 + q^2 (\tau^2 + x_\perp^2)} \,.
 }
(The components $u^\eta$ and $u^\phi$ vanish by symmetry.)  The ordinary differential equation satisfied by $\hat\epsilon(g)$ as a consequence of energy conservation is
 \eqn{epsilonDiffEQ}{
  \hat\epsilon'(g) - {8g/3 \over 1+g^2} \hat\epsilon(g) = 0 \,,
 }
where viscous corrections have been neglected.  The final result for the energy density, from \eno{epsilonForm} together with the solution of \eno{epsilonDiffEQ}, is
 \eqn{epsilonFinal}{
  \epsilon = {\hat\epsilon_0 \over \tau^{4/3}}
    {(2q)^{8/3} \over \left[ 1 + 2q^2 (\tau^2 + x_\perp^2) + 
       q^4 (\tau^2 - x_\perp^2)^2 \right]^{4/3}} \,.
 }
First order viscous corrections can be included without too much difficulty \cite{Gubser:2010ze}.  The four-velocity is unaffected by viscous corrections: recall that it is entirely determined by the symmetry principles.  The energy density does receive corrections which can be expressed in closed form in terms of a hypergeometric function.

\subsubsection{Conformally adapted coordinates}
\label{ADAPTED}

There is a useful coordinate system \cite{Gubser:2010ui} in which the $SO(3)$ symmetry discussed so far becomes more manifest.  If we start with coordinates $(\tau,\eta,x_\perp,\phi)$, where $\tau$ is Bjorken time, $\eta$ is spatial rapidity, $x_\perp$ is distance from the beamline, and $\phi$ is azimuthal angle around the beamline, then the new coordinate system is $(\rho,\theta,\phi,\eta)$, where we define $\rho$ and $\theta$ via the equations
 \eqn{RhoTheta}{
  \sinh \rho = -{1 - q^2 \tau^2 + q^2 x_\perp^2 \over 2q\tau} \qquad
   \tan \theta = {2q x_\perp \over 1 + q^2 \tau^2 - q^2 x_\perp^2}
 }
Clearly, $\rho$ is a new version of the symmetry-invariant, timelike coordinate $g$ that we made heavy use of in section~\ref{BjorkenVariant}.  The angular coordinate $\theta$ is {\it not} the usual angle from mid-rapidity: that information is still carried by the spatial rapidity $\eta$.  Instead, $\theta$ parametrizes distance from the beampipe in a $\tau$-dependent way.  The standard flat metric on the future wedge of ${\bf R}^{3,1}$ can be expressed as
 \eqn{WedgeMetric}{
  ds^2 = -d\tau^2 + \tau^2 d\eta^2 + dx_\perp^2 + x_\perp^2 d\phi^2 = 
    \tau^2 d\hat{s}^2 \,,
 }
where
 \eqn{deSitter}{
  d\hat{s}^2 = -d\rho^2 + \cosh^2 \rho (d\theta^2 + \sin^2 \theta d\phi^2) + d\eta^2
 }
is the metric of $dS_3 \times {\bf R}$, where $dS_3$ is three-dimensional de Sitter space.  Now the $SO(3)$ symmetry acts by ordinary rotations on the $S^2$ which forms the constant $\rho$ time-slice of $dS_3$.  Boost symmetry acts by translations in the $\eta$ direction.  As we pass from $dS_3 \times {\bf R}$ quantities (hatted) to flat space quantities (unhatted), we must include factors of $\tau$ that account for the fact that these two spacetimes are only conformally equivalent:
 \eqn{epsilonAndu}{
  u_\mu = \tau {\partial \hat{x}^\nu \over \partial x^\mu} \hat{u}_\nu \qquad
  \epsilon = {\hat\epsilon \over \tau^4} \,.
 }
The advantage of the new coordinates is that the fluid's velocity field is obvious in $dS_3 \times {\bf R}$:
 \eqn{hatUValues}{
  \hat{u}_\rho = -1 \qquad \hat{u}_\theta = \hat{u}_\phi = \hat{u}_\eta = 0 \,.
 }
In other words, we have a static fluid in a time-dependent geometry, and we map it using \eno{epsilonAndu} to a time-dependent fluid in ordinary flat space.

Several points are now worth noting:
 \begin{itemize}
  \item The part of $dS_3 \times {\bf R}$ that maps to the future wedge of ${\bf R}^{3,1}$ is the {\it contracting} Poincar\'e patch.  This is obvious if we write the flat space metric as
 \eqn{dsFlatAgain}{
  ds^2 = \tau^2 \left( {-d\tau^2 + dx_\perp^2 + x_\perp^2 d\phi^2 \over \tau^2} + 
    d\eta^2 \right) \,,
 }
because the first term in parentheses is precisely this contracting patch in standard Poincar\'e coordinates.  In contrast, the coordinate system $(\rho,\theta,\phi)$ covers all of global $dS_3$.
  \item Finite chemical potentials can be included, even in the presence of viscous corrections, and closed form expressions can still be found for the hydrodynamic stress tensor provided the equation of state obeys conformal invariance.
  \item A gravity dual of the fluid flow can be found, similar to the boost-invariant flow of \cite{Janik:2005zt}, but preserving the enhanced $SO(3)$ little group symmetry of colliding null geodesics at $X_1=X_2=X_4=0$.
 \end{itemize}

\subsubsection{Perturbations around the $SO(3)$-invariant flow}
\label{ThreePERTURB}

A complete linear stability analysis was performed in \cite{Gubser:2010ui}, including first-order viscous corrections, with the result that the flow \eno{hatUValues} is stable in the regime where hydrodynamics is applicable.  We are unaware of any comparable stability analysis for semi-realistic heavy-ion flows.  Stability is interesting because it gives some indication that turbulence is not necessarily involved in the hydrodynamic phase of collisions.  This conclusion is intuitive in terms of $dS_3 \times {\bf R}$: the claim is simply that a stationary fluid in this geometry is stable against small perturbations.  Technically, the tools are simple: a scalar quantity such as the variation in energy density is expressed in separated form as $\delta(\rho,\theta,\phi) = R(\rho) Y_{\ell m}(\theta,\phi)$ (assuming that boost invariance is preserved by the perturbations), and then one develops ordinary differential equations for the functions such as $R(\rho)$ that capture the time dependence.

Perturbations to the $SO(3)$-invariant, boost-invariant flow were studied phenomenologically in \cite{Staig:2011wj}, with Glauber initial conditions; see also the recent work \cite{Shuryak:2013ke}.  A power spectrum emerges which is reminiscent of acoustic oscillations in the early universe.  This even makes sense since in the $dS_3 \times {\bf R}$ frame, the perturbations of interest actually are acoustic oscillations within a cosmological geometry.  The upshot is that there is a maximum in the power spectrum for azimuthal quantum number $m=3$, then a minimum around $m=7$, and a smaller maximum near $m=9$.  Modes of higher $m$ feel the effects of viscosity much more strongly, so it is crucial to treat the perturbations using linear Navier-Stokes rather than the Euler equations.  Indeed, modes with wave-number $k$ are suppressed by a factor
 \eqn{Psuppress}{
  P_k = \exp\left( -{2 \over 3} {\eta \over s} {k^2 t \over T} \right) \,.
 }

\subsubsection{Action of a complexified boost symmetry}
\label{COMPLEX}

Because deformations of the $ISO(2)$ symmetries of the transverse plane into $SO(3)$ were successful at introducing finite size and radial flow, it is natural to inquire whether some deformation of boost symmetry might lead to a phenomenologically appealing rapidity structure.  Rapidity profiles \cite{Murray:2004gh,Steinberg:2004vy,Alver:2010ck} at RHIC are in fact an embarrassment to the Bjorken picture: although the distribution of particles in pseudorapidity shows a nearly flat central region, full particle identification reveals that the rapidity distribution is in essentially perfect agreement with the hydrodynamical Landau model, in which $dN/dy$ has a Gaussian profile.  Nevertheless, the Landau model is rightly criticized for assuming the validity of hydrodynamics far before any known process in QCD could achieve local thermalization.  If there were a symmetry principle which would enforce a Landau-like rapidity profile---independent of the validity of the hydrodynamic approximation---it would explain a lot.  To this end, a deformation
 \eqn{Bdef}{
  B_{(3)} \to b \equiv B_{(3)} + \mathfrak{t}_3 T_{(3)}
 }
was considered in \cite{Gubser:2012gy}.  If $\mathfrak{t}_3$ is real, then $b$ is just a boost around a different point in ${\bf R}^{3,1}$, namely the point $t = -\mathfrak{t}_3$ and $\vec{x}=0$.  But if instead $\mathfrak{t}_3$ is imaginary, it was explained in \cite{Gubser:2012gy} how to pass through a formal construction of a complex stress tensor whose real part is conserved and shows global properties which are essentially what one wants for heavy ion collisions.  The construction is simple because it hinges on uniformly sending $t \to t+\mathfrak{t}_3$ in the treatment of inviscid Bjorken hydrodynamics.  Thus we have
 \eqn{uCform}{
  u^{\bf C}_\mu &= \left( -{t+\mathfrak{t}_3 \over \sqrt{(t+\mathfrak{t}_3)^2 - x_3^2}}, 0, 0, 
    {x_3 \over \sqrt{(t+\mathfrak{t}_3)^2 - x_3^2}} \right)  \cr
  \epsilon^{\bf C} &= {\epsilon^{\bf C}_0 \over ((t+\mathfrak{t}_3)^2-x_3^2)^{2/3}}  \cr
  T^{\bf C}_{\mu\nu} &= \epsilon^{\bf C} u^{\bf C}_\mu u^{\bf C}_\nu + 
    {\epsilon^{\bf C} \over 3} (g_{\mu\nu} + u^{\bf C}_\mu u^{\bf C}_\nu) \,,
 }
where $\epsilon_0^{\bf C}$ is a (complex) constant.  As promised,
 \eqn{TmnReal}{
  T_{\mu\nu} = \Re \{ T_{\mu\nu}^{\bf C} \}
 }
is real and conserved; moreover, provided $\arg \epsilon^{\bf C}_0 = \pi/3$ when $\arg \mathfrak{t}_3 = \pi/2$, $T_{\mu\nu}$ satisfies appropriate positive energy conditions in the future light-wedge of the collision.

\begin{figure}[t]
  \centerline{\includegraphics[width=4in]{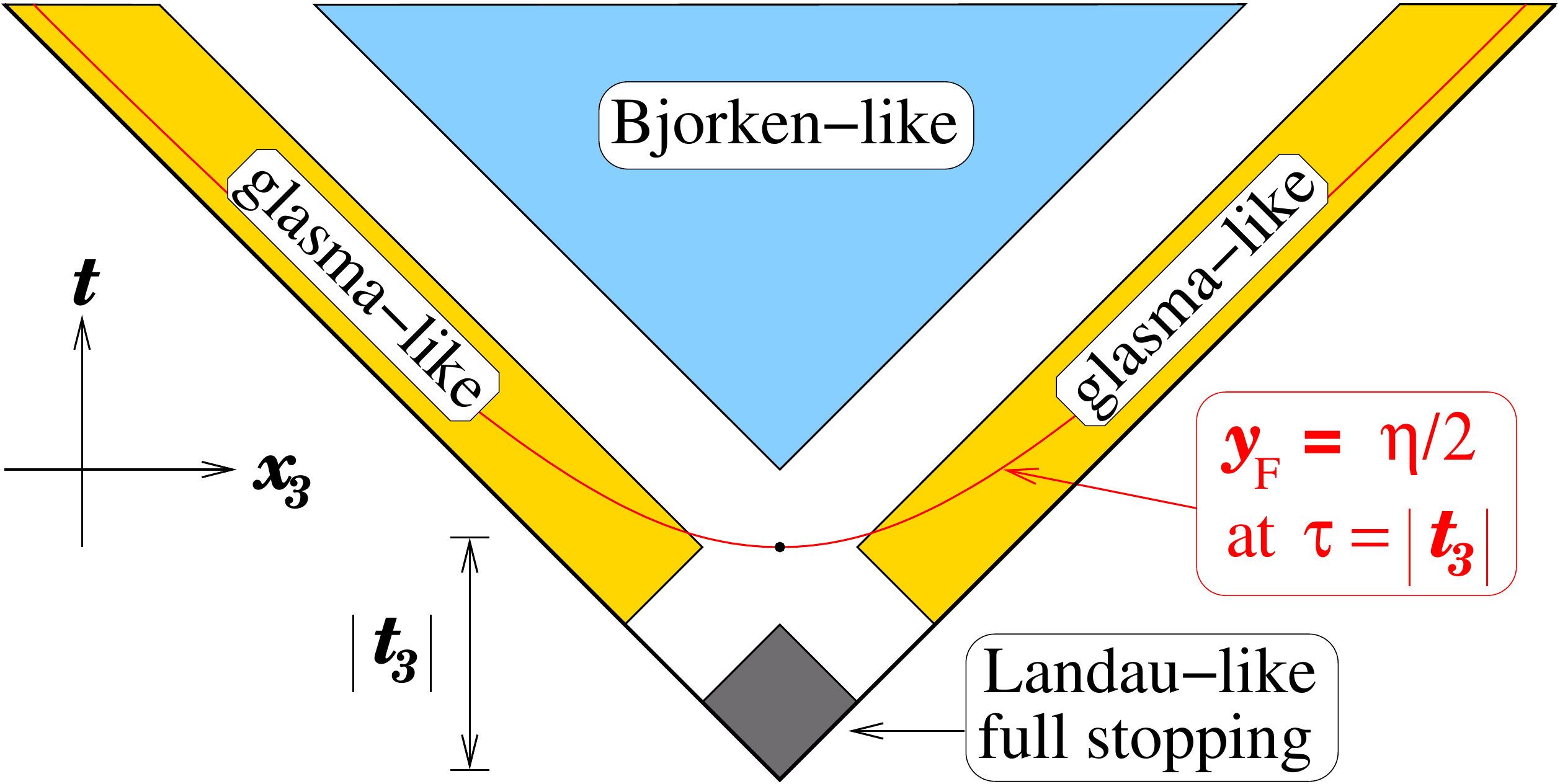}}
  \caption{A simplified cartoon of the rapidity structure of the complex deformation of Bjorken flow indicated in \eno{uCform}-\eno{TmnReal}, with $\arg\epsilon^{\bf C} = \pi/3$ and $\arg\mathfrak{t}_3 = \pi/2$.  From \cite{Gubser:2012gy}.}\label{CartoonWedge}
 \end{figure}

The future wedge divides up into regions illustrated in Fig.~\ref{CartoonWedge}: A Bjorken-like region at late times and central-to-moderate rapidities, with glasma-like regions at very forward rapidities and a Landau-like region at early times and rapidities not too large.  Bjorken-like means that standard Bjorken flow is recovered; glasma-like means that the stress tensor approaches the form $T^\mu{}_\nu = \diag\{ \epsilon,-\epsilon,-\epsilon,\epsilon \}$ characteristic of longitudinal color-electric and color-magnetic fields; and Landau-like means that there is almost full stopping, with hydrodynamic constitutive relations satisfied.  The Landau-like region is a surprise, and the validity of hydrodynamics doesn't last; instead it is asymptotically recovered in the Bjorken-like region, and not recovered at all in the glasma-like region.  Another interesting and simple feature of the stress tensor defined by \eno{TmnReal} is that the Landau frame may be defined everywhere in the future light-wedge, and doing so leads to a relation
 \eqn{yVersusEta}{
  y = {\eta \over 2} \qquad\hbox{when}\qquad \tau = |\mathfrak{t}_3|
 }
between the fluid rapidity $y$ and the spatial rapidity of its location in the wedge.  Compared to Bjorken's relation $y=\eta$, the result \eno{yVersusEta} indicates more clustering of particles near central rapidity.  Indeed, if one hadronizes the Bjorken region, with some assumptions spelled out in \cite{Gubser:2012gy}, the resulting $dN/dy$ is peaked at central rapidity with strong decay at forward rapidities.  The decay is in fact somewhat faster than it should be to make a really successful comparison with data.  However, it's possible that improved hadronization and proper inclusion of the glasma-like regions will improve the fit.

The construction \eno{TmnReal} is somewhat {\it ad hoc}; however, an appealing feature is that the complexified boost symmetry \eno{Bdef} commutes with wave operators.  This is because it is a linear combination of a boost and a translation, and hence in the complexified algebra of the Poincar\'e group.  It is interesting to inquire whether  solutions to wave equations can be usefully classified in terms of their content under complexified boost symmetry.

\section{Lessons from the AdS/CFT for non-equilibrium dynamics}
\label{noneqdynamics}

\def\st{\begin{equation}}
\def\stp{\end{equation}}
\def\ed{{\epsilon}}
\def\pr{{\mathcal P}}
\def\q{{\bm q}}
\def\k{{\bm k}}
\def\x{{\bm x}}
\def\O{{\mathcal O}}
\def\llangle{\left\langle}
\def\rrangle{\right\rangle}

\def\half{{\textstyle\frac{1}{2}}}
\def\third{{\textstyle\frac{1}{3}}}
\def\quarter{{\textstyle\frac{1}{4}}}
\def\fifth{{\textstyle\frac{1}{5}}}
\def\Eq#1{Eq.~(\ref{#1})}
\def\App#1{Appendix~\ref{#1}}
\def\Fig#1{Fig.~\ref{#1}}
\def\Sect#1{Section~\ref{#1}}
\def\Ref#1{Ref.~\cite{#1}}

From a theoretical perspective, it is 
important to understand the dynamics of the equilibration 
process, and the imprints of the non-equilibrium initial state 
on the  subsequent evolution of the system. It is here that holography can  
be a useful tool, offering a foil to perturbative descriptions
of the thermalization process. 

Of necessity, much of the literature on equilibration in holographic theories goes beyond analytically tractable models and relies upon numerical methods in the bulk gravitational theory.  So we will start in section~\ref{RELATIVITY} by reviewing three of the main numerical methods in active use.  We will then focus in on a set of work 
which simulates with holography the evolution of a non-equilibrium
boost invariant initial condition of infinite transverse extent \cite{Chesler:2009cy, Heller:2011ju}.  Initially there were two approaches to creating the 
non-equilibrium state: first by turning on a boost invariant source
on the boundary \cite{Chesler:2009cy}, and second by setting up an 
ensemble of initial conditions and studying the subsequent dynamics \cite{Heller:2011ju}.  We will focus on the latter approach \cite{Heller:2011ju} and then indicate the common conclusions.

\subsection{Numerical relativity in anti-de Sitter space}
\label{RELATIVITY}

\def\ssp{\phantom{a}}
\def\R{\mathcal R}

Recent years have seen increased effort to use methods of numerical relativity to extract useful information from the gauge-string duality for heavy-ion physics.  The recurring theme in this work has been rapid thermalization.  The aim of this section is to provide a summary of some of the different numerical schemes which have been employed and also to give some indications of results obtained.  We will focus on the generalized harmonic gauge approach of \cite{Bantilan:2012vu}, the null coordinate method of \cite{Chesler:2010bi}, and the ADM formalism of \cite{Heller:2012je}.

\subsubsection{Generalized harmonic gauge}
\label{HARMONIC}

The general harmonic gauge approach relies upon a specific method for making a gauge choice, together with two modifications of the Einstein equations which fix diffeomorphism freedom and result in numerically tractable hyperbolic differential equations.  To understand the gauge choice, consider first the identity 
 \eqn{xIdent}{
  \square x^\mu = -g^{\alpha\beta} \Gamma^\mu_{\alpha\beta} \,.
 }
Diffeomorphism freedom is realized by making coordinate shifts $x^\mu \to x^\mu + \xi^\mu$, where $\xi^\mu$ is allowed to vary over the spacetime.  At least locally, this is enough freedom to impose harmonic gauge, $\square x^\mu = 0$.  Once coordinates are chosen, the plan is to specify Cauchy data on the slice $t=0$, and then to evolve forward time-slice by time-slice.  This doesn't quite work because, in general, $t$ doesn't stay timelike.  A highly successful, broadly applicable method for simulating general relativity in asymptotically flat space \cite{Pretorius:2004jg,Pretorius:2005gq} has emerged from a relaxation of the harmonic gauge condition together with some further technical tricks.  We provide a brief summary of this method at the end of this subsection.  But first, let us survey the broad outlines of how the method is applied to geometries in $AdS_5$ involving black hole horizons, together with the conclusions reached in \cite{Bantilan:2012vu}.

Regions of strong curvature develop during any numerical study which involves gravitational collapse and/or black hole formation.  The strategy for dealing with this is to excise regions of the geometry which are behind black hole horizons.  More precisely, one looks for apparent horizons, defined as locations where one of the forward directed null directions points inward and the other points tangent to the apparent horizon.  On fairly general grounds, event horizons must enclose all apparent horizons (a fact used heavily in the trapped surface estimates of total multiplicity summarized in section~\ref{TRAPPED}).  The key property of an apparent horizon is that nothing that happens inside can propagate causally to the nearby outside geometry.  In generalized harmonic schemes, horizon excision doesn't have to be very precise: one needs only to make sure that enough geometry is excised to avoid excessively strong curvatures, while preserving the property that information about the excised regions is not needed in order to correctly evolve the geometry that remains.  In other words, the excision surface is chosen to be slightly inside the outermost trapped surface, such that all characteristics point inward.

Significant technical issues still have to be worked out in order to successfully apply the generalized harmonic gauge method in an asymptotically anti-de Sitter geometry.  The metric components diverge near the boundary, so it is advantageous to work not with the metric components themselves, but with quantities proportional to the deviation of the geometry from pure $AdS_5$.  Likewise, asymptotic behavior of source functions near the boundary must be treated carefully.  Another challenge is to specify initial conditions in a physically interesting way.  In \cite{Bantilan:2012vu}, an initial time-slice was set up which corresponds to a perfectly isotropic fluid on the boundary at $t=0$ which is very unevenly distributed over the boundary geometry, corresponding to a distorted black hole in $AdS_5$.  The initial time-slice in the bulk is conformal to $AdS_5$, and it is supported by a scalar field which is sufficiently focused at the center of $AdS_5$ that its subsequent disappearance behind a black hole horizon is very rapid.  The upshot of the study of \cite{Bantilan:2012vu} is that when the fluid begins in an isotropized state, it remains remarkably close to hydrodynamical throughout its evolution---particularly when first and second derivative corrections to the ideal hydrodynamic stress tensor are included.  These results support the notion that strongly coupled gauge theories are surprisingly good at maintaining near-hydrodynamic stress tensors even in dynamical situations that are far from equilibrium.  The simulations in \cite{Bantilan:2012vu} were performed in global $AdS_5$ with the imposition of an $SO(3)$ symmetry of the type discussed in section~\ref{BJORKEN}.  This means that the numerical problem is $2+1$-dimensional.  In principle, the underlying methods generalize immediately to less symmetrical situations.  The boundary of global $AdS_5$ is $S^3 \times {\bf R}$, but after a conformal mapping to Minkowski space ${\bf R}^{3,1}$, the initial state corresponds to a pancake which has finite extent in the transverse directions and is highly compressed in the longitudinal direction.  Altogether, the setup is similar to the Landau model \cite{Landau:1953gs}, but it is demonstrated rather than assumed that the system remains close to hydrodynamical equilibrium throughout its evolution.

Now let's turn back to an explanation of generalized harmonic gauge.  As previously mentioned, the difficulty with harmonic gauge is that there is no guarantee that $x^0 \equiv t$ will be everywhere timelike, and if it does not, standard numerical methods for advancing from one ``time''-slice to the next will break down.  The generalized harmonic gauge scheme starts by choosing five so-called ``source functions'' $H^\mu$, defining
 \eqn{Cdef}{
  C^\mu \equiv H^\mu - \square x^\mu \,,
 }
and demanding $C^\mu = 0$ as the gauge condition.  Note that $H^\mu$ need only be prescribed timeslice-by-timeslice; indeed, one need only have a definite set of equations that can be solved on each new time-slice for the source functions.  Much of the art of the generalized harmonic gauge method comes down to choosing these equations cleverly.  

The Einstein equations can be cast in the form
 \eqn{EinsteinReversed}{
  R_{\mu\nu} = \bar{T}_{\mu\nu} \equiv 
    8\pi \left( T_{\mu\nu} - {1 \over 3} g_{\mu\nu} T^\alpha{}_\alpha \right) \,,
 }
where we have set $G_5 = 1$ and absorbed the cosmological constant into the definition of $T_{\mu\nu}$.  More explicitly, \eno{EinsteinReversed} reads
 \eqn{ExplicitER}{
  -{1 \over 2} g^{\alpha\beta} g_{\mu\nu,\alpha\beta} - g^{\alpha\beta}{}_{,(\mu}
    g_{\nu)\alpha,\beta} - \nabla_{(\mu} \square x_{\nu)} - 
    \Gamma^\alpha_{\beta\mu} \Gamma^\beta_{\alpha\nu} = \bar{T}_{\mu\nu} \,,
 }
where $(\mu\nu) = {1 \over 2} (\mu\nu + \nu\mu)$.  If it weren't for the third term in \eno{ExplicitER}, the Einstein equations would take the form $\square g_{\mu\nu} + \ldots = \bar{T}_{\mu\nu}$, where $\ldots$ represents terms which involve at most first derivatives of the metric.  Such a form would be good for numerical work because it is linear in second derivatives and hyperbolic.  It can be achieved by subtracting $\nabla_{(\mu} C_{\nu)}$ from the Ricci tensor:
 \eqn{FirstModification}{
  R_{\mu\nu} - \nabla_{(\mu} C_{\nu)} = 
  -{1 \over 2} g^{\alpha\beta} g_{\mu\nu,\alpha\beta} - g^{\alpha\beta}{}_{,(\mu}
    g_{\nu)\alpha,\beta} - \nabla_{(\mu} H_{\nu)} - 
    \Gamma^\alpha_{\beta\mu} \Gamma^\beta_{\alpha\nu} = \bar{T}_{\mu\nu} \,.
 }
Using \eno{FirstModification} and the identity $\nabla^\mu R_{\mu\nu} = {1 \over 2} \partial_\nu R$, one can show that
 \eqn{SquareC}{
  \square C_\mu = -C^\mu \partial_{(\mu} C_{\nu)} - C^\mu \bar{T}_{\mu\nu} \,.
 }
We can arrange to have $C^\mu=0$ and $\partial_t C^\mu=0$ on the initial timeslice.  It follows then from \eno{SquareC} that $C^\mu = 0$ everywhere.  In other words, correct propagation of the constraint is implied by the equations of motion, in either the form \eno{FirstModification} or the form \eno{SecondModification}.  But it has been found that when \eno{FirstModification} is discretized, round-off error accumulates quickly (in fact exponentially) to make $C^\mu$ non-zero.  This is the phenomenon of constraint-violating modes.  The solution is to add some further terms to the Einstein equations:
 \eqn{SecondModification}{
  R_{\mu\nu} - \nabla_{(\mu} C_{\nu)} - \kappa \left( 2 n_{(\mu} C_{\nu)} - 
    (1+P) g_{\mu\nu} n^\alpha C_\alpha \right) = \bar{T}_{\mu\nu} \,.
 }
where $\kappa$ is a negative constant and $P \in [-1,0]$ is another constant.  In \cite{Bantilan:2012vu} the choices $\kappa=-10$ and $P=-1$ were made.  To obtain the final form of the evolution equations, one uses the definition \eno{Cdef} to eliminate $C_\mu$ in favor of $H_\mu$ in \eno{SecondModification}, as was done explicitly in \eno{FirstModification} above.  One then solves the evolution equations, together with the gauge evolution equations for the source functions, for $g_{\mu\nu}$ and $H^\mu$.  All the equations for metric components evidently have the schematic form $\square X = \ldots$, where $\ldots$ means terms involving at most first derivatives.  Of course, $C_\mu$ is required to be small, and to show signs of converging to $0$ with decreasing grid spacing, in order to have a trustworthy solution.  In \cite{Bantilan:2012vu}, the method of spatial discretization was second order finite-difference stencils, on which we will provide a bit more detail in section~\ref{nullcoord}.  Time-stepping was accomplished in \cite{Bantilan:2012vu} using an iterative Newton-Gauss-Seidel relaxation procedure.

\subsubsection{A null coordinate method}
\label{nullcoord}

An alternative approach, pursued for example in \cite{Chesler:2010bi}, is based on choosing the following specific form for the metric:
 \eqn{EFForm}{
  ds^2 = -A dv^2 + \Sigma^2 \left[ e^B d\vec{x}_\perp^2 + e^{-2B} dx_3^2 \right] + 
    2 dv (dr + F dx_3) \,,
 }
where $\vec{x}_\perp = (x_1,x_2)$ and $A$, $B$, $\Sigma$, and $F$ can be functions of $v$, $r$, and $x_3$.  Explicit forms of the equations of motion are then found.  Their form is sufficiently complicated as to be unenlightening to record in full.  We will comment on the broad outlines of the solution strategy for the equations of motion after presenting, in the next paragraph, a pedagogical example of a similar strategy applied to a simpler problem.

Consider a massless scalar in a fixed $AdS_5$-Schwarzschild background, which we write as
 \eqn{EFAdSSch}{
  ds^2 = {L^2 \over z^2} \left[ -\left( 1-{z^4 \over z_H^4} \right) dv^2 - 2 dv dz + d\vec{x}^2 \right] \,,
 }
where $\vec{x} = (x_1,x_2,x_3)$.  Let's set $L=z_H=1$ as an additional simplification.  To recover the more familiar form of $AdS_5$-Schwarzschild, one may define $t$ through
 \eqn{tDiffEQ}{
  dt = dv + {dz \over 1 - z^4} \,,
 }
and then eliminate $v$ in favor of $t$ throughout.  Note that setting $z=0$ inside the square brackets in \eno{EFAdSSch} results in the metric of ${\bf R}^{3,1}$, with $v$ playing the role of time.  Thus $v$ is in fact a perfectly good timelike variable on the boundary of $AdS_5$-Schwarzschild, and it is privileged among bulk extensions of boundary time by being a null coordinate which is constant on trajectories which fall inward, away from the boundary.  An additional favorable feature is that all metric components remain finite at the black hole horizon, which is located at $z=z_H$.

In the geometry \eno{EFAdSSch}, the scalar equation of motion $\square\phi = 0$ reads
 \eqn{EFSquarePhi}{
  (3 - 2z \partial_z) \partial_v \phi = 
    \left[ (3+z^4) \partial_z - z(1-z^4) \partial_z^2 - z \partial_{x_3}^2 \right] \phi \,.
 }
Notably, this is first order in $v$.  In broad terms, the plan is to specify $\phi$ along a $v=const$ slice of $AdS_5$-Schwarzschild and then find $\partial_v \phi$ in order to advance to the next timestep.  From \eno{EFSquarePhi} it's clear that in order to do this, we need only invert the differential operator $(3-2z\partial_z)$, which is to say we solve a linear {\it ordinary} differential equation for each value of $x_3$ and at each timestep.

In practice, there are some additional important details.  First, we must specify boundary conditions.  Assuming there is no deformation of the Lagrangian by the operator dual to $\phi$, the boundary conditions on $\phi$ are that it falls to $0$ as $z^4$ when $z \to 0$.  This fourth-order behavior is hard to see numerically, so it is more efficient to define
 \eqn{phiTildeDef}{
  \tilde\phi \equiv {\phi \over z^4} \,.
 }
An additional trick is to introduce
 \eqn{dPlusPhi}{
  d_+ \tilde\phi \equiv \partial_v \tilde\phi - {1-z^4 \over 2} \partial_z \tilde\phi \,,
 }
which is the partial derivative of $\tilde\phi$ along the lightlike direction in the $v$-$z$ plane other than the $v$ direction itself.  That is, we derive from \eno{EFSquarePhi} an equation of the form
 \eqn{EFtildeForm}{
  {\cal L} d_+ \tilde\phi = {\cal M} \tilde\phi \,,
 }
where ${\cal L}$ is a linear differential operator involving only $z$ and $\partial_z$, and ${\cal M}$ is a linear differential operator expressible in terms of $z$, $\partial_z$, and $\partial_{x_3}^2$.  To evolve forward one timestep, we first invert ${\cal L}$ to get $d_+ \tilde\phi = {\cal L}^{-1} {\cal M} \tilde\phi$---imposing appropriate boundary conditions on $d_+ \tilde\phi$ so that the inverse is uniquely determined---and then obtain $\partial_v \tilde\phi$ straightforwardly from the definition \eno{dPlusPhi}.

Before turning back to the full gravitational problem, let us briefly discuss numerical methods for handling differential equations of the form \eno{EFtildeForm}.  Because the differential equation is linear and there is translation invariance in the $x_3$ direction, we could decompose $\tilde\phi$ into Fourier modes proportional to $e^{ik_3 x_3}$ and solve the equations mode-by-mode.  Methods based on truncated Fourier series converge very quickly; however, they don't immediately generalize to problems with non-linear effects.  Therefore the method of choice is the pseudospectral representation of derivative operators.  Reviews of these methods can be found in \cite{Boyd00,Trefethen00}; however, we will give sufficient indications in the next paragraph to orient the reader unfamiliar with this approach.

Consider a function $f(x)$ of a periodic variable $x \sim x + L$ sampled at $N$ evenly spaced points (the so-called ``collocation points'')
 \eqn{CollationX}{
  x_n = na \qquad\hbox{where}\qquad a = {L \over N} \,.
 }
and $n=0,1,2,\ldots,N-1$.  Knowing the sampled values $f(x_n)$, how do we calculate the approximate derivative $f'(x_n)$?  The simplest method in active use is the second order stencil 
 \eqn{fnSecond}{
  f'(x_n) = {f(x_{n+1}) - f(x_{n-1}) \over 2a} + {\cal O}(a^2) \,.
 }
(For example, precisely this stencil was employed for spatial discretizations in \cite{Bantilan:2012vu}).  A fourth-order scheme would have remainders scaling as ${\cal O}(a^4)$ for small $a$.  Pseudospectral methods arrange for the remainder to scale to zero faster than any power of $a$, provided $f(x)$ itself is infinitely differentiable.  Conceptually, the pseudospectral method amounts to taking derivatives in momentum space and then passing immediately back to position space.  Explicitly, we first perform a discrete Fourier transform:
 \eqn{TildeF}{
  \tilde{f}_N(k_j) = 
    {1 \over N} \sum_{n=0}^{N-1} f(x_n) e^{-i k_j x_n} \qquad\hbox{for}\qquad 
    k_j = {2\pi j \over L} \,,
 }
where $j$ runs from $-N/2$ to $N/2$, with $N$ assumed even.  Next we take the derivative in Fourier space: $\tilde{f}'_N(k_j) \equiv i k_j \tilde{f}_N(k_j)$.  Finally we perform a discrete inverse Fourier transform to find 
 \eqn{BackToFp}{
  f'_N(x_n) = \sum_{j=-N/2}^{N/2} {1 \over c_j} i k_j \tilde{f}_N(k_j) e^{i k_j x_n} \,,
 }
where $c_j=1$ except for $c_{\pm N/2}=2$.  The true derivative $f'(x_n)$ equals $f'_N(x_n)$ plus corrections vanishing faster than any power of $a$ because of the convergence properties of discrete Fourier transforms.  Note that all the computational steps are linear operations; thus one can write
 \eqn{PseudospectralDerivative}{
  f'_N(x_n) = \sum_{m=1}^N {\cal M}_{nm}^{(N)} f(x_m) \,,
 }
where ${\cal M}_{nm}^{(N)}$ is a matrix which can be worked out in advance (and for which closed form expressions are known).  Remarkably, ${\cal M}_{nm}^{(n)}$ generically has all entries non-zero: thus differentiation is a non-local operation.  One might fear that this property is in conflict with the locality of relativistic theories; however, experience seems to show that convergence to the correct, local dynamics is not problematic.

The discussion above shows how to handle $x_3$ derivatives, since we can certainly compactify this direction on a circle.  Using pseudospectral methods in the $z$ direction, where boundaries are an essential part of the story, requires one more trick: instead of using discrete Fourier transforms, one expands a function $f(z)$ in a basis of $N$ Chebyshev polynomials, which have the form $T_k(z) = \cos k\theta$ where $\theta = \cos^{-1} (2z-1)$, so that the boundaries are at $z=0$ and $z=1$.  Instead of sampling at equally spaced points in $z$, one samples at points equally spaced in $\theta$.  Expressions analogous to \eno{PseudospectralDerivative} can be derived, and the convergence properties are again better than algebraic provided $f(z)$ is infinitely differentiable.  Both the horizon at $z=1$ and the boundary at $z=0$ are included explicitly in the list of collocation points: thus it is essential to define fields in such a way that they are explicitly finite at these endpoints.

To apply pseudospectral methods to the classical scalar field theory problem \eno{EFtildeForm}, one finds matrix versions of the differential operators ${\cal L}$ and ${\cal M}$ by replacing $\partial_{x_3}$ and $\partial_z$ by their pseudospectral approximations.  The matrix for ${\cal L}$ will be non-invertible because there is no information about boundary conditions included in it.  A standard approach for solving this is so-called boundary bordering, whereby the rows of ${\cal L}$ corresponding to $z=0$ are explicitly replaced by linear equations that enforce the desired boundary conditions.

Returning now to the problem of solving gravity in an asymptotically anti-de Sitter background: The form of the metric \eno{EFForm} allows a treatment remarkably similar to the scalar field.  One can express Einstein's equations in a form which reads, schematically,
 \eqn{EinsteinForm}{
  {\cal L}_\ell X_\ell = S_\ell \qquad\hbox{for}\qquad \ell = 2,3,4,5,6 \,.
 }
Here $X_\ell = (B,\Sigma,F,d_+\Sigma,d_+B,A)$ (so that $X_1 = B$, $X_2 = \Sigma$, etc.), and in general $d_+ X = \partial_v X + {1 \over 2} A \partial_r X$.  The source term $S_\ell$ depends only on the {\it previous} $X_\ell$ and their spatial derivatives: for example, $S_3$ depends only on $X_1$ and $X_2$, not $X_3$ or $X_4$.  In general, the linear operators ${\cal L}_\ell$ also have coefficients depending on the previous $X_\ell$.  The upshot is that once one specifies $B$ at a given timestep, one can solve the equations \eno{EinsteinForm} in order, with appropriate boundary conditions, and once they are all solved one can resort to the definition of $d_+ B$ to extract $\partial_v B$ and thereby take one timestep.  For actual computations, the equations \eno{EinsteinForm} should be rewritten in terms of a radial variable $z$ which runs from $0$ at the boundary to $1$ at the horizon, with field redefinitions such that all quantities are explicitly finite at the boundary.  A particularly subtle issue is the choice of boundary conditions at the horizon.  This is because the horizon is not known in advance, but must be determined timestep-by-timestep.  The approach used in \cite{Chesler:2010bi} is to find a differential equation satisfied at the apparent horizon (cast in terms of vanishing expansion) and use it as a boundary condition at $z=1$.

The null-coordinate method described in
this section has been used 
to simulate
the formation a thermal black hole after driving vacuum AdS out of 
equilibrium \cite{Chesler:2008hg}.  All of the results of \Sect{transverdynamics} investigating the 
transverse and longitudinal dynamics of the collision have also been
obtained with this method \cite{Chesler:2010bi,vanderSchee:2012qj}.

\subsubsection{An ADM formulation} 
\label{ADM}

Finally, we will discuss an ADM \cite{Arnowitt:1962hi}  formulation of numerical 
relativity that has been used to simulate the boost invariant 
plasma of infinite transverse extent, which will be  described in \Sect{Hydroization} \cite{Heller:2012je}. 
In the numerical study of this problem,
the initial conditions  for the gravitational fields at Bjorken $\tau=0$  are a 
function of the holographic coordinate only, $g_{\mu\nu}(0,u)$. Specifying these
initial conditions for the metric amounts to specifying the initial state  
of the field theory. The gravitational fields are then evolved in the temporal
coordinate $t$ (which equals the proper time $\tau=\sqrt{t^2-x_3^2}$ on the boundary), and eventually form  thermal AdS, the gravitational dual of equilibrated plasma.
The more refined variants of the ADM equations, known as the BSSN equations
\cite{Shibata:1995we,Baumgarte:1998te}, have so far not been used to simulate asymptotically AdS spacetimes. 
The BSSN formulation is generally competitive with the  generalized harmonic
coordinate schemes described in the previous subsection
\cite{Centrella:2010mx,Zilhao:2013gu}, and  both schemes were used in the
breakthroughs of 2005 \cite{Pretorius:2005gq,Campanelli:2005dd,Baker:2005vv}.

In an ADM formulation of numerical  relativity, 
the initial data is specified on a spatial surface, and the 
Einstein equations provide an update rule for temporal evolution.
The induced metric of each spatial slice is 
\st
   \gamma_{ab} = g_{ab} + n_{a} n_{b} \, , 
\stp
where $n^{a}$ is the normal to the  hypersurface, $n^2 = -1$, and  
the extrinsic curvature  of the hypersurface is
\st
K_{ab} = -\frac{1}{2} \mathcal L_{n} \gamma_{ab} =  -\gamma_{a}^{\;c} \gamma_{b}^{\; d} \nabla_{(c} n_{d)} \, ,
\stp
where $\mathcal L_n$ is the Lie derivative along the normal
direction, $n^a$.
Introducing coordinates $y^i = (\eta,x^1,x^2,u)$ for the spatial slice, the metric in the ADM formalism is parametrized by the lapse and shift,
$\tilde \alpha$ and $\beta^i$,  
\st
 ds^2 = -\tilde \alpha^2 dt^2  + \gamma_{ij} (dy^i + \beta^i dt) (dy^j + \beta^j dt) \, .
\stp
$t$ is the temporal coordinate in the  gravitational
theory, and $\alpha$, $\beta^{i}$ and $\gamma_{ij}$ are functions 
of $t$ and the holographic coordinate, $u$.
The lapse and shift describe how the spatial slices
fit together to provide a foliation of space-time. 
In \Ref{Heller:2012je} the shift was set to zero $\beta^i=0$, though
it should be remarked that most of the current advances in numerical
relativity using the BSSN formalism exploit the shift   
to select a numerically reasonable gauge  based on
the local derivatives of the induced metric, {\it e.g.} $\gamma^{lm} \Gamma_{lm}^i$ \cite{Alcubierre:2002kk,Campanelli:2005dd,Baker:2005vv}. Thus, in more refined BSSN treatment of boost invariant plasma, $\beta^{u}$ would 
be non-zero and could adapt the temporal coordinate to changes in 
the bulk geometry.

Given the symmetries of boost invariant Bjorken flow,
the metric, the extrinsic curvature, and the lapse are parametrized by 
seven functions of $t$ and $u$ 
\begin{subequations}
\label{metric_adm}
\begin{align}
   \gamma_{ij} =& {\rm diag}\left[\gamma_{\eta\eta}, \gamma_{x_1x_1}, \gamma_{x_2x_2}, \gamma_{uu} \right] \, , \\
    =& {\rm diag}\left[ \frac{t^2 a^2(u) b^2(t,u)}{u}, \frac{c^2(t,u)}{u}, \frac{c^2(t,u)}{u}, \frac{d^2(t,u) }{4u^2} \right] \, , \\
   K_{ij} = &{\rm diag}\left[K_{\eta\eta}, K_{x_1x_1}, K_{x_2x_2}, K_{uu} \right] \, , \\
   =& {\rm diag}\left[ \frac{t a(u) L(t,u)}{\sqrt{u}}, \frac{M(t,u}{\sqrt{u}}, \frac{M(t,u)}{\sqrt{u}}, \frac{P(t,u)}{4 u\sqrt{u} } \right] \, \, ,  \\
   \tilde \alpha(t,u) =& \frac{a(u) \alpha(t,u) }{\sqrt{u} } \, .
\end{align}
\end{subequations}
This parametrization is chosen so that  the functions  $a, b, c, d, L, M,
P$ are regular at the boundary.  We note that a time independent function
$a(u)$ has been factored out of the lapse for reasons described below.  Our
next goal is to record the evolution equations for these functions following
the standard ADM-York treatment. 

The five dimensional curvature tensor, $R^{a}_{\ssp bcd}$,  can be 
decomposed into a four dimensional curvature tensor describing the intrinsic
geometry of a spatial slice and products of the extrinsic curvature \cite{BigBlackBook}
\st
\gamma^{a_1}_{\ssp a_2} \gamma^{b_2}_{\ssp b_1}  \gamma^{c_2}_{\ssp c_1} \gamma^{d_2}_{\ssp d_1} \, R^{a_2}_{\ssp b_2c_2d_2} = \mathcal R^{a_1}_{\ssp b_1c_1d_1} + K^{a_1}_{\ssp c_1} K_{b_1 d_1} - K^{a_1}_{\ssp d_1} K_{c_1 b_1}  \, ,
\stp
which is known as the Gauss equation.
$\R^{a}_{\ssp bcd}$ is the curvature intrinsic to the spatial hyper-surface, and can be  determined from the induced metric. 
Using this identity in the Einstein equations yields two constraint equations, known as  the Hamiltonian and momentum constraints, which are analogous to Gauss' law in electrodynamics
    \begin{align}
       \mathcal R + K^2 - K_{ab} K^{ab} =& 16\pi G_N \rho \, , \\
       D_b K^{b}_{\ssp a}  - D_a K  =& 8\pi G_N j_{a} \, .
    \end{align}
In these formulas $\rho$ and $j_{a}$ are the projections of  the 
stress tensor, $\rho \equiv T_{ab} n^{a} n^b$ and $j_{c} \equiv -T_{ab} n^a \gamma^{b}_{\ssp c} $,  and $D_{a}$ denotes the  covariant derivative 
with respect to the induced metric, $D_{a} \equiv \gamma_a^{\ssp b}\nabla_b$.

The remaining evolution equations for the extrinsic curvature and 
the induced metric  are found by projecting  their Lie derivatives 
in the temporal coordinate\footnote{Note, we have assumed that the shift $\beta^i = 0$  when writing these equations.}
\begin{align}
   \partial_t \gamma_{ab} =&  -2\tilde\alpha K_{ab}  \, ,   \\
   \partial_t K_{ab} =&  -D_{a} D_{b} \tilde\alpha + \tilde\alpha(\R_{ab}  - 2 K_{ac} K^c_{\ssp b} + K_{ab} K) -8\pi G_{N} \tilde \alpha \left[ S_{ab} + \frac{\rho - S}{d-1} \gamma_{ab} \right] \, , 
\end{align}
where $S_{ab} \equiv T_{cd}\gamma^{c}_{\ssp a} \gamma^{d}_{\ssp b}$ with  $S\equiv g^{ab} S_{ab}$, and $d=3+1$ is the number of space-time dimensions in the boundary theory.
So far the setup  parallels the standard treatments of the ADM-York 
formalism except that the five dimensional 
stress tensor $T_{ab}$ is determined by the radius of curvature of the 
AdS space
\st
T_{ab} =  \frac{d(d-1)}{16\pi G_N L^2} g_{ab} \, . 
\stp
As we will see, both the boundary and the infrared require special treatment 
when simulating asymptotically AdS geometries.

In general,  the strong curvature near the boundary did not seem to create
difficulties.  It was
necessary, however, to analyze the equations analytically near the boundary in
order to determine temporal update rules for the boundary fields. 
The infrared cutoff in the bulk required additional considerations.
Indeed, the authors implemented the IR-cutoff by parametrizing the lapse as
\st
\tilde \alpha(t,u) =  \frac{a(u) \alpha(t,u) }{\sqrt{u} } \, ,  \quad \mbox{with} \quad a(u) = \cos\left(\frac{\pi}{2}\frac{u}{u_o} \right)  \, .
\stp
The leading factor, $\cos(\pi u/2 u_o)$,  freezes space-time
 at $u = u_o$, since initial data with $u> u_o$ does not influence the  simulation.
As long as $u_o$ is within the event horizon this choice is acceptable for all times, and
the simulation can be run until a singularity is formed in the computational domain.
In practice, $u_o$ was first chosen somewhat arbitrarily, and then the geometry was evolved for modest simulation times. This exploratory run   was
used to estimate the ($u$-coordinate) location of the true event horizon at $t=0$, which was set to $u_o$ in subsequent runs.
In this way a singularity which is inside the event horizon does not form within the computational domain, and 
the late (boundary) time dynamics can be studied.
Presumably, an excision technique could have been used to excise the singular geometry, and obviate this two step procedure.

The final factor $\alpha(t,u)$ was  chosen by trial and error as the standard
choices of the BSSN scheme, {\it e.g.} $ 1 + \log \det \gamma_{ij}$, did not lead to stable results.
The lapse measures how the spatial slices at fixed time fit together to foliate the space-time geometry, and choosing the lapse amounts to selecting a numerically satisfactory coordinate system  based on the induced metric.  
The simplest reasonable choice (for example ${\it lapse 4}$ of \Ref{Heller:2012je}),  takes
$\alpha(t,u) = d(t,u)/d(0,u)$.  This choice is somewhat ad hoc
and will require further investigation.  
Clearly, the holographic direction (which is parametrized by $d(t,u)$)
plays an essential role in deciding the appropriate  foliation.

\begin{figure}
   \includegraphics[width=0.48\textwidth]{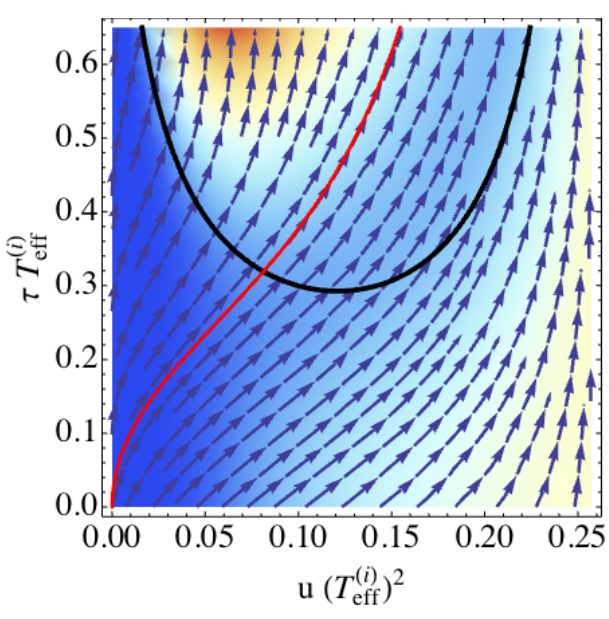} 
   \includegraphics[width=0.49\textwidth]{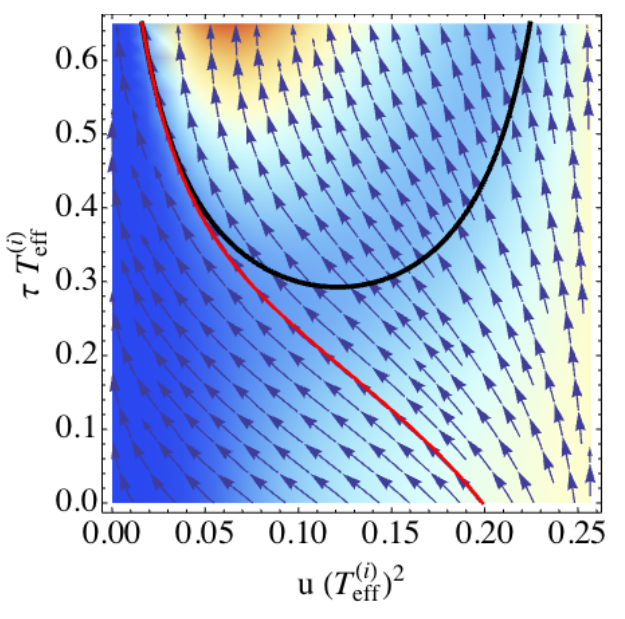} 
   \caption{ 
      The formation of  thermal AdS from a given initial state for 
      a boost invariant plasma \cite{Heller:2012je}. The background
      colors correspond to the scalar curvature as a function 
      of the Bjorken time $\tau$ and the holographic coordinate $u$, with blue  corresponding to regions of small curvature.  (The boundary is at $u=0$, the left hand side of the plot, and all coordinates are measured  in terms of the initial effective temperature $T_{\rm eff}^i \propto (\epsilon^i)^{1/4}$.)
      (a) In the left figure, the arrows indicate inward going lightlike geodesics.  The red line is a radial null geodesic sent from the boundary  into the
      bulk.
      (b) In the right figure, the arrows indicate outward going lightlike geodesics. The red line indicates the event horizon. 
      Here the parameter of the lapse (which cuts off the geometry in the IR) is $u_0\simeq 0.26\,(T_{\rm eff}^{(i)})^2$.
      In subsequent runs the event horizon estimate $u_{o}^{EH}\simeq 0.20\, (T_{\rm eff}^{(i)})^2$ is used for  $u_o$.
   \label{HellerBHFig} 
   }
\end{figure}

\Fig{HellerBHFig} shows  the formation of  thermal AdS from an  arbitrary initial condition (initial condition 23) in the numerical 
simulations of \Ref{Heller:2012je}.
Similar figures are produced by a variety of initial conditions.   
The boundary physics of these gravitational solutions will be described in the next section, \Sect{Hydroization}.

\subsection{Equilibration  and the onset of hydrodynamics}
\label{Hydroization}

\subsubsection{Equilibration of a Bjorken expansion} 
\label{Bj01 }

In the previous section we described an ADM formulation of numerical 
relativity. To date this 
has been used to study the 
equilibration of a boost invariant 
plasma initialized at $\tau =0^+$, and the approach to hydrodynamics at late times \cite{Heller:2011ju}.
Here we will review the boundary physics extracted from this holographic simulation. 

We first note that the 
energy density at mid-rapidity is  finite as $\tau\rightarrow 0^+$, and thus the energy density  can be used to define an initial effective temperature:
\st
\ed^i = \frac{3}{8} \pi^2 N_c^2 \left(T_{\rm eff}^i\right)^4   \, .
\stp
After specifying the energy density (but not its time derivatives), the only other non-zero component
of the stress tensor not determined by boost invariance, transverse
translational invariance, and conformal invariance  is the longitudinal
pressure $\pr_L$: 
\st
\label{bstress}
(T^{\tau\tau}, T^{xx}, T^{yy} , \tau^2 T^{\eta\eta}) \equiv (\ed, \pr_T, \pr_T, \pr_L)  \, ,   \qquad \pr_T = \half(\ed - \pr_L) \, .
\stp
Specifying the stress tensor does not completely specify the initial conditions for the gauge theory even at large $N_c$. 
Rather it specifies the one point function, and does
not specify the full variety of momentum space configurations
that lead to this one point function. 
In the gravitational set-up this ambiguity means that there is considerable
freedom in choosing a five-dimensional asymptotically AdS metric 
which has the boundary stress tensor specified by \Eq{bstress} \cite{Heller:2012je}.  
After specifying initial conditions for the 
five dimensional metric functions (as will be detailed shortly), the Einstein equations determine
the subsequent evolution. 
The evolution of the boundary stress tensor satisfies
the boundary conservation laws, $\partial_{\mu} T^{\mu\nu}=0$, 
which reduce to an ordinary differential equation
for boost invariant plasma
\st
\label{bjpdV}
\frac{\dd \ed(\tau)}{\dd \tau} = -  \frac{\ed(\tau) + \pr_L(\tau)}{\tau}  \, . 
\stp
Thus, the longitudinal pressures and transverse pressures are completely determined  by the energy density and its time derivative.
Physically, this equation says that the energy density per space-time rapidity, $\tau e(\tau)$, decreases due to longitudinal ${\mathcal P_L}\,dV$ work of
the expansion.

\begin{figure}
\begin{center}
   \includegraphics[width=0.49\textwidth]{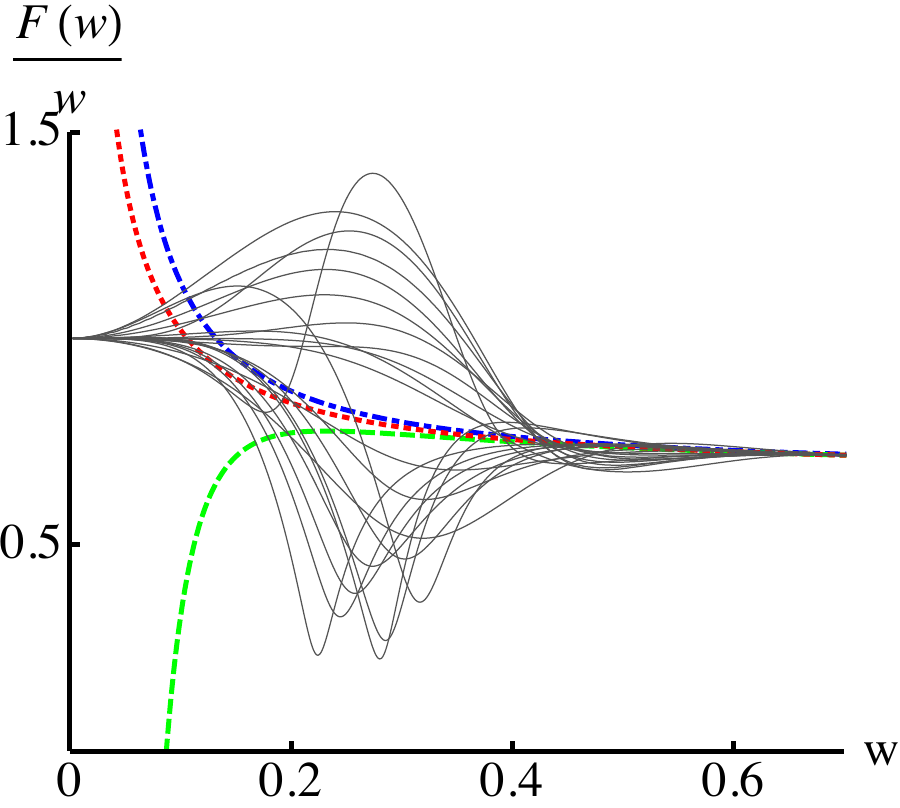} 
   \hfill
   \includegraphics[width=0.49\textwidth]{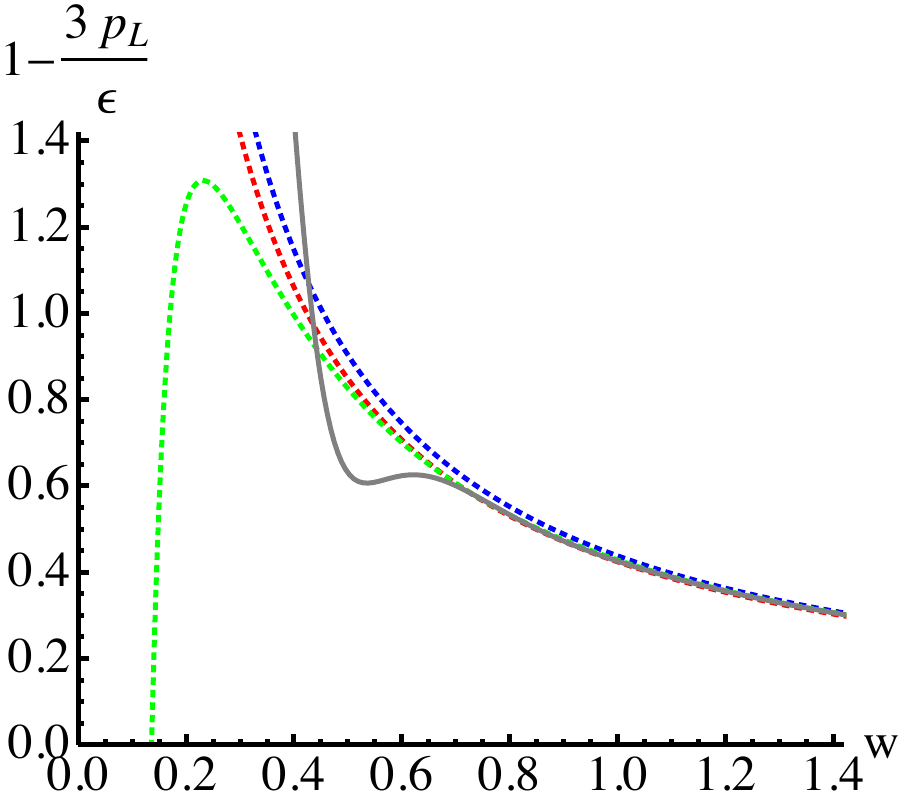} 
\end{center}
\caption{Left: The logarithmic time derivative of  $w=\tau T_{\rm eff}$,
   $\dd(\log w)/\dd(\log\tau) \equiv F(w)/w$, versus $w$ for different
   initial conditions (the grey curves) from \Ref{Heller:2011ju}. The (red) dotted, (blue) dash-dotted and 
(green) dashed curves show the predictions of hydrodynamics at first, second
and  third order.
Right: Evolution of the pressure anisotropy, $1 - 3\pr_L/\ed$, as a
function of $w$ from \Ref{Heller:2011ju}.  Only one representative initial condition is shown, together
with the hydrodynamic curves.
\label{Hellerfig}
}
\end{figure}

In the holographic setup, the short time behavior of boost invariant plasma prepared in an arbitrary initial state is determined by the expansion  of the metric near the boundary at $\tau=0$ \cite{Beuf:2009cx}.
In Fefferman-Graham coordinates the five dimensional metric 
compatible with the boundary stress, \Eq{bstress},  can be parametrized by
\st
\dd s^2 =  \frac{-e^{a_{FG}(\tau,z)} \dd\tau^2 + \tau^2 e^{b_{FG}(\tau,z)} \dd \eta^2 + e^{c_{FG}(\tau,z)} \dd \x_\perp^2 + \dd z^2 }{z^2} \, .
\stp
where metric parameters $a_{FG}, b_{FG},$ and $c_{FG}$ are related
through a coordinate transformation to the metric parameters defined in 
the previous section, \Eq{metric_adm}.
With this parametrization, the Einstein equations relate the 
$\tau$ derivatives of $\ed(\tau)$ at $\tau=0$ to the $z$ derivatives of 
$a(0,z)$. Through second order these relations read \cite{Beuf:2009cx}
\st
a_{FG} (\tau,z) =  -\ed(\tau) \, z^4 + \left( -\frac{\ed'(\tau)}{4\tau} - \frac{\ed''(\tau) }{12} \right) z^6 +  \ldots \, . 
\stp
Demanding regularity of the metric functions  near the boundary together with 
finite energy density as $\tau\rightarrow 0$,  leads to the requirement that
the first derivative $\dd e/\dd \tau$ vanish as $\tau\rightarrow 0$. In more
physical terms, this implies (from \Eq{bjpdV}) that the longitudinal pressure is
negative and the transverse pressures satisfy 
\st
\left( \pr_T, \pr_T, \pr_L \right) = \left(\ed, \ed, -\ed \right) \,.
\stp
This initial condition should be compared with the initial conditions used in boost invariant classical Yang Mills simulations of heavy ion collisions \cite{Kovner:1995ts},
\st
  \left( \pr_T, \pr_T, \pr_L \right) = \left(\half \ed, \half \ed, 0 \right) \, .
\stp
After specifying the initial gravitational fields ({\it i.e.} $a_{FG}(0,z)$),
the Einstein system is evolved numerically with 
the ADM  formulation of \Sect{ADM}
in order to see the formation of thermal AdS at late times \cite{Heller:2012je,Heller:2011ju}.  
A large variety of initial metric functions $a_{FG}(0,z)$ 
were considered and evolved to equilibrium. 
The results are summarized in \Fig{Hellerfig},  which plots the longitudinal
pressure versus $\tau$ in a slightly rewritten form motivated by hydrodynamics. More precisely, defining
$w(\tau)=\tau T_{\rm eff}$, the time derivative $w(\tau)$ is determined by \Eq{bjpdV}, 
\st
\frac{\tau}{w} \frac{\dd w}{\dd\tau } = \frac{1}{4} \left(3 - \frac{\pr_L}{\ed} \right)  \equiv \frac{F(w) }{w} \, ,
\stp
and the LHS is plotted for a wide range of gravitational initial conditions.
\Fig{Hellerfig} shows that  each different initial condition gives rise to a different early time behavior. The goal of non-equilibrium studies in gauge-gravity duality is to characterize the common  features of these  curves.

Indeed, a striking feature of these curves is their approach
to a universal hydrodynamic limit at late times. 
Close to equilibrium, the hydrodynamic constituent
relation determines  the value of $\pr_L$ (and hence $F(w)/w$) as a function of the energy density and flow velocity, and their spatial derivatives, 
which may be directly evaluated for a Bjorken flow.
In conformal hydrodynamics this constituent relation
is 
\st
T^{\mu\nu} = \ed u^{\mu} u^{\nu} + \pr \left(\eta^{\mu\nu} +u^{\mu} u^{\nu} \right)
   -  \eta \sigma^{\mu\nu} + \mbox{2nd order} + \ldots   \, ,
\stp
where $-\eta\sigma^{\mu\nu}$ is the viscous stress tensor.   We will not write out the form
of the second  order  spatial gradient terms explicitly, but refer to the original  
work \cite{Baier:2007ix,Bhattacharyya:2008jc} and  the pedagogical review  \cite{Teaney:2009qa}. For the Bjorken flow these spatial derivatives are 
evaluated easily,  yielding powers of $1/\tau$ that determine the longitudinal
pressure through second order
\st
\label{constituent}
\pr_{L} =  \pr - \frac{4}{3} \frac{\eta}{\tau} + (\lambda_1 -\eta\tau_\pi) \frac{8}{9 \tau^2} + \ldots \, .
\stp
With the  known transport coefficients  of ${\mathcal N}=4$ SYM theory \cite{Baier:2007ix,Bhattacharyya:2008jc}, the hydrodynamic expansion for the longitudinal pressure, or $F(w)/w$,  reads\footnote{ 
Specifically, we substitute the equation of state $\pr=\ed/3$, the  
first order transport coefficient $\eta/(\ed + \pr) = 1/4\pi T$, 
and the second order transport coefficients,  $\lambda_1 = 2 \,\eta^2/(\ed + \pr)$, and $\eta\tau_\pi = (4-2\log 2) \,  \eta^2/(\ed + \pr) $ into \Eq{constituent}. }
\st
\frac{F(w)}{w} = \frac{2}{3} + \frac{1}{9\pi w} + \frac{1 - \log 2}{27\pi^2w^2} + \ldots   \, , 
\stp
where each additional term is suppressed by a power of $w\equiv 1/\tau T_{\rm
eff}$. The third order term has been given explicitly \cite{Booth:2009ct}, and recently
the first 240 orders of the hydrodynamic expansion have been computed numerically to
investigate the asymptotics of the hydrodynamic expansion \cite{Heller:2013fn}.

\Fig{Hellerfig}(a) compares the evolution of $F(w)/w$ from a variety of initial conditions to the hydrodynamic prediction at first, second, and third order.
The figure shows that although hydrodynamics 
determines the late time evolution, it can yield qualitatively wrong results
at early time, and underestimate the magnitude of non-equilibrium corrections. 
There have been various attempts
to resum higher order terms in the hydrodynamic expansion in order to reproduce
some non-equilibrium physics \cite{Lublinsky:2009kv,Martinez:2010sc,Florkowski:2010cf}. If such resummations were 
applicable as opposed to genuine non-equilibrium dynamics, 
the evolution in \Fig{Hellerfig}(a) would be described  a single universal curve, $F(w)/w$ \cite{Heller:2011ju}.
However, the curves in this figure  do not exhibit a universal behavior until 
late times.
Indeed, the corrections to 
hydrodynamics are determined by the
decay time of ``non-hydrodynamic" modes, rather than still higher order terms in
the gradient expansion. Thus, the utility of such hydrodynamic resummations  seems limited.
Similar conclusions were originally reached by studying 
plasma formation with boost invariant kinematics \cite{Chesler:2009cy}, and subsequently 
at weak coupling by comparing  kinetic theory simulations for a
heavy quark propagating through plasma to the corresponding hydrodynamic
expansion \cite{Hong:2011bd}.  

The rapid approach to viscous hydrodynamics, 
sometimes called ``hydroization",
is extremely characteristic of gauge-gravity duality and has been found in
essentially all holographic non-equilibrium studies of strongly coupled plasmas
\cite{Chesler:2007sv,Gubser:2007ga,Chesler:2009cy,Chesler:2010bi,Heller:2011ju}.
However, it is important to note that viscous corrections to the equilibrium
isotropic state are large \cite{Chesler:2009cy,Chesler:2010bi,Heller:2011ju}.  \Fig{Hellerfig}(b) shows the pressure anisotropy $1 - 3\pr_L/\ed$ as a function of $w=\tau T_{\rm eff}$ for a representative initial condition, together with  hydrodynamics at various orders.   Clearly, although the system is 
well described by viscous hydrodynamics at $w=0.8$, the pressure tensor is rather anisotropic, $(\pr_T - \pr_L)/(\pr_T + \pr_L)\simeq \third$, and remains  anisotropic until late times.  This anisotropic regime
is out-of equilibrium,  but the stress tensor is well characterized by gradients around the equilibrium state.  Presumably, other measures of equilibration such as the emission rates studied in \Sect{twopnt} are also  corrected gradients \cite{Rebhan:2011ke,Baier:2012tc,Chesler:2012zk,Balasubramanian:2012tu,Erdmenger:2012xu}, although the form of these corrections remains to be fully clarified.  The holographic description of Bjorken flow also offers a venue for studying non-equilibrium phenomena outside of hydrodynamics, for example the behavior of a quark condensate near a holographic chiral phase transition \cite{Evans:2010xs}.

\subsubsection{Transverse and longitudinal collision dynamics at strong coupling}
\label{transverdynamics}

The non-equilibrium dynamics described in the  previous section has been
extended in several ways. Both the  longitudinal \cite{Chesler:2010bi} and
transverse dynamics \cite{vanderSchee:2012qj} of the collision have been studied and simulated in gauge-gravity duality using the null coordinate method described in \Sect{nullcoord}.

The longitudinal dynamics  was studied by colliding two shock waves which
are infinite and flat in the transverse direction, and of finite width in the
longitudinal direction \cite{Chesler:2010bi}.  Notable results included a finite
speed $\beta = 0.86$ of the outgoing maxima, and the behaviors of the
longitudinal and transverse pressures  which become quantitatively similar to
hydrodynamics  after time that is a  finite multiple of the width of the
original colliding shocks.  Boldly extrapolating these results to conditions
appropriate to collisions at top RHIC energies, one expects hydrodynamical
behavior to set in  at $\tau \sim 0.35\,{\rm fm}/c$.  However, the strong attenuation of
stress energy outside a rapidity window $|y| \lesssim 1$ is unlike real heavy ion collisions, and may be unfortunately  characteristic of the strong coupling limit.

\begin{figure}
\begin{center}
   \includegraphics[width=0.6\textwidth]{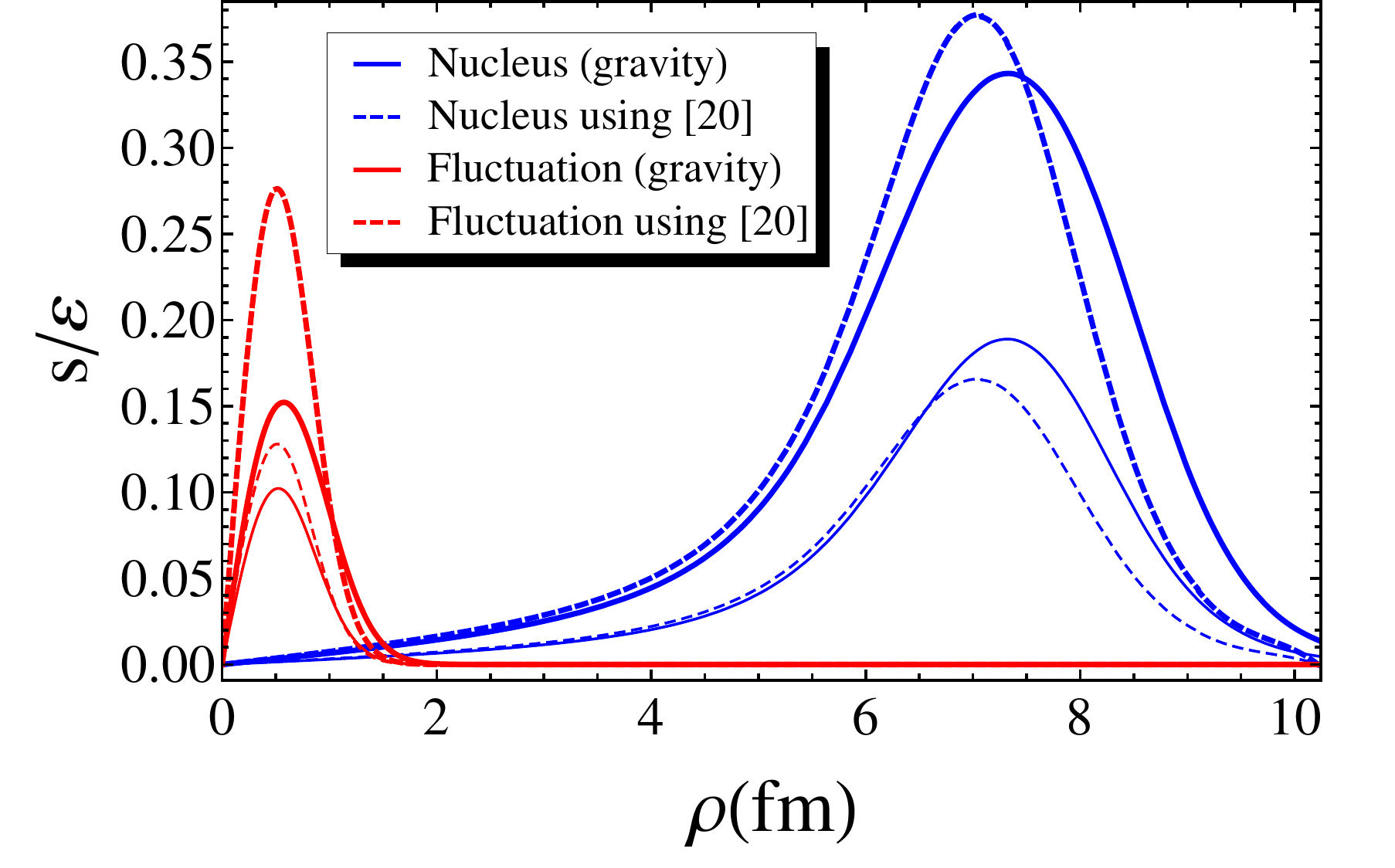}
\end{center}
\vskip -0.2in
\caption{
   The transverse momentum density relative to the
   energy density,  $\mathcal S/\mathcal E$, for a non-equilibrium
   cylindrically symmetric Bjorken expansion  simulated in gauge
   gravity duality in \Ref{vanderSchee:2012qj}.  The blue curves 
   result from a  Wood-Saxon profile, while the red curves 
   are for a small perturbation which is not discussed in this review.
   The thick curves show $\mathcal S/\mathcal E$ at $\tau=0.75\,{\rm fm}$ 
   while the thin curves show the result at $\tau=0.4\,{\rm fm}$. 
   The dash curves (labeled $[20]$) show the Vredevoogd and Pratt estimate (\Eq{pratt}) for ``pre-flow" \cite{Vredevoogd:2008id}.
   \label{vanderschee}
}
\end{figure}

At later times $\tau \sim R/c$, 
the finite transverse size also influences the evolution of the system.
 An important practical question is how large is the transverse flow 
generated  by the early time non-equilibrium dynamics, and  how this ``pre-flow" influences the subsequent hydrodynamic expansion.
A reasonable estimate of pre-flow due to Vredevoogd and Pratt  finds that the initial
momentum density 
after a time  $\Delta \tau\equiv \tau-\tau_o$ is determined by 
local gradients of the energy density \cite{Vredevoogd:2008id}. Specifically,
for an initial energy density profile $\epsilon_0(\x_\perp, \tau_o)$  the transverse momentum density $\mathcal S^{i} \equiv T^{0i}$ relative to energy density $\mathcal E$ increases in time as
\st
\label{pratt}
\frac{\mathcal S^{i} }{\mathcal{E}} \simeq   -
   \frac{\partial^i\epsilon_0}{2 \epsilon_0 } \Delta \tau\, .
\stp
This estimate is supported by gauge gravity simulations. 
To simulate a boost invariant expansion for a finite sized nucleus, Wilke van der Schee initialized a finite nucleus geometry using the AdS/CFT correspondence \cite{vanderSchee:2012qj}, {\it i.e}   the  non-trivial coordinates are $\tau,\rho,$ and $z$,  where $\rho$ is the transverse radius, and $z$ is the holographic coordinate. Solving 
the $2+1$ dimensional Einstein equations using the  null coordinate method (see
\Sect{nullcoord}) he determined the stress energy tensor in boundary theory at 
late times. In \Fig{vanderschee} the resulting pre-flow is compared to  
 \Eq{pratt}, which  reasonably describes the 
transverse flow velocity generated 
during the earliest moments.   Since this
amount of ``pre-flow"  is modest compared to later evolution
\cite{Vredevoogd:2008id}, the non-equilibrium evolution is not expected to
dramatically change the results of hydrodynamic simulations. However, such
``pre-flow" contributions can constitute an important correction when quantifying the uncertainties in current hydrodynamic simulations \cite{Heinz:2013th,Luzum:2012wu}.

\subsection{Equilibration of two-point functions}
\label{twopnt}
The previous  section described  the thermalization of the average stress tensor, and the approach to the hydrodynamic limit.  This only partially characterizes
the equilibrium state.  In this section we will discuss other measures 
of equilibration focusing  on two point functions. So far the equilibration
of two point functions has been studied in particularly simple geometries
such as Vaidya metric \cite{Lin:2008rw,Baier:2012tc,Balasubramanian:2010ce, Balasubramanian:2012tu},  and a numerical background initially realized
by Chesler and Yaffe \cite{Chesler:2008hg,CaronHuot:2011dr,Chesler:2012zk}.

\subsubsection{The Fluctuation-Dissipation Theorem}

Equilibrated  two-point functions obey detailed balance; the 
emission rate of weakly coupled quanta with energy $q^0$  is related
to the  absorption  rate by the Fluctuation-Dissipation relation
\st
\frac{\mbox{emission rate}}{\mbox{absorption rate}} = e^{-q^0/T } \, .
\stp
More formally, the emission and absorption rates  of a (bosonic) field weakly coupled to an operator $\O$  are given by the Wightman correlation 
functions $G^<(Q)$  and $G^>(Q)$ respectively
\begin{align}
   G^<(q^0, \q) =& \int d^4X \, e^{-iQ\cdot X} \left\langle \O(0) \, \O(X) \right\rangle  \, , \\
   G^>(q^0,\q) =& \int d^4X \, e^{-iQ\cdot X} \left\langle \O(X) \, \O(0) \right\rangle  \, .
\end{align}
In equilibrium  these two Wightman functions  are related by the  Fluctuation-Dissipation Theorem (FDT)
\st
\label{fdt}
G^<(q^0,\q) = e^{-q^0/T} G^>(q^0,\q) \, ,
\stp
where here and below capital letters  denote four momenta
$Q = (q^0, \q)$, and  space-time coordinates $X  = (x^0, \x)$ in flat spacetime.
Out  of equilibrium, these two correlation functions 
are unrelated to each other. 
The emission of quanta from an equilibrating system is dual to the emission of
quanta from  an equilibrating black hole. Thus, the study of thermalization in
strongly coupled systems is directly connected to the equilibration of Hawking radiation in  non-equilibrium gravitational backgrounds. 

Often, in semi-classical contexts (such as  gauge-gravity duality) it is more
convenient to express the FDT in terms of the spectral density 
and the symmetrized correlation function
\begin{align}
   \rho(Q) =& \int d^4X \, e^{-iQ\cdot X } \llangle \left[ \O(X), \O(0) \right] \rrangle  = G^{>}(Q) - G^<(Q)  \, ,  \\
   G_{\rm sym}(Q) =& \int d^4X \, e^{-iQ\cdot X} \, \half \llangle \left\{ \O(X), \O(0) \right\} \rrangle = \half (G^>(Q) +  G^<(Q) ) \, . 
 \end{align}
For instance, the evolution of the occupation numbers $n_\k$ of a quantum
electromagnetic field weakly interacting with strongly coupled plasma satisfies a Boltzmann type equation
\st
\label{boltz}
\left[ \partial_t  +  \hat \k \cdot \partial_x \right]  n_\k(t,\x) = \frac{G^<(K)}{2\omega_k}  (1+ n_\k (t,\x))  - n_\k(t,\x) \frac{G^>(K)}{2\omega_k}   \, , 
\stp
where $G^<(K)$ and $G^>(K)$ are the lightlike current-current correlators, and 
$K=\omega_k (1, \hat\k)$ is a Fourier mode  of the electromagnetic field.
For a classical field with $n_\k(t,\x) \gg 1$,  we see that 
\st
\left[ \partial_t  +  \hat \k \cdot \partial_\x \right]  n_\k(t,\x) =  - n_\k (t,\x)  \frac{\rho(K)  }{2\omega_k} \, , 
\stp
and thus  the spectral density describes
the absorption rate of a classical field 
by plasma.  
A very worthwhile (and very short) exercise starting with \Eq{boltz} shows that if the 
FDT is satisfied  and the spectral density is positive,  then the
interactions between the electromagnetic field  and the strongly
coupled plasma will drive the occupation numbers to equilibrium, 
$n_\k = 1/(e^{\omega_k/T} - 1)$.

While the spectral density (which describes classical absorption)  is easily computed using the classical methods 
of the correspondence, the symmetrized correlation function (which describes
the quantum-statistical fluctuations)  can  only be
computed through an analysis of the Hawking flux \cite{deBoer:2008gu,Son:2009vu,CaronHuot:2011dr}.

\subsubsection{Non-equilibrium two-point functions}
\label{2pnt}

\begin{figure}
   \begin{center}
      \begin{minipage}[c]{0.49\textwidth}
   \includegraphics[width=\textwidth]{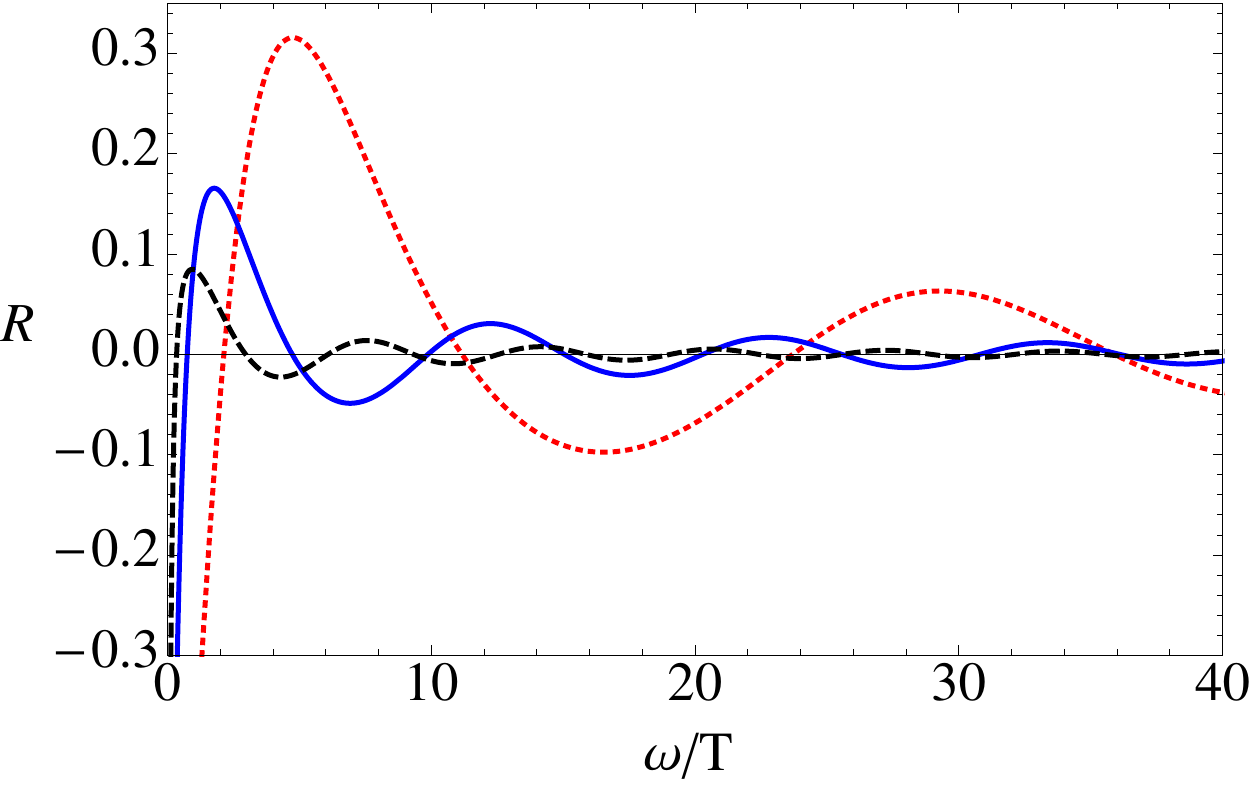} 
\end{minipage}
\begin{minipage}[c]{0.50\textwidth}
   \includegraphics[width=\textwidth]{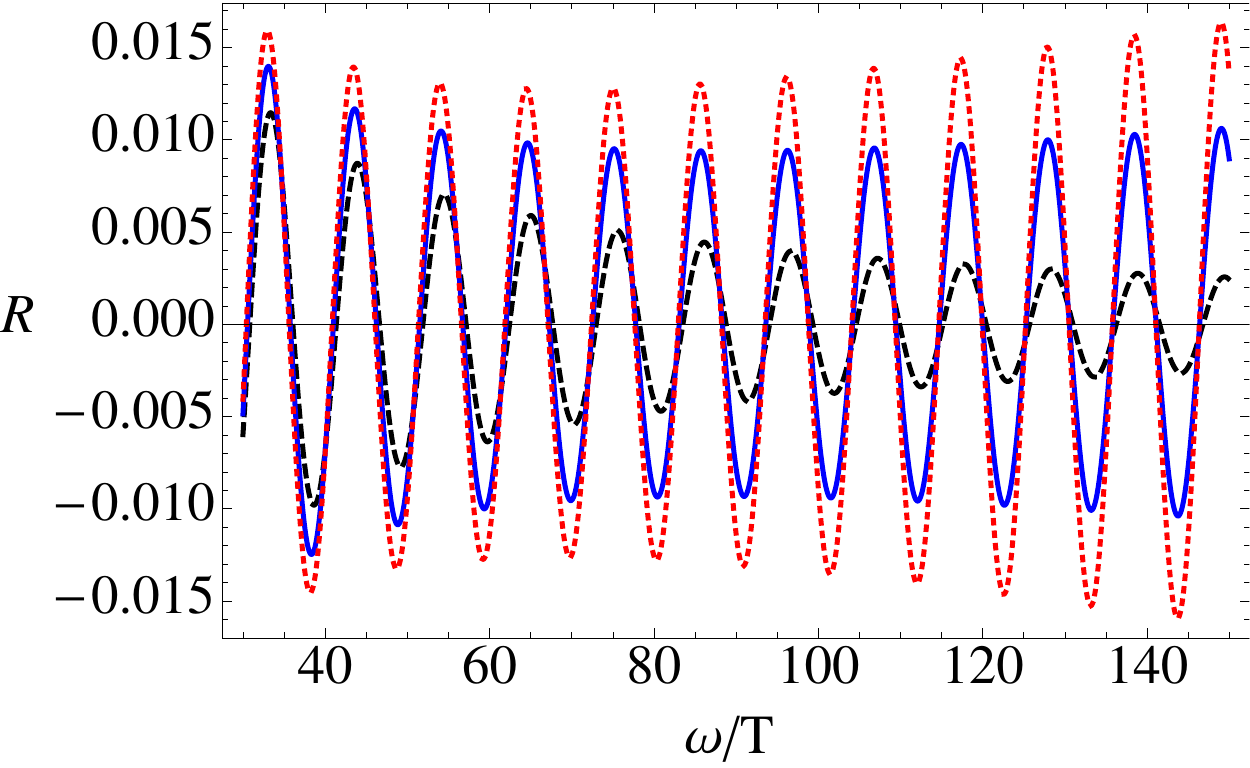} 
\end{minipage}
\end{center}
\caption{ 
   (a) The relative deviation of the photon spectral function from its thermal limit, $R$ (see\Eq{Rdef}),  
for various values of the shell height and the frequency
from \Ref{Baier:2012ax}. The black, blue, and red curves
are for $r_s/\pi T_{\rm f} = 1.001, \,  1.01, \, 1.1$. 
(b) The relative deviation $R$ for $r_s/\pi T_{\rm f}=1.01$ and $\lambda =
\infty, 500, 350$ with the amplitudes decreasing with coupling from \Ref{Steineder:2012si}. While in the
$\lambda=\infty$ case the amplitude of the oscillations gets damped at large
$\omega$, for all finite values of $\lambda$ it first decreases but ultimately
starts growing linearly with $\omega$. 
\label{spectralfig1} 
}
\end{figure}

First, we will analyze the non-equilibrium evolution and equilibration of the spectral density, $\rho(Q)$,
which has been studied in non-equilibrium contexts by several groups \cite{Lin:2008rw,Erdmenger:2011jb,Baier:2012tc,Baier:2012ax,Steineder:2012si}. The spectral density
records the absorption rate external classical field by the non-equilibrium
plasma.
For definiteness, we will consider the ``falling-shell" geometry first
considered  by Shu and Shuryak \cite{Lin:2008rw}, as a prototype non-equilibrium geometry.   In
this case the radial coordinate of the shell follows the 
trajectory $r_s(v)$, and the metric takes a specific form  
\st
\label{shu_metric}
ds^2 = -A(r,r_s(v)) dv^2  + 2 dv dr +   r^2 d\x^2 \, ,
\stp
where 
\st
A(r, r_s(v)) = \begin{cases}
   r^2  & r < r_s(v) \\
   r^2 \left(1 -  \left(\frac{4\pi T_{\rm f}}{d \, r}\right)^d \right) & r > r_s(v) 
\end{cases} \, , 
\stp
{\it i.e.} inside of a shell ($r < r_{s}(v))$
the metric is vacuum AdS, while outside the shell ($r> r_s(v)$) the metric is  thermal
black hole AdS. Here $d=4$ is the number of space time dimensions. As an extreme limit,  one can  profitably consider an infinitely thin shell falling at the speed of light in 1+1 dimensions \cite{Balasubramanian:2010ce}, yielding the Vaidya space time metric
\st
\label{Vaidya_metric}
ds^2 =  - r^2 \left(1 - \theta(v) \left(\frac{2 \pi T_{\rm f}}{r}\right)^2 \right)  dv^2 + 2 dv dr + r^2 dx^2 \, .
\stp
In the original prototype shell geometry given in \Eq{shu_metric}, the shell falls
at a specified rate, which is  determined by solving the Israel junction conditions \cite{Lin:2008rw}.  The
spectral density can be found from the retarded and advanced Green's
functions in this time dependent geometry.
In general,  once the geometry is time dependent,
Fourier transforms  are of limited use, and the equations to be solved
are partial differential equations rather than ordinary equations.
However, when the frequency of the spectral density $\omega$  is large compared to  all 
other time scales in the problem, the position of the 
probe brane  may be considered constant over a  time
scale of $1/\omega$. In this case,  Fourier transforms can be introduced, and
the solution to the wave equation in the static, but discontinuous, metric
can be found by solving the attendant differential equations above and
below the falling shell, and matching the solution across the discontinuity at $r=r_s$.
Then, the infalling and outgoing solutions  for $r\rightarrow 0$ can  be selected,
and these solutions determine the retarded and advanced  Green's functions 
in the non-equilibrium metric. The spectral density  is a function of the shell height $r_{s}$ and the frequency $\omega$.  
A particularly interesting spectral density to study is the {\it light-like}
R-charge current-current correlator, 
\st
\rho(Q) = \int_X e^{-i Q\cdot(X-Y)} \llangle\left[ J^{\mu} (X) J_\mu(Y) \right] \rrangle\, ,   \quad \mbox{with} \quad Q^{\mu} =  \omega \, (1, \hat \q) \, ,
\stp
which provides a gravitational dual  for the absorption of a classical electromagnetic wave by plasma.
\Fig{spectralfig1}(a) shows relative deviation  of the spectral density
from equilibrium
\st
\label{Rdef}
R = \frac{ \rho(r_s, \omega) - \rho_{\rm thermal}(\omega) }{\rho_{\rm thermal}(\omega) } \, , 
\stp
for the lightlike R-charge current current correlator. The results
show a characteristic feature of non-equilibrium AdS \cite{Balasubramanian:2010ce}, {\it i.e.}\  the spectral density  first attains its equilibrium form at  high
frequencies \cite{Balasubramanian:2010ce,Baier:2012ax}.  The
analysis has been pushed to next to leading order in $\lambda$ (\Fig{spectralfig1}(b))  with the
satisfying result  that finite $\lambda$ corrections increase the deviation
from equilibrium as $\omega \rightarrow \infty$ \cite{Steineder:2012si}.

Two calculations have gone beyond the quasi-static approximation, and
computed the spectral  density in a time dependent background for dilaton correlators \cite{Balasubramanian:2012tu,Chesler:2012zk}.
A general approach is to recognize that the  bulk spectral density satisfies
homogeneous equations of motion, together with initial conditions
specified by the canonical commutation relations.  
Once the bulk spectral density is determined by solving the
equations, the correlator can be brought 
to the boundary, determining the spectral density in the field theory \cite{CaronHuot:2011dr,Iqbal:2008by}. 
Specifically, for a bulk scalar field, $S= -K\int \half(\partial\phi)^2$, with spatial momentum $\k$ in a homogeneous, but time 
dependent background, the spectral density of the bulk gravitational
theory, 
\st
\rho(v_1r_1|v_2 r_2) = \llangle \left[ \phi_{\bm k}(v_1, r_1) ,\phi_{\bm k}(v_2, r_2) \right] \rrangle \, , 
\stp
satisfies the causal equations of motion
\st
\label{eom}
\left[  \partial_{\mu} K\sqrt{g} g^{\mu\nu} \partial_{\nu}  -  K \sqrt{g} g^{ij} k_i k_j  \right]  \rho(v_1 r_1 | v_2 r_2 )  = 0 \, , 
\stp
where $\mu,\nu$ run over the space-time indices $v_1, r_1$, and $i,j$ run over
the spatial directions $x^1, x^2,x^3$.
With the canonical commutation relations,
\st
\label{canonical}
\lim_{v_1 \rightarrow v_2} K \sqrt{g} g^{v\nu} \partial_{\nu}  \rho(v_1r_1| v_2 r_2) = \delta(r_1 - r_2)   \, ,
\stp
the PDE system can be solved in $v_1r_1\otimes v_2r_2$ to determine the spectral density  at all times. 
This technically challenging procedure has been followed \cite{Chesler:2012zk,Balasubramanian:2012tu},
and the resulting spectral densities show a  top down pattern of thermalization that is similar to the quasi-static example discussed above.

The calculation of the emission rate, $G^<(Q)$, and 
the statistical fluctuations, $G_{\rm sym}(Q)$, amounts to computing 
the Hawking radiation in a time dependent geometry. 
The physics probed by these fluctuations is markedly different from the 
spectral density.
To illustrate the essential differences, we will sketch a somewhat unusual  derivation of Hawking radiation for the equilibrium black brane geometry 
\st
ds^2 = -r^2 \left(1 - \left(\frac{\pi T}{r}\right)^4 \right) dv^2 + 2 \,dr dv  + r^2 d\x^2 \, ,  
\stp
 and then we will generalize these ideas to non-equilibrium backgrounds \cite{CaronHuot:2011dr,Ebrahim:2010ra}.  

 For the symmetrized function
 arbitrary initial conditions are specified in the distant
 past, and evolved to the future with the homogeneous equations of motion.  This should 
be contrasted with the spectral density, where the initial conditions
are completely fixed by the canonical commutation relations (\Eq{canonical}).
However, even in the symmetrized case, the initial conditions 
are not completely arbitrary. In particular, the Hadamard conditions
require that at vanishing coordinate separation,
$v_1 r_1 \rightarrow v_2 r_2$,
the symmetrized correlator should be asymptotic with the 
flat space results for  a free  scalar field  in 1+1 dimensions:
\st
\label{hadamard}
\lim_{X_1 \rightarrow X_2} \half \llangle \left\{ \phi (X_1) \, \phi (X_2) \right\}  \rrangle    
=   -\frac{1}{4\pi K} \log\left| \mu \eta^{\mu\nu} \Delta X_{\mu} \Delta X_{\nu} \right|  \, ,
\stp
where $\Delta  = X_1 - X_2$. Such coincident point singularities describe 
ultraviolet vacuum fluctuations
in the  five dimensional geometry.   Hawking radiation  is emitted when
stochastic UV fluctuations close to the horizon
 are inflated by the diverging geodesics of the near horizon geometry.
The symmetrized two-point function  in the bulk
records the statistics of these redshifted fluctuations emerging 
from the black hole.

\begin{figure}
   \begin{center}
   \includegraphics[width=0.6\textwidth]{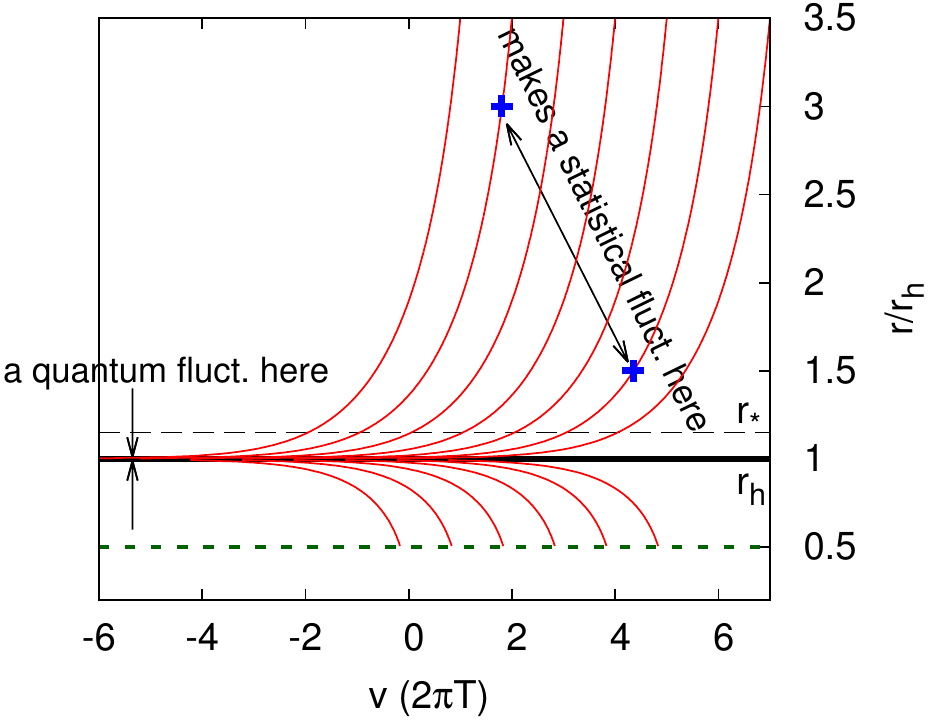} 
\end{center}
\vspace{-0.2in}
   \caption{A schematic figure illustrating the physics of Hawking radiation
      from an equilibrium black hole in Eddington-Finkelstein coordinates
      \cite{CaronHuot:2011dr}. Red lines show outgoing null geodesics. The
      stretched horizon is at $r_{*}$ (the dashed line), while the true event
      horizon is at $r_{h} = \pi T$ (the thick solid line).  The
      blue markers indicate two space-time points $v_1 r_1$ and $v_2 r_2$ 
      where  dilaton fluctuations are correlated 
      through quantum fluctuations close to
      the horizon.
   \label{hawk}
   }
\end{figure}

A schematic of the Hawking emission process is illustrated in \Fig{hawk}. 
In the distant past initial conditions for the  symmetrized two point functions are specified that  satisfy  Hadamard constraints (\Eq{hadamard}).  
Examining the near-horizon geodesics in equilibrium, one
concludes that the bulk statistical correlation function in the far future is
determined by initial data exponentially close to the horizon, 
which is completely fixed by these UV constraints. 
This UV-singular initial data is propagated  
along
outgoing null geodesics from the event horizon at $r_h=\pi T$  to the stretched
horizon at $r_*=r_h + \epsilon$, using a WKB or geodesic approximation.
Such geodesics are diverging exponentially near the horizon as
\st
\label{geo}
r(v)- r_h  = (r_o - r_h) e^{2\pi T (v-v_o)} \, ,
\stp
and this  divergence is responsible for the gravitational redshift.
Finally, the solution above the stretched horizon (where the
eikonal approximation  is no longer valid due to the red shift), can be written as a convolution of
retarded propagators from the stretched horizon to the bulk, and an effective stretched source that 
describes the quantum fluctuations below $r_{*}$
\st
G_{\rm sym}(v_1r_1 | v_2 r_2) = \int \dd v_1' \dd v_2' \,  G_R(v_1 r_1| v_1' r_*)  G_R(v_2 r_2 |v_2' r_*) \left[ G_{\rm sym}^h(v_1' | v_2') \right] \, . 
\stp
The horizon source is 
\st
\label{horizon_correlator}
G_{\rm sym}^h(v_1 | v_2) =   - \frac{K}{\pi} \sqrt{g} \, \partial_{v_1} \partial_{v_2} \log \left| e^{2\pi T v_1 } - e^{2\pi T v_2 } \right| \, ,
\stp
which clearly reflects the logarithmic initial conditions, \Eq{hadamard}, together with the transport of this initial data along the geodesics given in \Eq{geo}.   
Additional analysis shows that this horizon symmetrized correlator  and the 
corresponding horizon spectral density satisfy the FDT.  In the context of heavy
quarks, the near-horizon statistical fluctuations codified by
\Eq{horizon_correlator} ultimately determine the spectrum of random forces
experienced by a heavy quark in strongly coupled plasma \cite{deBoer:2008gu,Son:2009vu,CaronHuot:2011dr}. The 
validity of the FDT on the horizon then guarantees the Einstein relation between the
drag and momentum diffusion coefficients in the boundary theory.

\begin{figure}
\begin{center}
   \includegraphics[width=0.49\textwidth]{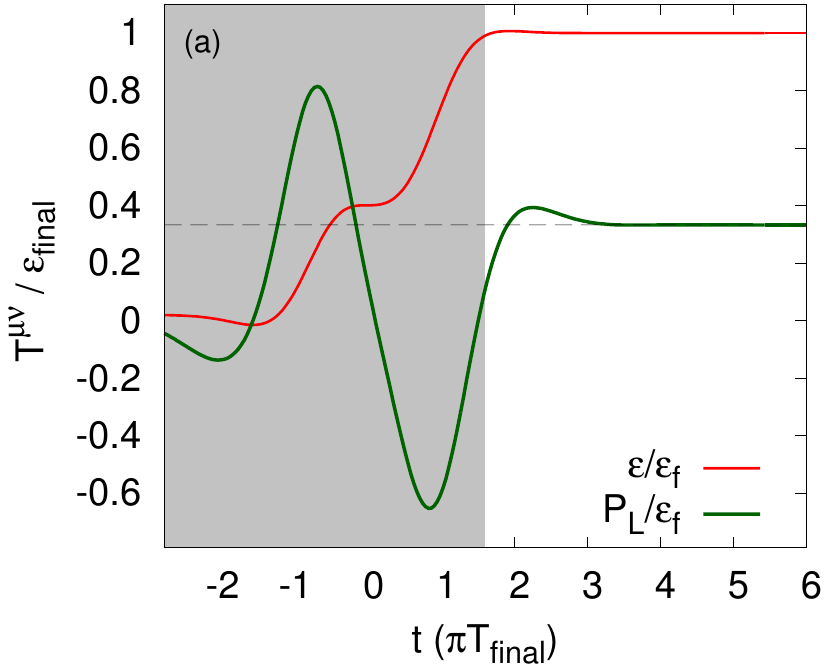}
   \includegraphics[width=0.49\textwidth]{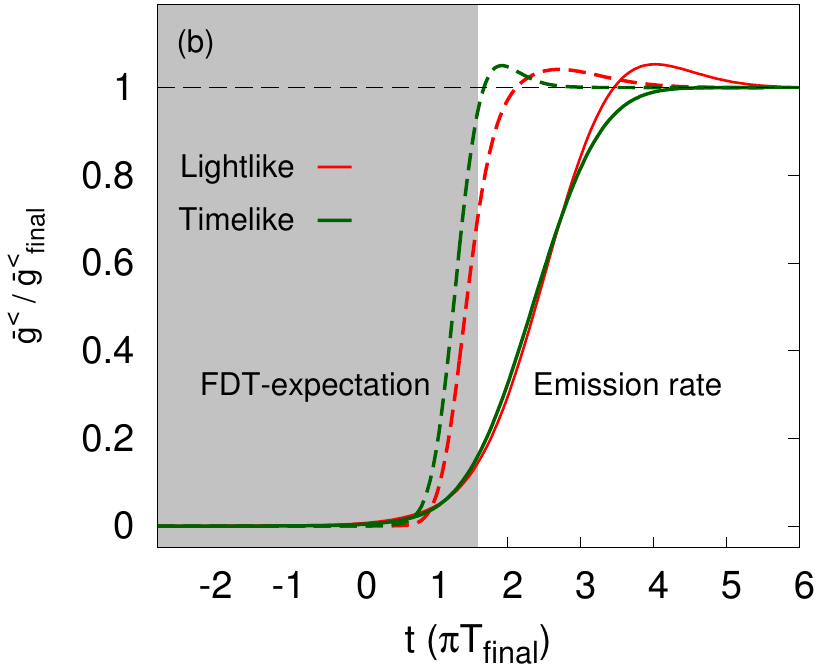}
\caption{
   (a) The SYM stress tensor $T^{\mu\nu}$ relative to the final energy density $\mathcal E_f$ as a function of time in 
the boundary theory. The shaded
band indicates when the energy density is changing due to
the work done by the external source in the boundary gauge theory \cite{Chesler:2008hg}.
(b) A non-equilibrium emission rate, $\bar G^<(\bar t,\omega;\q)/\bar G^<_{\rm final}(\omega; \q)$.
The emission rate is exhibited for time-like momenta, with $q^0 = 8\pi T_{\rm f}$ and $\q = 0$, and for light-like momenta, 
with $q_\perp = q_\parallel = q^0/\sqrt{2}$ and $q^0 = 8\pi T_{\rm f}$. The dashed lines show the FDT-expectation for the
emission rate, {\it i.e.} the rate derived using the absorption rate $\bar G^>(\bar t,\omega;\q)$ and the FDT result \cite{Chesler:2012zk}.
\label{pauldt}
}
\end{center}
\end{figure}

Given this  dynamical picture of the Hawking emission process, which is based
on solving equations of motion rather than a Fourier decomposition, it is
conceptually straightforward (if technically challenging) to compute the
emission two point function in an out of equilibrium geometry, such
as   the Chesler-Yaffe background \cite{Chesler:2008hg}. In this
background a non-equilibrium state is created by turning on a time-dependent gravitational source in the
boundary theory.  The resulting numerical 5d metric in the bulk bears some structural
similarity to the Vaidya space time metric, and can be parametrized as
\st
ds^2 = - A dv^2 + 2 dv dr + \Sigma^2  \left[e^{B} d\x_\perp^2 + e^{-2B} dx_\parallel^2 \right] \, ,
\stp
where $A$, $B$, and $\Sigma$ are functions of $r$ and $v$. 
\Fig{pauldt} (a) shows the stress tensor  $T^{\mu\nu}$ for this numerical background, which indicates that the stress tensor isotropizes shortly after the source
is turned off.

The equilibration of the system is quantified by monitoring the emission and absorption rates
in a given frequency band $\Delta \omega$ as a function of time \cite{Chesler:2012zk}. 
Clearly, the frequency and time resolutions of this analysis are limited by the 
uncertainty principle, $\Delta \omega\Delta t \geq \half$.
To achieve the best possible  resolution in both frequency and time
we will compute a windowed Fourier transform of $G(t|t')$ known 
as the Gabor Transform\footnote{We will suppress the three momentum $\q$ of the correlation function, $G(t,t') = G(t,t';\q)$. }
\begin{align}
   \bar G^<(\bar t,\omega) \equiv& \frac{1}{\sqrt{\pi \sigma^2}}\int \dd t\, \dd t' \, e^{-\frac{(t-\bar{t})^2}{2\sigma^2} } \, e^{-\frac{(t'-\bar{t})^2 }{2\sigma^2} } \, e^{i\omega(t-t') } G^<(t|t') \, , \\
   \bar G^>(\bar t,\omega) \equiv& \frac{1}{\sqrt{\pi \sigma^2} }\int \dd t \, \dd t' \, e^{-\frac{(t-\bar{t})^2}{2\sigma^2} } \, e^{-\frac{(t'-\bar{t})^2 }{2\sigma^2} } \, e^{i\omega(t-t') } G^>(t|t')  \, ,
\end{align}
where in practice the smearing width is tied to the final temperature,  $\sigma=1/\pi T_{\rm f}$.
The Gabor transform equals the Wigner transform  $G(t,\omega)$  
averaged over time and frequency with a minimal uncertainty  wave packet of
temporal width $\Delta t=\sigma/\sqrt{2}$, and frequency width $\Delta \omega = 1/\sqrt{2}\sigma$. If the FDT is satisfied, then
the Gabor transforms of $G^>$ and $G^<$ satisfy
\st
   \label{smearedFDT}
   \bar G^<(\bar t,\omega) = e^{-\omega \beta_{\rm eff}(\bar t) + \beta^2_{\rm eff}(\bar t) /4\sigma^2}  \bar G^>(\bar t, \omega - \beta_{\rm eff}(\bar t)/2\sigma^2) \, ,
\stp
where $\beta_{\rm eff}=1/T_{\rm eff}$ with $T_{\rm eff} \propto \epsilon^{1/4}$.
\Fig{pauldt}(b) shows the Gabor transform of $G^<$ for the time dependent
geometry together with FDT expectation. Clearly the emission two point
function, $G^<$, equilibrates after the corresponding one point function and 
the absorption rate, $G^>$.  Heuristically, this is because  the horizon
geometry must equilibrate before it emits an equilibrium flux, which can traverse
the bulk and reach the boundary.

\section{Black brane geometries: static and linear response properties}
\label{blackbranes}

We have thus far discussed non-equilibrium processes, associated to non-stationary gravity backgrounds. We turn now to the study of equilibrium and linear response properties of  
strongly-coupled, large-$N_c$ gauge theories at nonzero temperature and density, which can be probed via static black brane geometries.
These spacetimes may be thought of as generalizations of the AdS-Schwarzschild geometry \eno{AdSSchw}; four-dimensional Lorentz invariance is broken only by the selection of a time coordinate by the black hole horizon function, leaving three-dimensional rotations and translations preserved. They thus describe a homogeneous medium at nonzero temperature. The models we consider will all break conformal invariance by some additional feature, such as a nontrivial radial profile for a scalar field.
Thermodynamic properties come immediately from properties of the geometry, and transport coefficients  may be calculated from the response to  perturbations. Here we will largely consider ``bottom-up" models that do not come from any known string theory construction, and as a consequence the precise duality map is not known; nonetheless the models may be engineered to incorporate various desired properties, such as the beta function and the thermodynamics of QCD.

\subsection{Black branes and thermodynamics}

While generalizations can be considered, here we will content ourselves with a gravity theory containing the metric $g_{\mu\nu}$, a scalar field $\phi$ and a gauge field $A_\mu$, with Einstein-Maxwell-scalar Lagrangian,
\eqn{LwithF}{
  {\cal L} = {1 \over 2\kappa^2} \left[ 
    R - {1 \over 2} (\partial\phi)^2 - V(\phi)- {f(\phi) \over 4} F_{\mu\nu}^2  \right] \,.
 }
We take the scalar potential term to include the cosmological constant,
\eqn{Vassumptions}{
  V(\phi = 0) \equiv -{12 \over L^2}\,,
   }
and then the AdS and AdS-Schwarzschild geometries \eno{AdS}, \eno{AdSSchw} reviewed in the introduction are solutions with $\phi = A_\mu = 0$.
We will generalize these solutions to a class of asymptotically $AdS$ black brane geometries of the form,
\eqn{BHgeom}{
  ds^2 = e^{2A(r)} \left[ -h(r) dt^2 + d\vec{x}^2 \right] + {e^{2B(r)} \over h(r)} dr^2 \,,
 }
 where $h(r)$, $A(r)$ and $B(r)$ are functions of the radial coordinate that approach the AdS solution at large $r$. The function $B(r)$ can be adjusted arbitrarily by transformations of the radial coordinate, and thus may be set to a convenient form. 
 
Geometries with $h(r) = $ constant lack a horizon, while those with a zero $h(r_H) = 0$ have a horizon at the radius $r = r_H$. The former class may still be associated with a state in thermodynamic equilibrium if the Euclidean continuation is given a periodicity $\beta = 1/T$ in imaginary time, where $T$ is identified with the temperature of the dual gauge theory; such a geometry may be considered a ``thermal gas". For the latter class with a horizon, the temperature $T$ and entropy density $s$ associated to the black brane geometries are given by
\eqn{TandS}{
T = {1 \over 4 \pi} h'(r_H) e^{A(r_H) - B(r_H)} \,, \quad \quad
s = {2 \pi \over \kappa^2} e^{3A(r_H)} \,,
}
corresponding as usual in black hole thermodynamics to the surface gravity and area of the horizon, respectively. To translate the five-dimensional gravitational constant $\kappa$ appearing in the expression for the entropy density into field theory terms, we note first that in the case of $AdS_5 \times S^5$ dual to $SU(N_c)$ ${\cal N}=4$ super-Yang-Mills, one can relate the gravitational constant and the AdS radius to the string coupling $g_s = g^2_{\rm YM}/2\pi$ and the Regge slope $\alpha'$:
\eqn{}{
{1 \over \kappa^2} = {L^5 \over 64 \pi^4 g_s^2 {\alpha'}^2} \,, \quad \quad
L^4 = 4 \pi g_s N_c {\alpha'}^2\,.
}
Field theory quantities should not involve the string theory parameter $\alpha'$, which drops out of the combination
\eqn{LKappa}{
{L^3 \over \kappa^2} = {N_c^2 \over 4 \pi^2} \,,
}
and indeed the entropy \eno{TandS} must be proportional to this on dimensional grounds.  For AdS/CFT models not based on a known string theory construction, the precise coefficient in \eno{LKappa} is not determined; however, the $N_c^2$ dependence is expected to remain the same for any gravity dual of a four-dimensional gauge theory. As a result the black brane solutions have an entropy that is $O(N_c^2)$.  The thermal gas solutions, on the other hand, have no horizons and hence any entropy is subleading in $1/N_c$; one expects it to be of $O(1)$.

\begin{figure}\centering
\includegraphics[scale=0.1]{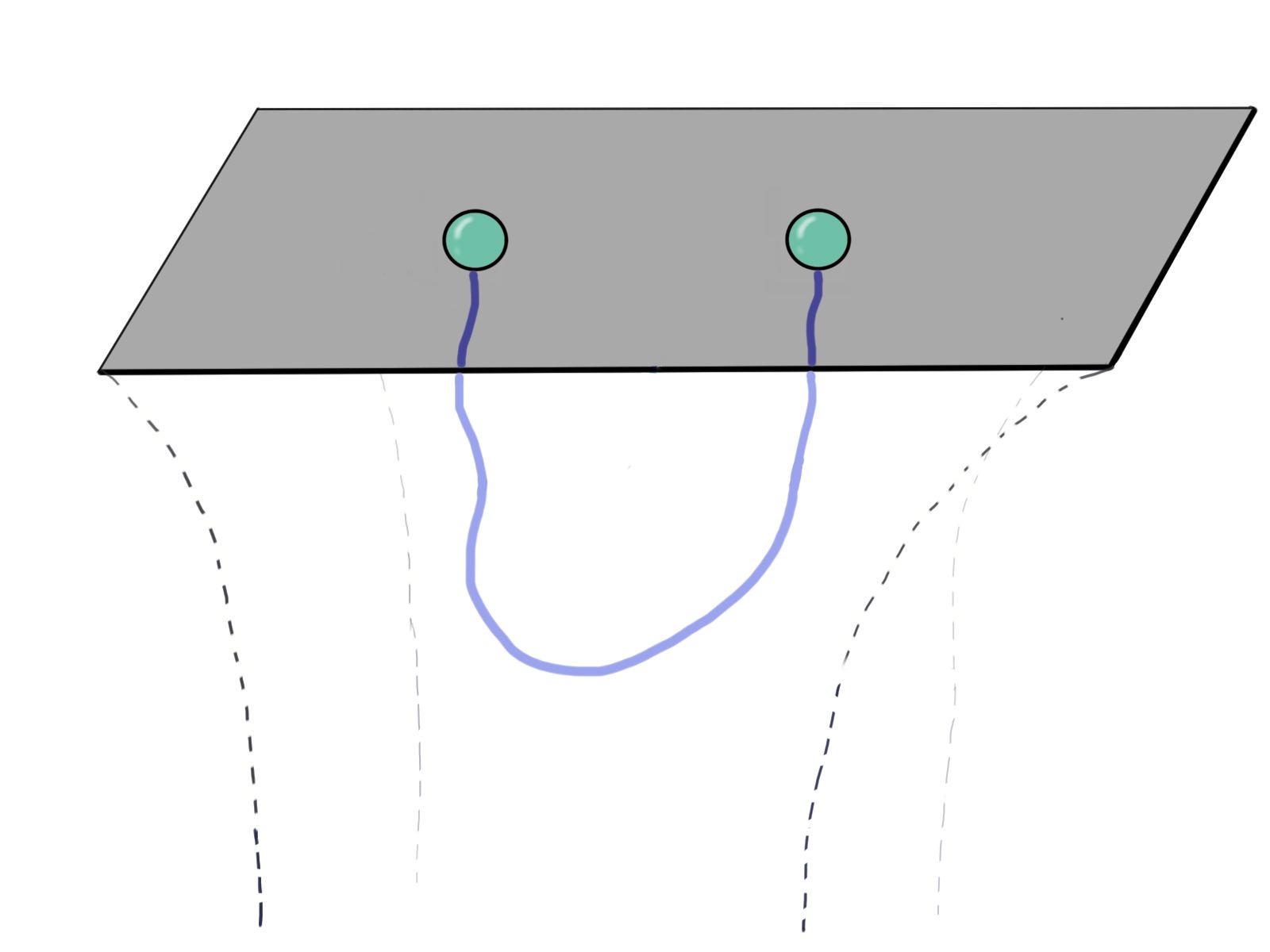}
\includegraphics[scale=0.1]{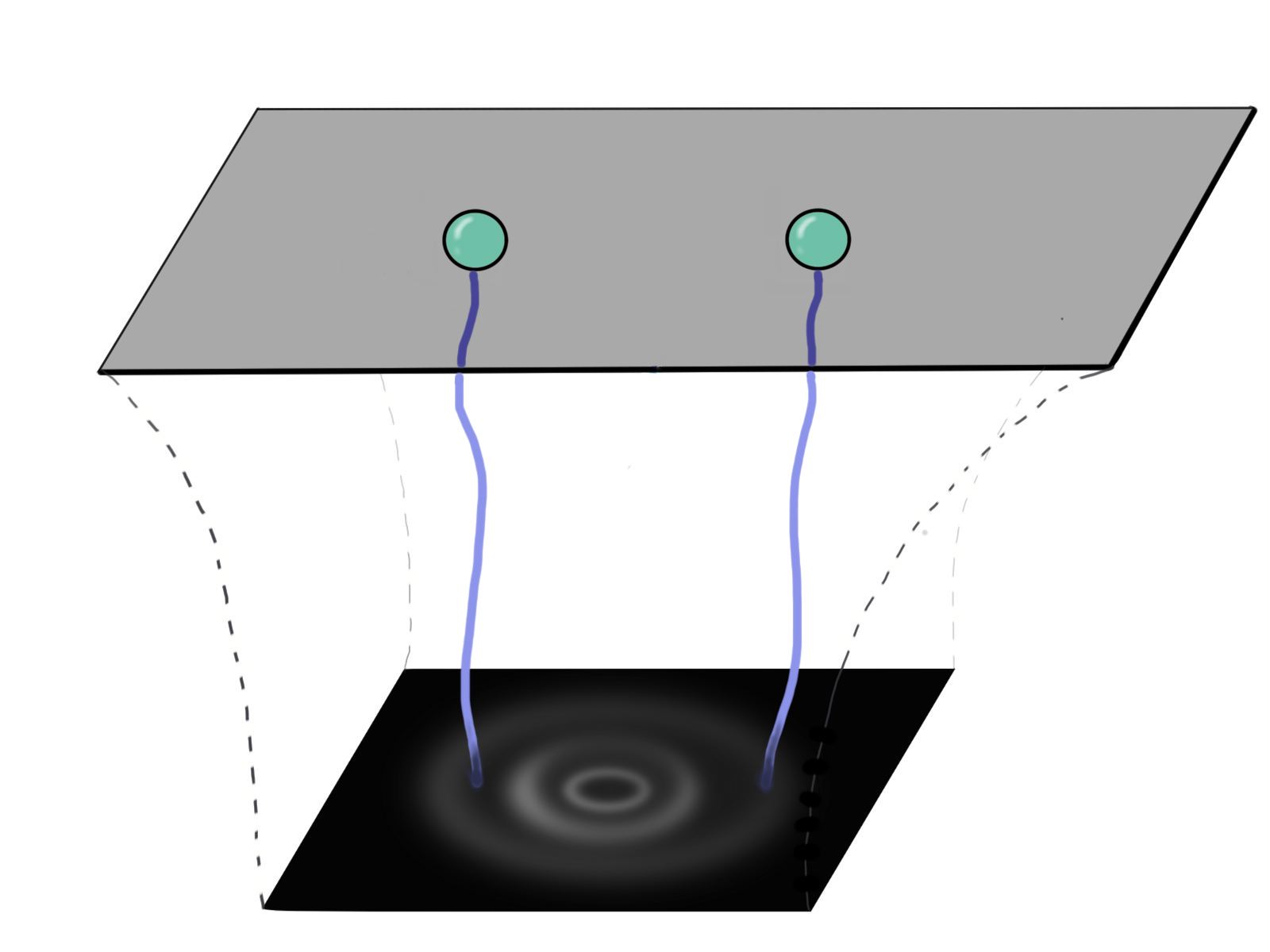}
\caption{\label{fig:conf}The potential between two test quarks in the boundary theory is holographically related to the behavior of the string world sheet connecting them. Confining geometries (left) allow the string to dip arbitrarily low in the bulk geometries radial direction. Deconfinement (right) is realized when the world sheet  encounters a black hole horizon,  effectively severing the string. }
\end{figure}

We may interpret the difference between the thermal gas solution and the black brane background as being that the former is dual to a confining gauge theory, while the latter is deconfined \cite{Witten:1998zw}. This can be ascertained by studying, for example, the potential between two test quarks in the boundary theory \cite{Kinar:1998vq}. This potential can be computed gravitationally by minimizing the area of the string worldsheet that connects one quark to the other.  Heuristically, the existence of the black brane horizon provides a hole the world sheet can fall in to, leaving its ``quark" endpoints agnostic of one another. These ideas are summarized in figure \ref{fig:conf}.

These geometries may be accompanied by a radial profile for the scalar,
\eqn{}{
\phi = \phi(r) \,.
}
The scalar field is holographically dual to an operator ${\cal O}_\phi$ whose conformal dimension $\Delta_\phi$ is determined by the mass term in the potential:
\eqn{}{
V(\phi) = - {12 \over L^2} + {1 \over 2} m_\phi^2 \phi^2 + \ldots\,, \quad \quad
m^2 L^2 \equiv \Delta_\phi( \Delta_\phi - 4) \,.
}
Near the boundary the equations of motion constrain the profile of $\phi$ to have the form,
\eqn{PhiProfile}{
\phi(r) = \phi_{(4 - \Delta)} r^{\Delta - 4} + \ldots + \phi_{(\Delta)} r^{-\Delta} + \ldots \,,
}
where the value of $\phi_{(4 - \Delta)}$ encodes adding a source for ${\cal O}_\phi$ to the dual Lagrangian,
\eqn{}{
\Delta S = \int d^4x \,\phi_{(4 - \Delta)} \, {\cal O}_\phi \,,
}
and $\phi_{(\Delta)}$ controls the one-point function,
\eqn{}{
\langle{\cal O}_\phi \rangle \sim \phi_{(\Delta)}\,.
}
Thus, a generic asymptotically AdS black brane geometry has the interpretation of ${\cal N}=4$ super-Yang-Mills deformed by the addition of ${\cal O}_\phi$ to the Lagrangian.

If the scalar is massless, the dual operator has $\Delta_\phi = 4$ and is exactly marginal. The archetype of this case is the type IIB dilaton, which in the case of  compactification on $AdS_5 \times S^5$ is holographically dual to the exactly marginal gauge coupling of ${\cal N}=4$ Super-Yang-Mills. Even in bottom-up models, the term ``dilaton" may be understood to refer to such a scalar, whose nontrivial profile corresponds to the running of the gauge coupling of the dual theory.
 
For a massive scalar, we have $\Delta_\phi \neq 4$ and the existence of a source $\phi_{(4 - \Delta)}$  introduces a mass scale,
\eqn{LambdaScale}{
\Lambda \equiv \left(\phi_{(4 - \Delta)} \right)^{1 \over 4 - \Delta} \,,
}
which explicitly breaks conformal invariance to the dual quantum field theory. As we shall see, it can be useful to model the running of the physical QCD coupling, which is not conformal, by replacing the massless dilaton with a massive scalar with potential suitably chosen to approximate aspects of the beta function. Irrelevant operators correspond to fields with positive mass-squared, while relevant operators are dual to fields with negative mass-squared. Unlike in flat space, negative mass-squared does not lead to an instability as long as the Breitenlohner-Freedman bound is satisfied, 
\eqn{}{
m^2 L^2 \geq -4 \,. 
}
The introduction of a $U(1)$ gauge field is dual to the presence of a global conserved current in the field theory dual. A profile for the gauge field of the form,
\eqn{Vector}{
A_\mu dx^\mu = \Phi(r) dt \,,
}
is consistent with the symmetries of the black brane ansatz. Near the boundary the function $\Phi(r)$  has the expansion
\eqn{}{
\Phi(r) = \mu L- \rho \, {L^2 \kappa^2   \over r^2}  + \ldots \,,
}
where $\mu$ and $\rho$ are respectively the chemical potential and charge density of the conserved $U(1)$ current. In the case of lattice calculations, introducing a nonzero chemical potential is difficult, as it generates a complex action vastly complicating the numerical sampling of the path integral, the so-called ``sign problem". Adding a chemical potential in a holographic dual, on the other hand, just involves adding a single field without any introduction of problems of principle. This makes holographic techniques valuable probes of finite density physics, as we shall see.

\subsection{Transport coefficients}

Transport coefficients encode dynamical processes in the medium, arising as coefficients parametrizing the derivative expansion of the long wavelength (hydrodynamic) behavior of the system. 
The shear viscosity $\eta$ and bulk viscosity $\zeta$ characterize energy-momentum transport, with the shear viscosity describing the fluid's response to shearing stresses, and the bulk viscosity roughly characterizing  the response of the system to expansion or contraction; we also consider the conductivity $\lambda$ of the conserved current.  These can be calculated in field theory via Kubo formulae of the form
\eqn{}{
\lim_{\omega \to 0} {1 \over \omega} {\rm Im}\, G^{\rm R}(\omega, {\bf 0}) \,,
}
where $G^R(\omega, {\bf 0})$ is the zero-momentum retarded Green's function for appropriate polarizations of the energy-momentum tensor or conserved current. Because these are real-time quantities, they are not easily calculated in lattice gauge theory, which is formulated in Euclidean time.
The gravity dual has no corresponding difficulty.

In the holographic dual, such a Green's function is calculated by solving the linearized fluctuation equation for the corresponding  gravity mode, with both infalling boundary conditions imposed at the horizon of the black brane geometry, as well as boundary conditions reflecting a canonically normalized source at the boundary. 
In general, the precise calculation of the full Green's function $G^R$ requires the machinery of holographic renormalization \cite{Bianchi:2001de, Bianchi:2001kw, Skenderis:2002wp}. Fortuitously however, the imaginary part of the Green's function is determined by a so-called ``conserved flux" associated to each linearized fluctuation equation, a combination of solutions guaranteed to be $r$-independent by Abel's identity (for more details of these calculations see \cite{DeWolfe:2011ts}), which does not require holographic renormalization to compute.

Taking time-dependent fluctuations of the form, 
\eqn{Flucts}{
ds^2 &= ds^2_{0} +  {\rm Re}( e^{2A(r)} e^{- i \omega t} h_{\mu\nu}(r)) dx^\mu dx^\nu\,, \cr
A_\mu dx^\mu &= \Phi(r)\, dt + {\rm Re} ( e^{ - i \omega t} a_\mu(r) )dx^\mu  \,, \quad \quad \phi = \phi(r)+{\rm Re} ( e^{ - i \omega t} \tilde\phi(r))  \,,
}
where $ds_0^2$ is the background black brane metric \eno{BHgeom}, the relevant fluctuations are then any of the five traceless $h_{ij}$ with $i, j = x, y, z$ for the shear viscosity $\eta$, any of the three $a_i$ for the conductivity, and for the bulk viscosity the combination of the graviton trace and the scalar fluctuation
\eqn{CurlyH}{
{\cal H} \equiv {1 \over 3} ( h_{xx} + h_{yy} + h_{zz}) - {2 A' \over \phi'} \tilde\phi \,.
}
We note that in the case that $\phi(r)$ does not include a source term $\phi_{(4 - \Delta)}$, the conformal invariance of the theory is not explicitly broken, and the bulk viscosity is identically zero.

Infalling boundary conditions are imposed by taking the near-horizon behavior for $X = h_{ij}, a_i, {\cal H}$:
\eqn{XSoln}{
X(r \to r_H) = (r - r_H)^\alpha (x_0 + x_1 (r - r_H) + \ldots) \,, \quad \quad \alpha \equiv -i \omega  {e^{B(r_H) - A(r_H)} \over h'(r_H) } \,,
}
valid as long as $h(r)$ has a simple zero at the horizon and the other background functions are regular there. The near-boundary conditions are simply
\eqn{BoundaryBC}{
h_{ij}(r \to \infty),\, a_i(r \to \infty), \,{\cal H}(r \to \infty) \to 1 \,.
}
The formulae for the transport coefficients are then
\eqn{Transport}{
\eta = - {1 \over 2 \kappa^2} \lim_{\omega \to 0} {1 \over \omega} {\cal F}_{h_{ij}} \,, \quad\quad
\lambda = -  {L^2 \over 2 \kappa^2} \lim_{\omega \to 0} {1 \over \omega} {\cal F}_{a_i} \,, \quad \quad
\zeta = - {2 \over 9 \kappa^2} \lim_{\omega \to 0} {1 \over \omega} {\cal F}_{\cal H} \,,
}
where the conserved fluxes are 
\eqn{Fluxes}{
{\cal F}_{h_{ij}} = h e^{4A-B} \, {\rm Im} \, (h_{ij}^* h_{ij}') \,, \quad 
{\cal F}_{a_i} = h f(\phi) e^{2A-B} \, {\rm Im} \, (a_i^* a_i') \,, \quad 
{\cal F}_{\cal H} = {e^{4A-B} h {\phi'}^2 \over 4 {A'}^2} {\rm Im}\, ({\cal H}^* {\cal H}') \,,
}
with no sum over indices, given in terms of the solutions $h_{ij}$, $a_i$ or ${\cal H}$ with the boundary conditions imposed to the appropriate fluctuation equations,
\eqn{hhydroEqn}{
h_{ij}'' + \left( 4 A' - B' + {h' \over h} \right) h_{ij}'  + { e^{2B-2A}\over h^2} \, \omega^2 h_{ij} = 0 \,.
}
\eqn{aEqn}{
a_i'' + \left( 2A' - B' + {h' \over h} + { \phi' f'(\phi) \over f} \right)a_i' + {e^{-2A} \over h}  \left( {e^{2B} \over h} \, \omega^2 - \, f(\phi) {\Phi'}^2 \right)a_i = 0\,,
}
and
\eqn{HEqn}{
{\cal H}'' + \left( 4A' -  B' + {h' \over h} + {2 \phi'' \over \phi'} - {2 A'' \over  A'} \right) {\cal H}' + 
 \left( {e^{2B-2A} \over h^2}  \omega^2+ \Sigma_{\cal H}(r) \right) {\cal H} = 0 \,,
 }
 with
  \eqn{}{
 \Sigma_{\cal H} = {h' \over h} \left({A'' \over A'} - {\phi''\over \phi'} \right) + {e^{-2A} \over h \phi'} \left( 3 A' f' - f \phi' \right) {\Phi'}^2 \,.
 }
Although the values of the fluxes are independent of $r$, in general one cannot determine the transport coefficients without solving over the entire space, since boundary conditions must be imposed on both ends. 

The shear viscosity equation, however, is simple enough that it can be solved precisely, giving the famous universal value of $\eta/s$ for these geometries \cite{Buchel:2003tz,Kovtun:2004de}.
We may demonstrate this by using an $\omega \to 0$ expansion to analytically match the $r \to r_H$ and $r \to \infty$ limits; 
we take our discussion from \cite{Gubser:2008sz}.
 In the $\omega \to 0$ limit  the term in \eno{hhydroEqn} with no derivatives vanishes, and we are left with
\eqn{}{
\partial_r (\log h_{ij}') = -\partial_r (4A -B + \log h) \,,
}
which has the solution
\eqn{}{
h_{ij} = a_0 + b_0 \int_r^\infty dr\,  {e^{-4A+B} \over h} \,.
}
The second term is  not allowed strictly at $\omega =0$ due to a logarithmic divergence; it may be kept for very small $\omega$, but for us it is enough to note that matching to the near-horizon expansion
\eqn{}{
h_{ij}(r) \approx h_0 (r - r_H)^{\alpha } = h_0  (1 +\alpha  \log (r - r_H) + \ldots) \,,
}
we have $h_0= a_0$; however as $r \to \infty$ the boundary condition \eno{BoundaryBC} requires $a_0 = 1$.
Thus in this case the near-boundary condition directly controls the near-horizon condition. Evaluating the conserved flux \eno{Fluxes} near the boundary, the shear viscosity then becomes
\eqn{}{
\eta = {1 \over 2 \kappa^2} e^{3 A(r_H)} \,,
}
implying the shear viscosity to entropy density ratio takes the  universal form for this class of theories,
\eqn{}{
{\eta \over s} = {1 \over 4 \pi} \,.
}
In more generality, this result holds for any two-derivative gravity theory \cite{Buchel:2003tz,Kovtun:2004de}.

The other transport coefficients are not universal in this way; from a practical standpoint, imposing the boundary condition at infinity will not impose a universal constraint on the near-horizon behavior. The modes may be determined by solving the complete differential equation, in general numerically, and the values of the conductivity $\lambda$ and bulk viscosity $\zeta$ are different for different geometries, as we shall see in certain examples.

The various Kubo formulae are all proportional to the five-dimensional gravitational constant $1/\kappa^2$, which as discussed is proportional to $N_c^2$.  As a result, all three transport coefficients go like $N_c^2$ in the large-$N_c$ limit defined by the dual gauge theory, and the ratios $\eta/s$ and $\zeta/s$ are of order 1.

\subsection{Adding Flavor}
\label{FLAVOR}

While ${\cal N}=4$ Super-Yang-Mills shares a number of properties with QCD, it remains a theory containing only adjoint matter. It is natural to extend the duality to cases involving matter in fundamental representations. In the string theory constructions that inspire AdS/CFT, the field theory arises from the dynamics of open strings on a collection of D-branes, with ${\cal N}=4$ Super-Yang-Mills being the worldvolume theory of a stack of D3-branes. Adding additional branes of other dimensionalities and configurations introduces fundamental matter as the dynamics of open strings stretched between the brane collections. In the AdS/CFT correspondence, this corresponds to adding $N_f$ ``flavor branes" inside $AdS_5 \times S^5$; this was first studied for D7-branes in \cite{Karch:2002sh}. For $N_f \ll N_c$, the backreaction of these branes may be neglected, and their impact can be studied in the so-called probe, or quenched, limit. The world-volume action of a single (p+1)-dimensional Dp-brane is
\eqn{Brane}{
S_p = - T_p \int d^{p+1}\xi \, e^{-\phi} \sqrt{\det(P[g]_{ab} + 2 \pi \alpha' F_{ab})} \,,
}
where $T_p \equiv \mu_p = (2 \pi)^{-p} {\alpha'}^{-(p+1)/2}$ is the brane tension, $P$ denotes the pullback of the spacetime metric to the brane worldvolume, and $F$ is the gauge field localized to the brane, holographically dual to the conserved global U(1) ``baryon number" associated to the fundamental matter. Alternatively, one can move beyond the quenched limit and access dynamical quark processes by solving the fully back reacted system. One approach to this end is to work in the so-called ``Veneziano limit" \cite{Veneziano:1979ec}, where $N_c,\,N_f\to \infty$ but $x$ is held finite. 

To date, both avenues have been used to investigate the properties of chiral symmetry breaking in strongly coupled gauge theories. In the quenched approximation, one approach (outlined in section \ref{PhaseFund}) uses the details of the embedding of probe $D7$-branes to study a broken ``chiral symmetry" $U(1)$ associated with the positions of the branes in the bulk; this case preserves ${\cal N}=2$ supersymmetry and hence the ``quarks" include fundamental scalars as well as fermions. One may move beyond the quenched approximation by including the backreaction of the D7-branes, which we will touch on as well. In a different realization of the unquenched approach, parallel stacks of $D4$ and $\overline{D4}$-branes are are aligned in the bulk, breaking supersymmetry entirely. The worldvolume theory of this set-up contains an open string tachyon stretching between the branes and transforming under the $U(N_f)\times U(N_f)$ gauge symmetry of the system, reflecting the instability of the system, and is holographically identified with the chiral condensate in the boundary theory. More details on this program, titled ``VQCD", can be found in section \ref{VQCD}.

\subsection{Hard probes}
\label{HardProbes}

An important class of observables in heavy ion collisions is related to the rapid energy loss by high-momentum particles traveling through the quark-gluon plasma.  Three approaches to energy loss which have been pursued in an AdS/CFT context are: heavy quark drag, reviewed in section~\ref{HEAVY}; light quarks and gluons as falling strings, reviewed in section~\ref{LIGHT}; and the jet-quenching parameter as determined by lightlike Wilson loops, also reviewed in section~\ref{LIGHT}.

\subsubsection{Heavy quarks}
\label{HEAVY}

A framework for understanding the drag force and stochastic forces on heavy quarks moving through a thermal plasma of ${\cal N}=4$ super-Yang-Mills theory was built up in a series of articles \cite{Herzog:2006gh,CasalderreySolana:2006rq,Gubser:2006bz,Gubser:2006nz,CasalderreySolana:2007qw}.

 \begin{figure}
 \begin{center}
  \includegraphics[width=4.5in]{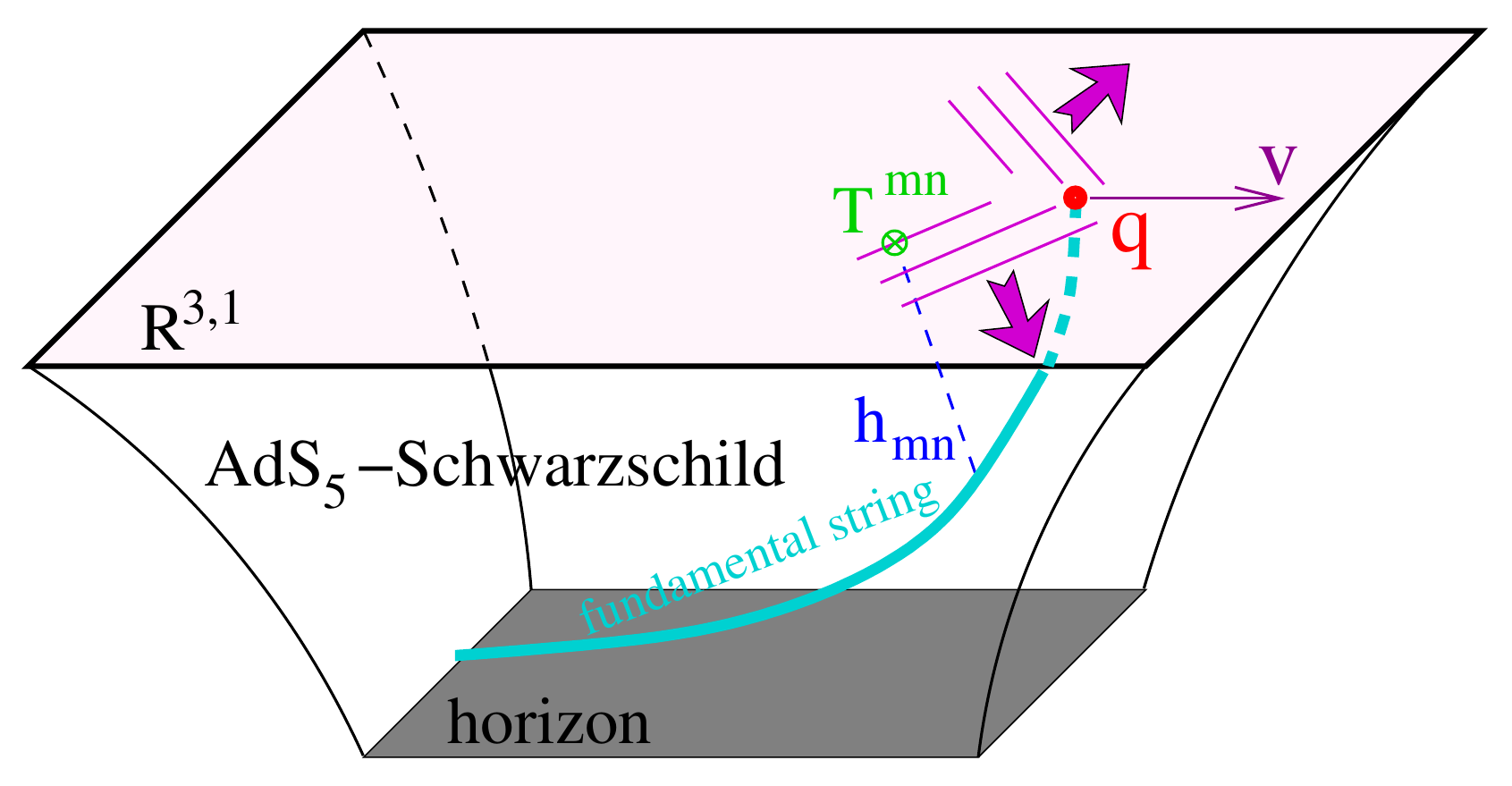}
  \caption{The trailing string.  If the quark's position is $x = vt$, then the shape of the string in the steady state solution is given by $x = vt + {L^2 v \over 2r_H} \left( \tan^{-1} {r_H \over r} - \tanh^{-1} {r_H \over r} \right)$.  From \cite{Friess:2006fk}.}\label{TrailingString}
 \end{center}
 \end{figure}
The string theory construction is shown in cartoon form in figure~\ref{TrailingString}.  The key idea is that the endpoint of the string acts as a source for the gauge-theoretic color fields whose location can be prescribed.  The shape of the string indicates how the theory responds to the motion of the endpoint.  The total energy of the string is infinite because of a divergence near the boundary: this is dual to the infinite energy of a Coulombic field around a perfectly pointlike charge.  This divergence occurs no matter what the motion of the endpoint is, but it can be cut off by introducing a brane into the bulk as described in section~\ref{FLAVOR}.  With such a cutoff, the string can be understood as dual to a massive quark.  As the cutoff is removed, the quark mass becomes infinite.  This is why trailing string calculations generally should be understood as corresponding to the dynamics of quarks whose mass is substantially larger than the temperature of the medium.

When the velocity $v$ of the quark is constant, the shape of the string can be straightforwardly computed, and the drag force it exerts on the endpoint is easily seen to be
 \eqn{DragForce}{
  {dp \over dt} = -{\pi \sqrt\lambda \over 2} T_{SYM}^2 {v \over \sqrt{1-v^2}} \,.
 }
For a quark whose mass is large but finite, corrections to the formula $p/m = v/\sqrt{1-v^2}$ are negligible, so we may re-express \eno{DragForce} as
 \eqn{DragForceAgain}{
  {dp \over dt} = -{p \over \tau_Q} \qquad\hbox{where}\qquad 
    \tau_Q = {2m_Q \over \pi T_{SYM}^2 \sqrt\lambda} \,,
 }
where $\lambda = g_{\rm YM}^2 N$ is the 't Hooft coupling.  It was argued in \cite{Gubser:2006qh} that a good comparison between ${\cal N}=4$ super-Yang-Mills theory and QCD can be made at fixed energy rather than fixed temperature, and with $\lambda \approx 5.5$ to match to lattice results on the static quark-anti-quark potential.  With approximately these parameters, a phenomenological study \cite{Akamatsu:2008ge} of non-photonic electrons from top-energy RHIC collisions showed fairly good agreement with the string theory predictions.  Altogether, the AdS/CFT calculations of $\eta/s$ and heavy quark drag force yield fairly impressive agreement with data from top-energy RHIC collisions \cite{Noronha:2009vz}.

In a related line of inquiry, the detailed spatial pattern of energy loss was calculated \cite{Friess:2006aw,Friess:2006fk,Yarom:2007ap,Gubser:2007nd,Yarom:2007ni,Gubser:2007ga,Chesler:2007an,Chesler:2007sv} across a wide range of length scales and was argued in \cite{Noronha:2008un,Betz:2008wy} to exhibit stronger high-angle emission than in perturbative QCD.  This high-angle emission does {\it not} come from the hydrodynamic region; indeed, it has been shown fairly generally in this regime that the diffusion wake is as strong as the Mach cone \cite{Gubser:2007ni}.  Instead, the dominant high-angle emission comes from a region near the quark, where the fields deviate from the Coulombic fields but are far from local hydrodynamic equilibration.  The form of the stress tensor in this ``neck region,'' as it was called in \cite{Noronha:2008un}, was found through a combination of asymptotic expansions \cite{Yarom:2007ap,Gubser:2007nd,Yarom:2007ni} and numerics \cite{Gubser:2007ga}.

Thus far, LHC results on hard probes seem to fit reasonably well to perturbative QCD expectations \cite{Zapp:2011ek}.  However, variants of the trailing string construction remain interesting theoretical tools, and we will close this section by summarizing two directions of further work.  First, detailed calculations of stochastic forces can be carried out \cite{CasalderreySolana:2006rq,Gubser:2006nz,CasalderreySolana:2007qw}, resulting in diffusion coefficients
 \eqn{KappaValues}{
  \kappa_T = \sqrt{\lambda\gamma} \pi T_{SYM}^3 \,, \qquad
  \kappa_L = \sqrt\lambda \gamma^{5/2} \pi T_{SYM}^3 \,,
 }
for transverse and longitudinal momentum.  The latter result indicates stochastic forces which grow more quickly with velocity than the Einstein relation allows.  The stochastic forces emerge on the gravity side from Hawking radiation from a causal horizon on the string worldsheet which is well above the event horizon of the ambient spacetime when the motion of the endpoint is relativistic.

Another direction pursued recently is the study of how synchrotron radiation propagates in a strongly coupled gauge theory, both without \cite{Athanasiou:2010pv} and with \cite{Chesler:2011nc} the presence of a thermal medium.  In addition to expected features such as the production of sound waves and rapid attenuation of high-energy gluons, interesting findings are that the angular spread of the radiation does not increase with attenuation, nor does the momentum distribution of gluons shift appreciably toward longer wavelengths during the attenuation process.

\subsubsection{Light quarks and gluons}
\label{LIGHT}

One of the two main approaches in AdS/CFT to energy loss by light quarks and
gluons is to consider string falling in an $AdS_5$-Schwarzschild background.
This is motivated in part by \cite{Gubser:2008as}, in which it was argued that
gluon scattering at zero temperature could be described by a string worldsheet
in $AdS_5$ where the asymptotic gluon states have both ends of the string
stretching down into $AdS_5$.  At finite temperature, a sensible adaptation of
this picture is to represent an energetic off-shell gluon as a string doubled
over in $AdS_5$-Schwarzschild with both its ends going through the black hole
horizon \cite{Gubser:2008as}.  Energy loss is associated with the rest of the
string falling into the horizon.  It was argued in \cite{Gubser:2008as} that
the stopping distance of an energetic gluon takes the form $x_{\rm stop}
\leq C/T \, (E/\sqrt{\lambda} T)^{1/3} $, where the dimensionless constant
$C$ is of order unity.  It was further remarked in \cite{Gubser:2008as} that
light quarks could be treated by replacing the doubled string by an open
string.  (The scaling $x_{\rm stop} \propto E^{1/3}$ was independently
proposed in \cite{Hatta:2008tx,Dominguez:2008vd} based on somewhat different considerations.)

A difficulty in the proposal of \cite{Gubser:2008as} is that the initial state
of the string representing the gluon (or light quark) is highly
underdetermined, making it difficult to provide an unambiguous evaluation of
the constant of proportionality in the stopping distance relation.  
Some bounds were provided in \cite{Gubser:2008as}. 
Significant refinements were developed
in subsequent numerical work \cite{Chesler:2008wd,Chesler:2008uy}, which also
considered a different initial state: a color singlet state where an open
string starts in a pointlike configuration in $AdS_5$-Schwarzschild and then
expands lengthwise as it falls down toward the horizon.  
This
numerical work varied the undetermined initial conditions  to 
achieve the maximum stopping distance, determining the constant of proportionality for light quarks,
$x_{\rm stop}^{\rm max} = 0.526/T\,  (E/\sqrt{\lambda} T)^{1/3}$.  

The physical meaning of the ambiguity in the initial conditions in these
calculations was clarified in subsequent work \cite{Arnold:2010ir,Arnold:2011qi}, where
a jet was formed by turning on a localized source in the boundary theory.  A 
jet is characterized  by an energy  $E$ and virtuality $Q^2$ with $E^2 \gg Q^2$. Thus,  the four
momentum of the jet is  $Q^{\mu} = (E+\sigma, 0, 0, E-\sigma)$ and the
virtuality is $Q^2 \sim  4 E \sigma$, where $\sigma$  is 
related to the inverse formation length of the jet, $\sigma \sim 1/L$. In the 
setup of \cite{Arnold:2010ir,Arnold:2011qi},
$L$ is the duration  the source in the boundary theory, which is directly 
related to the initial ``depth" in $z$ of the lightlike geodesic 
that is dual to the jet propagating in the boundary theory.
Then the stopping distance for jets having energy $E$ and virtuality $Q^2$ is of order 
$x_{\rm stop}  \sim  1/T\,  (E^2 /Q^2)^{1/4} \sim 1/T \, (E L)^{1/4}$.  The maximum
stopping distance can be estimated by setting the formation length of the
jet equal to the stopping distance itself, $L =x_{\rm stop}^{\rm max}$, in order to  achieve the smallest possible virtuality.  
With this requirement, a short instructive exercise shows that $x_{\rm stop}^{\rm max} \propto E^{1/3}$, as  determined previously.

 \begin{figure}
 \begin{center}
  \includegraphics[width=4in]{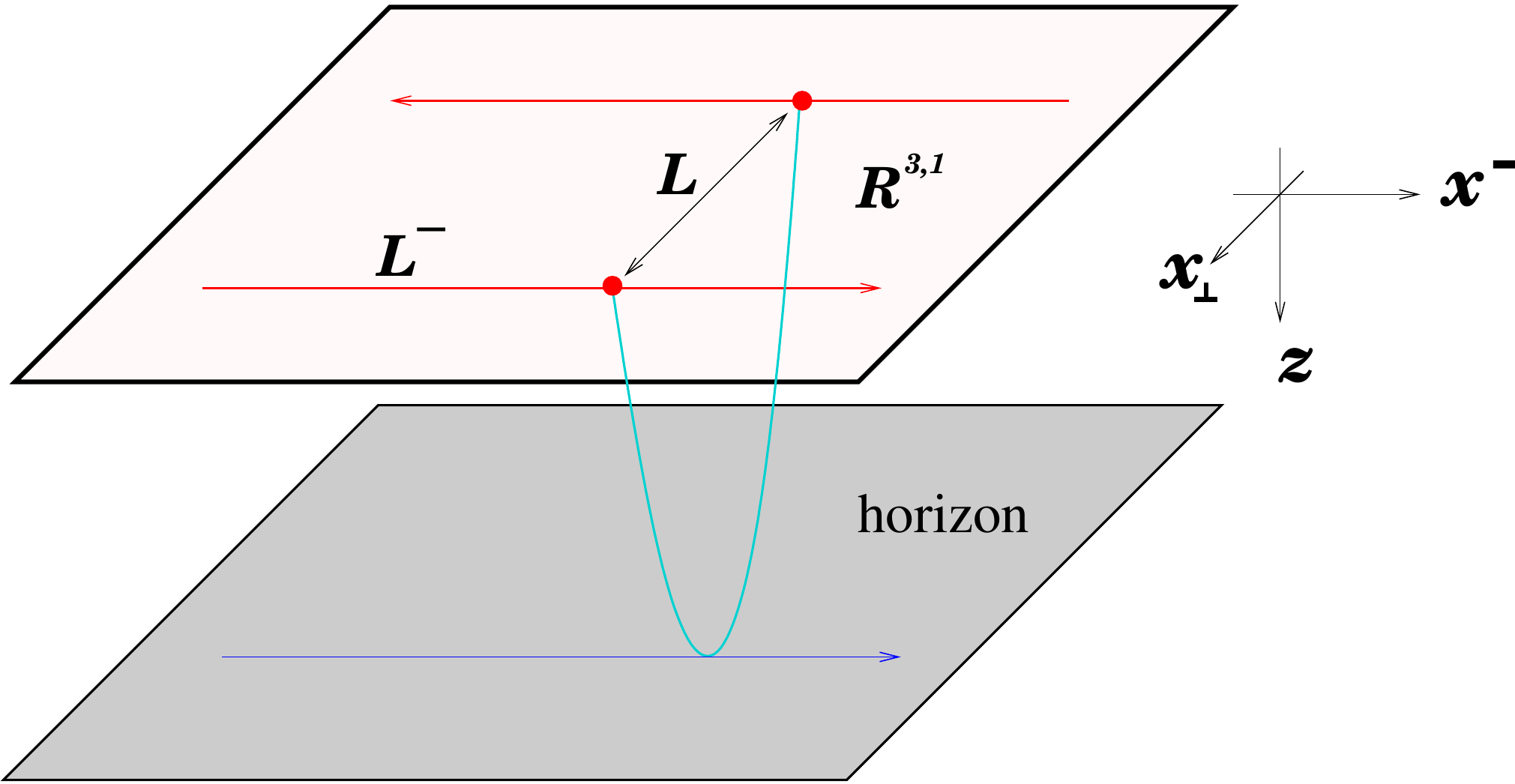}
  \caption{The string worldsheet underlying the calculation of $\hat{q}$.  The expectation value of the partially lightlike Wilson loop is $\langle W^A({\cal C}) \rangle \approx \exp\left\{ -{1 \over 4\sqrt{2}} \hat{q} L^- L^2 \right\}$.  The lightlike coordinate is $x^- = t-x$, and the transverse coordinates are $x_\perp = (y,z)$.}\label{qhatFigure}
 \end{center}
 \end{figure}
 An orthogonal line of work on light quark energy loss
 \cite{Liu:2006ug,Liu:2006he} seeks to determine the jet-quenching parameter
 $\hat{q}$ of the BDMPS formalism \cite{Baier:1996sk,Zakharov:1997uu} in terms
 of properties of lightlike Wilson loops in $AdS_5$-Schwarzschild.  The
 particular Wilson loop considered connects two points separated by a spatial
 distance $L$ and continues down into the bulk to just graze the horizon: see
 figure~\ref{qhatFigure}.  
 The result of evaluating the Wilson loop is
 \eqn{qHatValue}{
  \hat{q} = {\pi^{3/2} \Gamma(3/4) \over \Gamma(5/4)} \approx
    3.6 {{\rm GeV}^2 \over {\rm fm}} \left( {T_{SYM} \over 280 {\rm MeV}} \right)^3 \,,
 }
where in the last approximate equality, we used $\lambda = 6\pi$ as preferred
by the authors of \cite{Liu:2006ug}.  
 The worldsheet configuration has been called into
 question in \cite{Argyres:2006yz} but was later reexamined in
 \cite{DEramo:2010ak}. The configuration that was chosen in 
 the original calculation approaches the light cone from below, {\it i.e.} 
 for $x = v t$ with $v=1+\epsilon$. At weak
 coupling this prescription can be justified, and  has been used to determine the jet-quenching
 parameter beyond leading order \cite{CaronHuot:2008ni}.  
The result \eno{qHatValue} is for ${\cal
N}=4$ super-Yang-Mills theory, and some reduction should be made when
attempting a comparison with QCD.  In \cite{Liu:2006he}, based on a study of
other holographic theories, it was suggested that the reduction factor is about
$0.63$, corresponding to the ratio $\sqrt{s_{\rm QCD} / s_{SYM}}$.

\section{Improved Holographic QCD}
\label{IHQCD}

\renewcommand{\(}{\left(}
\renewcommand{\)}{\right)}

\newcommand{\wdg}{\wedge}
\newcommand{\hdg}{\star}
\newcommand{\Apb}{\mathbb{A}}

\subsection{Improving Holographic Models of QCD}

Improved Holographic QCD is a program developed in \cite{Gursoy:2007cb,Gursoy:2007er}, taking a ``bottom-up" approach to modeling QCD across many energy scales. A gravity-scalar system is constrained by requiring the radial profile of the scalar ``dilaton" to mimic the development of the QCD beta function and by matching lattice results for the thermodynamics of the confinement/deconfinement phase transition.

The Lagrangian is taken to be
\begin{equation}\label{eq:sihqcd}
S = -M_p^3 N_c^2\int \dd^5x\sqrt{g}\(R-\frac{4}{3}(\partial\Phi)^2+V(\Phi)\) \,,
\end{equation}
fitting into the class of \eno{LwithF} with no gauge field and a rescaled scalar; the gravitational constant is written to explicitly highlight the $N_c^2$ scaling of the action, with the dimensions encoded in the five-dimensional Planck scale $M_p$.

The dilaton encodes the running 't Hooft coupling via the relationship $\lambda = e^{\Phi}$. This running is in turn governed by the potential $V(\Phi)$, which can be chosen judiciously to reproduce the desired characteristics of the dual field theory. More explicitly, through careful engineering of the potential, it is possible to produce a gravitational theory whose dual is confining in the IR and has a coupling $\lambda$  which runs towards zero in the UV.  This is because the potential appearing in (\ref{eq:sihqcd}) is in a one to one correspondence with the $\beta$-function of the dual gauge theory. This matching is a critical aspect of the IHQCD construction, and what distinguishes it most markedly from other Einstein-dilaton models used in gauge/gravity duality. The functional form of the potential is unspecified in this bottom-up approach, but a suitable choice can be inferred from asymptotic considerations. 

Roughly, the idea is to engineer a bulk relationship between the gauge theory coupling $\lambda$ and its energy scale $\mu$. The latter can be accomplished by noting that a field theory observer living at the boundary of the bulk theory will measure energies redshifted from the interior like
\begin{equation}
\mu = \sqrt{g_{tt}}\,\mu_{\mathrm{bulk}} \,,
\end{equation}
which is to say $\log\mu \sim\frac{1}{2}\log g_{tt}$. From this, one makes the following identification between bulk solutions and the gauge theory $\beta$-function:
\begin{equation}
\beta(\lambda) \equiv \frac{\dd\lambda}{\dd \log\mu} \leftrightarrow 2\,\frac{g_{tt}}{g_{tt}{}'}\,\Phi'\,e^{\Phi} \,,
\end{equation}
where it is assumed that the dilaton and metric functions depend only on the radial variable $r$, and that a ``prime" denotes a derivative with respect to this variable. In an asymptotically free theory, the UV has $\beta\sim -\lambda^2$ and so the coupling at large energies runs like 
\begin{equation}\label{eq:rgc}
\lambda \sim \frac{1}{\log\mu} \leftrightarrow e^{\Phi}\sim \frac{2}{\log g_{tt}} \,.
\end{equation}
It turns out that this behavior can be ensured by requiring that the potential $V(\lambda)$ is asymptotically AdS near the boundary, corrected by a term linear in $\lambda$ which is responsible for the logarithmic running in (\ref{eq:rgc}).

At low energies, confinement, a linear glueball spectrum, and bulk fluctuations that obey well-posed spectral problems require the potential to take the form
\begin{equation}
V_{\mathrm{IR}}\sim \lambda^{4/3}\sqrt{\log\lambda} \,.
\end{equation}
This somewhat non-obvious limiting behavior can in part be motivated by the fact that generic dilaton potentials in non-critical string theories behave like $V\sim \lambda^{4/3}$ in five dimensions. The root of the log is somewhat more obscure, and is required by confinement at low energies.

In \cite{Gursoy:2007cb} it was found that a potential like
\begin{equation}\label{eq:ihqcdV}
V(\lambda) = \frac{12}{L^2}\left[1+V_0\lambda+V_1\lambda^{4/3}\sqrt{\log\Big(1+V_2\lambda^{4/3}+V_3 \lambda^2\Big)}\right] \,,
\end{equation}
provides a holographic description of a large $N_c$ Yang Mills theory which confines in the IR, produces a phenomenologically favorable glueball spectrum, and forces $\lambda\to0$ in the UV. The various parameters that appear in (\ref{eq:ihqcdV}) are not strictly independent. For example, in the limit $\lambda\to 0$, matching to the Yang Mills $\beta$-function constrains $V_2$ in terms of $V_0$ and $V_1$.  In fact, one can show that the only phenomenologically tunable parameters remaining in (\ref{eq:ihqcdV}) are $V_1$ and $V_3$. Loosely, it turns out that $V_1$ controls the speed with which the thermodynamic densities (such as the entropy density $s/T^3$) approach the free field limit as functions of $T/T_c$, while $V_3$ governs the latent heat. Once values for these constants are chosen, the gravity theory is fully specified, and solutions to the system represent predictions of the model.

By studying the thermodynamics of these solutions, and in particular the free energy as a function of temperature, one can identify the thermodynamically favored solutions and search for discontinuities in thermodynamic susceptibilities, signifying phase transitions. 

\begin{figure}
\centering
\includegraphics[scale=0.15]{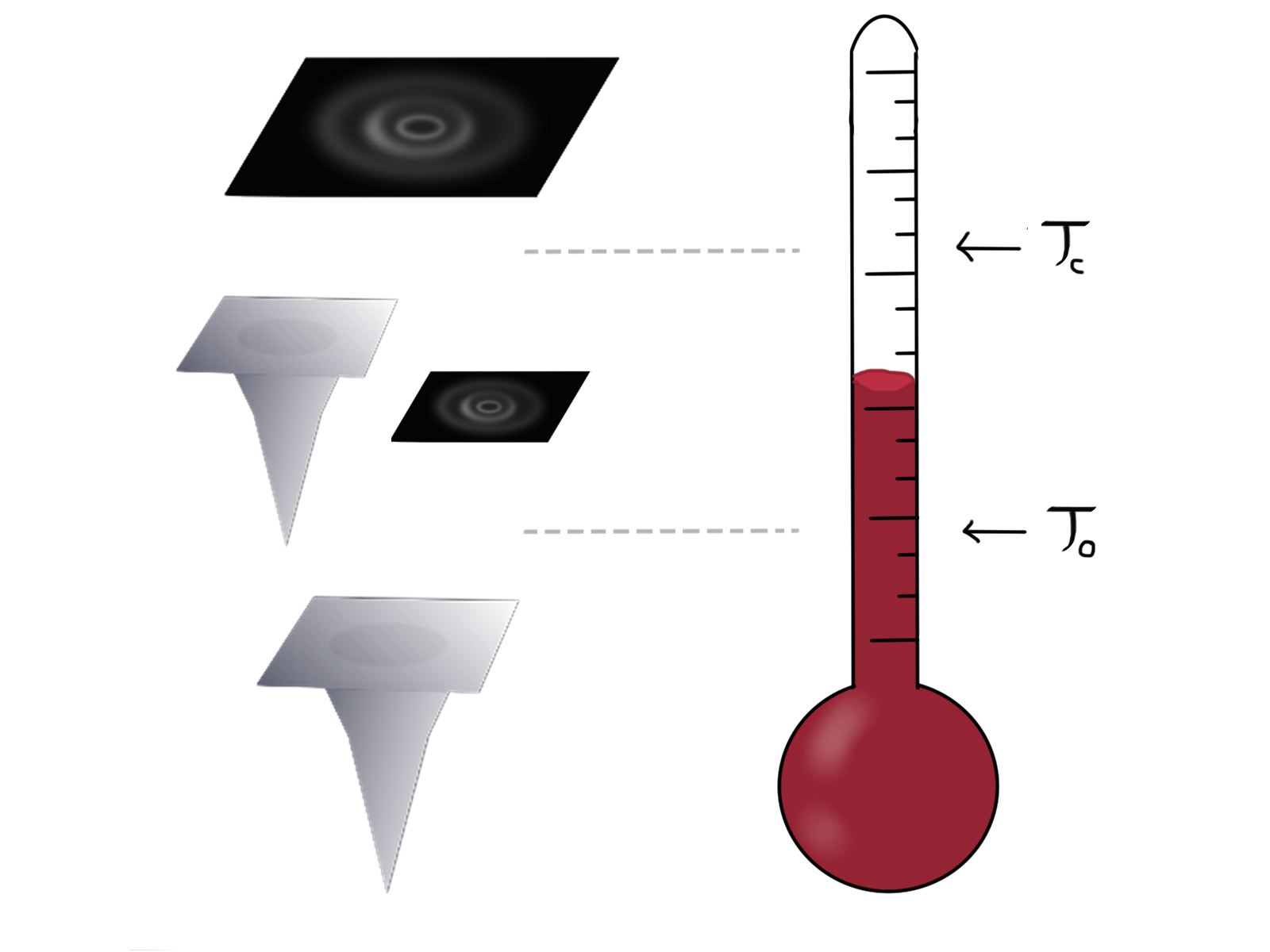}
\caption{\label{fig:ihqcdpd} The phase structure of Improved Holographic QCD. A first order phase transition occurs at $T=T_c$, signaling the thermodynamic dominance of the black brane solution.}
\end{figure}

Broadly, in the bulk theory of (\ref{eq:sihqcd}), (\ref{eq:ihqcdV}) above, the thermodynamics can be summarized by the cartoon in figure \ref{fig:ihqcdpd}. At low temperatures, 
the only solution to the equations of motion is the thermal gas. As the temperature increases, the thermal gas persists as the dominant solution (lowest free energy), but black brane solutions begin to appear. Finally, at the critical temperature $T = T_c$ 
there is a first order phase transition from the thermal gas to the black brane geometry, which remains the thermodynamically favored solution for all $T>T_c$.

Through careful choice of the potential parameters $V_1$ and $V_3$, one can arrange for this transition to mimic that of pure Yang-Mills. This is accomplished by determining the thermodynamics for each numerical solution, 
and studying the behavior of various (dimensionless) thermodynamic densities: $s/T^3$, $p/T^4$, and $\varepsilon/T^4$  where $p$ is the pressure and $\varepsilon$ is the energy density. These functions are constrained by the first law of thermodynamics, as $\varepsilon = p+sT$. The tunable parameters $V_1$ and $V_3$ are varied systematically, until quantitative agreement with lattice results is reached.

\begin{figure}[h!]\centering
\includegraphics[scale=1]{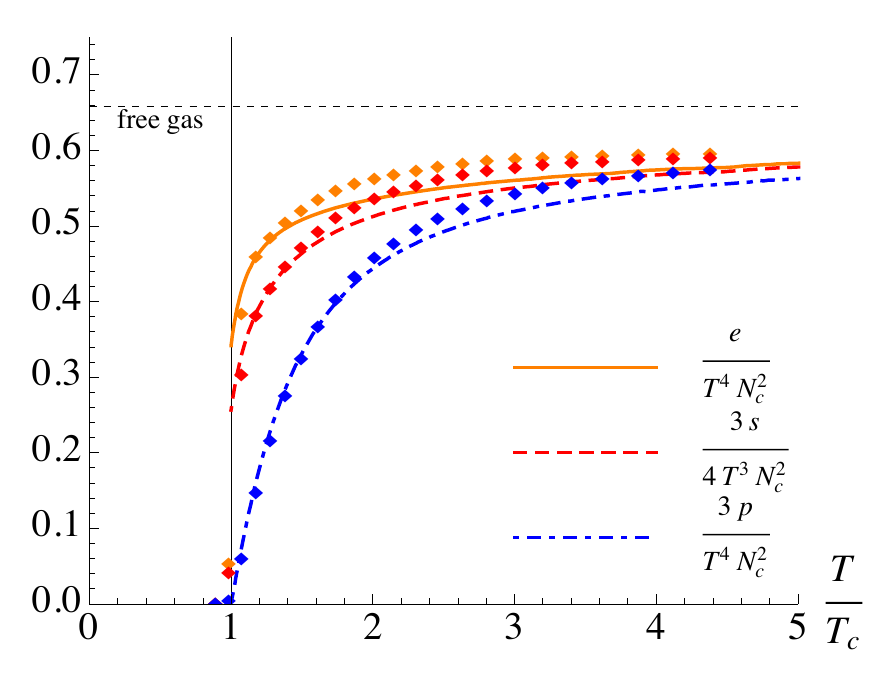}
\caption{\label{fig:splot}Comparison between IHQCD predictions (curves) for thermodynamic densities and the corresponding lattice results (points) for $N_c = 3$. Figure from \cite{Gursoy:2009jd}. More recent lattice results for Yang-Mills theory at large $N_c$ are reviewed in \cite{Lucini:2012gg}. }
\end{figure}

In \cite{Gursoy:2007cb} it was found that the best agreement with lattice data came from the parameter choice
\begin{equation}\label{eq:params}
V_1 = 14 \qquad \mathrm{and} \qquad V_3 = 170 \,,
\end{equation}
which produces the predictions displayed in figure \ref{fig:splot}. As is evident from the figure, the predictions from the IHQCD model are in excellent agreement with the lattice in the vicinity of the phase transition at $T=T_c$, and deviate primarily in their approach to the free gas.

With the potential parameters fully fixed via phenomenological fits to lattice data, it is interesting to explore the predictions this model makes for various observables of the dual gauge theory. An important first step is the spectrum of fluctuations about the zero temperature backgrounds. Specifically, under holography, the spectrum of normalizable fluctuations of the $T=0$ Einstein-dilaton system are dual to massive spin-0 and spin-2 glueballs.

\begin{figure}
\centering
\includegraphics[scale=1.2]{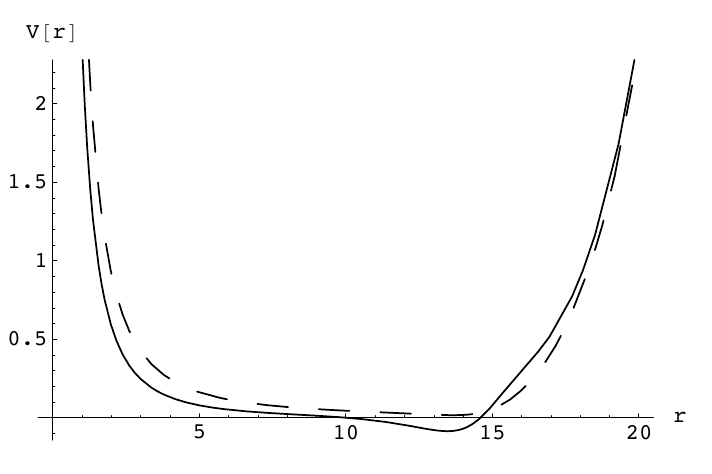}
\caption{\label{fig:pot} Effective Schrodinger potential for scalar (solid) and tensor (dashed) glue balls in IHQCD. Figure from \cite{Gursoy:2007er}.}
\end{figure}

As an illustrative example, it is useful to consider the lone gauge invariant spin-0 mode \eno{CurlyH}, whose fluctuation equation can be recast as 
a Schr\"odinger equation of the form
\begin{equation}
-\mathcal{H}''+V(r)\mathcal{H} = m^2\mathcal{H}
\end{equation}
and $m$ determines the glueball mass and $V(r)$ is an effective potential which depends on the various metric functions in a complicated way. An example of this potential evaluated on a zero temperature background is shown in figure (\ref{fig:pot}). Elementary quantum mechanics considerations immediately show that the spectrum is discrete, and suggest the presence of a mass gap.

\begin{figure}
\centering
\includegraphics[scale=0.32]{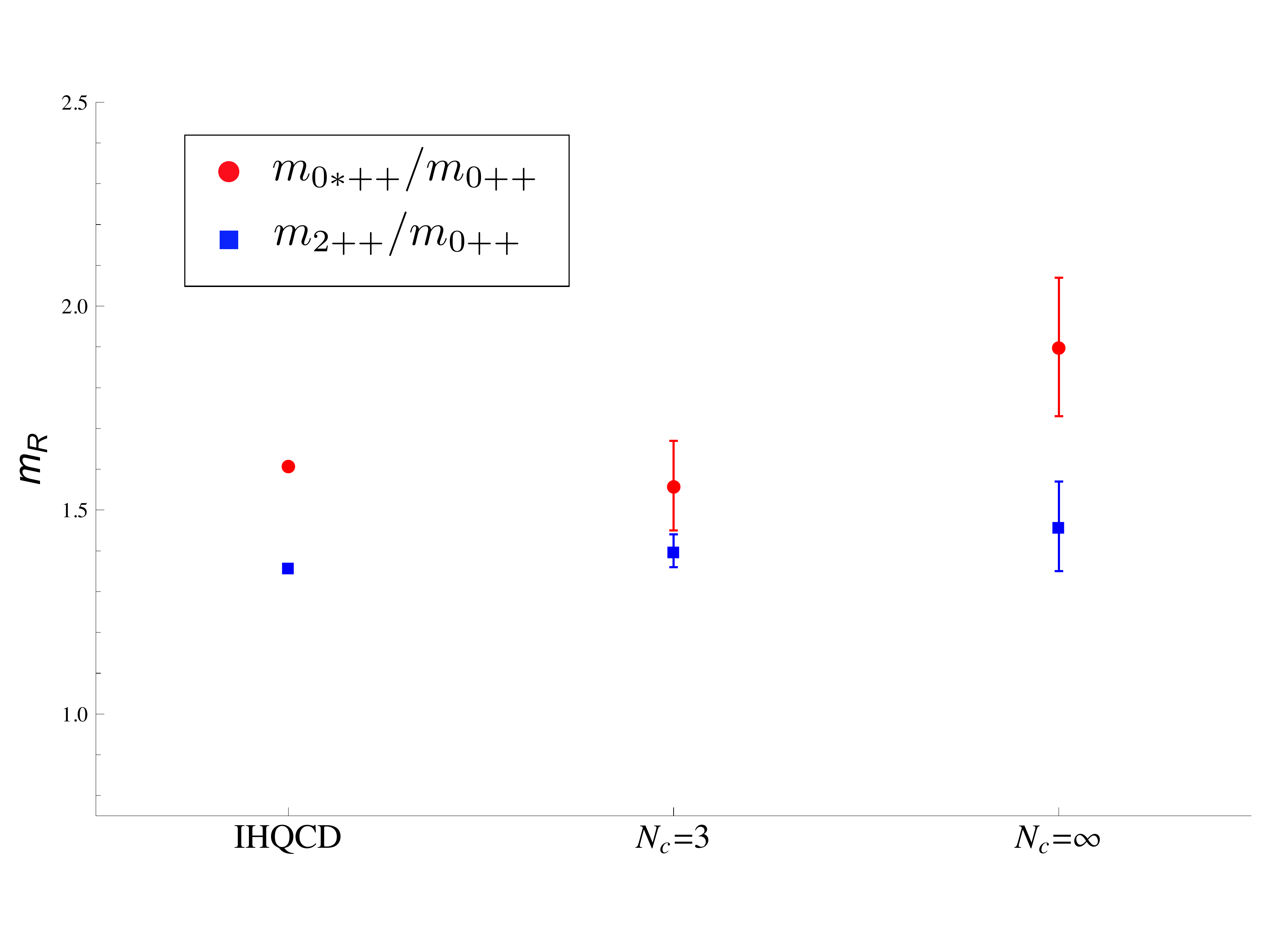}
\caption{\label{fig:gbspec}Glueball mass ratios in IHQCD and on the lattice. Lattice results from \cite{Morningstar:1999rf,Lucini:2001ej}.}
\end{figure}

Numerically solving the Schr\"odinger eigenvalue problem is equivalent to computing the spin-0 glueball spectra in the dual gauge theory. Operationally, one simply scans over $m$ values until a normalizable mode develops in the bulk. The corresponding mass $m$ is dimensionful, but because the non-critical theory does not descend from a known string theory, its precise numerical value is an ambiguous prediction. To circumvent this, one instead computes mass ratios between different glueball states.  These ratios are of course dimensionless, and can be compared directly to the lattice. This comparison is shown in figure \ref{fig:gbspec}. 

\begin{figure}
\centering
\includegraphics[scale=1.2]{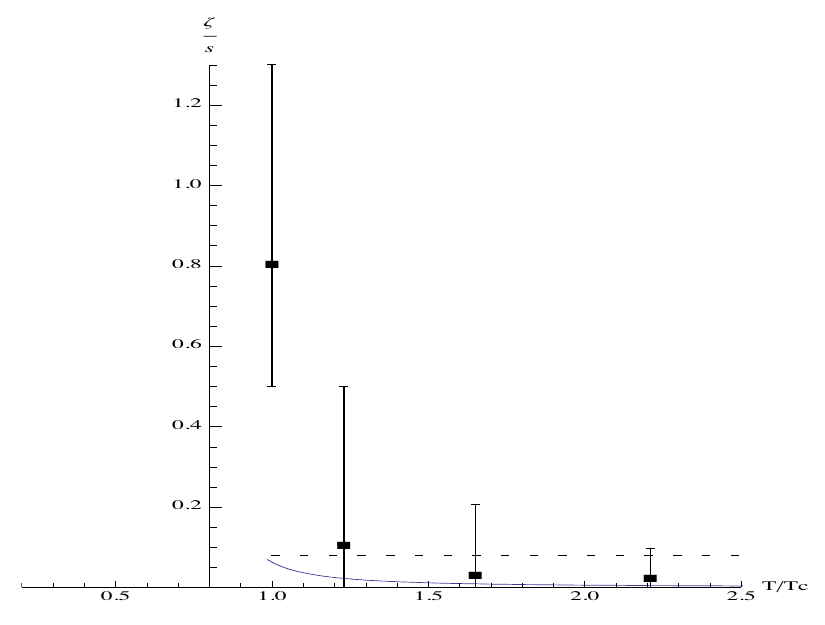}
\caption{\label{fig:bvis} The bulk viscosity divided by entropy density of  IHQCD (solid curve) compared to the universal value of shear viscosity per entropy density (dashed line) and lattice data for $SU(3)$ gluodynamics. Figure from \cite{Gursoy:2009kk}.}
\end{figure}

As the glueball masses provide an extra dimensionful parameter of the theory, it is possible to use them to measure quantities like the critical temperature in physical units (MeV). One simply fixes the lowest lying glueball mass in the holographic theory to the lattice result, and reworks the various dimensionless ratios computed via holography, restoring units systematically. For the parameter choice of (\ref{eq:params}), one finds
\begin{equation}
T_c = 247 \,\, \mathrm{MeV} \,.
\end{equation}
One can also compute transport coefficients; the shear viscosity takes the universal value of $\eta/s = 1/4\pi$, but the bulk viscosity is nontrivial to calculate. Evaluating it 
in a manner analogous to that described previously, one finds the results shown in figure \ref{fig:bvis}. It is noteworthy that the bulk viscosity in this model is always smaller than the shear viscosity, as well as the fact that there is a sudden rise in the bulk viscosity near the first order transition at $T_c$. Similar behavior was also noted in \cite{DeWolfe:2011ts}, as will be described in the next section.

\begin{figure}
\centering
\includegraphics[scale=1.2]{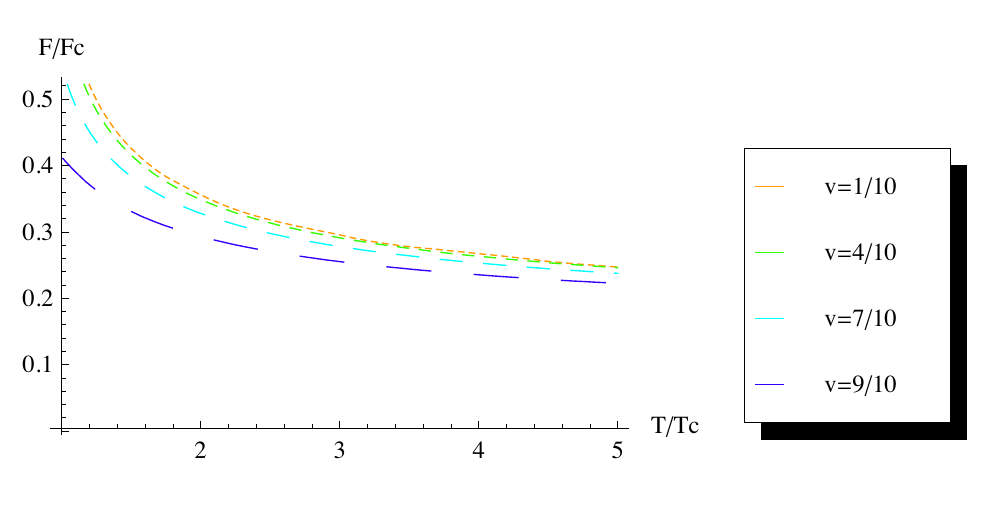}
\caption{\label{fig:Fd} The drag force as a function of temperature for IHQCD divided by the $\mathcal{N}=4$ result. Plot from \cite{Gursoy:2009kk}.}
\end{figure}

As already mentioned, these viscosities characterize the response of the system to perturbations of its stress-energy tensor. A related yet distinct quantity is the response of a system to an external ``test quark" dragged through it, as described in section~\ref{HardProbes}. The basic idea is to affix a string to the boundary of the space-time, and pull it through the plasma at constant velocity $v$.  As the test quark is dragged through the medium, the string will lose momentum by depositing it into the horizon on its world sheet. The drag force, identified with the rate of momentum loss, is found to be
\begin{equation}
F_d = -\frac{1}{2\pi l_s^2}v \,e^{2(A+\frac{2}{3}\Phi)} = -\frac{1}{2\pi l_s^2}v \,\lambda^{\frac{4}{3}}\,e^{2A}
\end{equation}
and is independent of the radial location at which it is evaluated (i.e. $\dd F_r/\dd r=0$). The drag force is plotted as a function of temperature and normalized to the conformal value for several speeds in figure \ref{fig:Fd}.

\subsection{Adding Flavor: VQCD}
\label{VQCD}

Moving beyond the one dimensional phase diagram described in the previous section can be accomplished by adding flavor. In particular, one can extend the study of large-$N_c$ Yang-Mills theory to a theory with many colors and $N_f$ flavors of ``quarks". This approach was pioneered by Veneziano in the '70's \cite{Veneziano:1979ec}, and accordingly is known as the Veneziano Limit:
\begin{equation}
N_c,N_f\to\infty \qquad x \equiv\frac{N_f}{N_c} \to \mathrm{finite} \qquad \lambda = g_{\mathrm{YM}}^2 N_c\to \mathrm{finite} \,.
\end{equation}
Because $x$ is held fixed in this limit, one is no longer confined to the quenched approximation characteristic of large-$N_c$ gauge theories, in which quark loops are suppressed. This in turn leaves open the possibility for the exploration of dynamical quark processes. Of special interest is the (zero temperature) quantum phase transition from a chirally broken phase at small $x$ to a conformal phase at larger $x$.  

In the gravitational theory, one can add flavor by throwing into the five dimensions space filling stacks of $D4$- and $\overline{D4}$-branes. These branes have among their excitations an open string tachyon which transforms under the brane/anti-brane $U(N_f)\times U(N_f)$ and can be identified holographically with the chiral condensate \cite{Bigazzi:2005md}. By combining the tachyon DBI action with (\ref{eq:sihqcd}) and (\ref{eq:ihqcdV}) above, one obtains a bulk theory that is dual to large-$N_c$,  large-$N_f$ Yang-Mills in the Veneziano limit, a theory which has been dubbed VQCD \cite{Jarvinen:2011qe}.

The flavor sector of the theory is defined by the action
\begin{equation}\label{eq:svqcd}
S_f = -x M^3 N_c^2\int \dd^5x\,V_f(\lambda, \tau)\sqrt{-\det{\Apb}} \,,
\end{equation}
where $V_f$ is the tachyon potential, $\tau$ is the tachyon, and 
\begin{equation}
\Apb_{ab} = g_{ab}+\kappa(\lambda,\tau)\partial_a \tau\partial_b \tau \,,
\end{equation}
is the pull back of the metric and tachyon kinetic term to the world volume of the flavor branes, with the $U(N_f)\times U(N_f)$ gauge fields set to zero. The kinetic function $\kappa(\lambda, \tau)$ ensures that the string frame tachyon action has been properly transformed to Einstein frame, where this action has been written.

Explicit choices for the tachyon potential and kinetic function are provided in \cite{Jarvinen:2011qe}. As before, their functional form is constrained by asymptotic considerations. In this case, the important points are that the tachyon potential does not interfere with the IR fixed points  controlled by (\ref{eq:ihqcdV}), and that the $\beta$-function matches that of the Veneziano limit field theory in the UV. A suitable choice is
\begin{equation}
V_f = V_{f0}(\lambda)\,e^{-a(\lambda)\tau^2} \qquad \mathrm{where} \qquad V_{f0} = W_0+W_1\lambda+W_2\lambda^2,\qquad a(\lambda) = \frac{3}{22}(11-x) \,, 
\end{equation}
with
\begin{equation}
W_0 = \frac{12}{11},\qquad W_1 = \frac{4(33-2x)}{99\pi^2},\qquad W_2=\frac{23473-2726x+92x^2}{42768\pi^4} \,, 
\end{equation}
and
\begin{equation}
\kappa(\lambda,\tau) = \frac{1}{\(1-\frac{3}{4}\kappa_1\lambda\)^{4/3}}\qquad\mathrm{with}\qquad \kappa_1=-\frac{115-16x}{216\pi^2} \,.
\end{equation}
While not unique, these potentials have the desired asymptotic behaviors, and represent one reasonable parametrization of the ignorance inherent to the bottom up approach taken here.
The properties of this model  have been studied in both zero and finite temperature backgrounds \cite{Jarvinen:2011qe,Alho:2012mh}. In both cases, a novel phase structure develops, generically inline with various field theory expectations.

At zero temperature, one numerically solves the equations of motion for a background \eno{BHgeom}  with constant horizon function. The tachyon, generically, can vanish or vary as a function of $r$. Under duality, the near-boundary behavior of the tachyon sources the dual quark mass operator, and determines the vacuum expectation value of the chiral condensate. Thus, a trivial tachyon in the bulk can be interpreted as a boundary state in which a massless quark enjoys chiral symmetry. 

For small $x<x_c\sim 4$, it was found in \cite{Jarvinen:2011qe} that two branches of tachyon solution can arise. The chirally symmetric $\tau=0$ solution, and the $\tau=\tau(r)$ solutions which signal broken chiral symmetry. To determine which solution is thermodynamically preferred, one requires that the free energy is minimized. It turns out that in this regime, the solutions signaling broken chiral symmetry have the smallest free energy.

When $x\ge x_c$ the trivial tachyon solution persists, and is in fact the only regular solution dual to massless quark operators. This tachyon contributes to a background solution that is asymptotically AdS in the IR as well as the UV, and corresponds to a dual gauge theory that flows towards an IR fixed point. This conformal window persists for $x_c\le x < 11/2$.

Interestingly, very near (but below) $x_c$ a ``walking" region exists. These backgrounds are identified by noting that the solution is almost, but not quite, conformal in the IR. More specifically, the walking solutions are characterized by a dilaton that is approximately constant over a large radial range, narrowly missing the IR fixed point as a consequence of the growing tachyon profile.  

\begin{figure}
\centering
\includegraphics[scale=0.35]{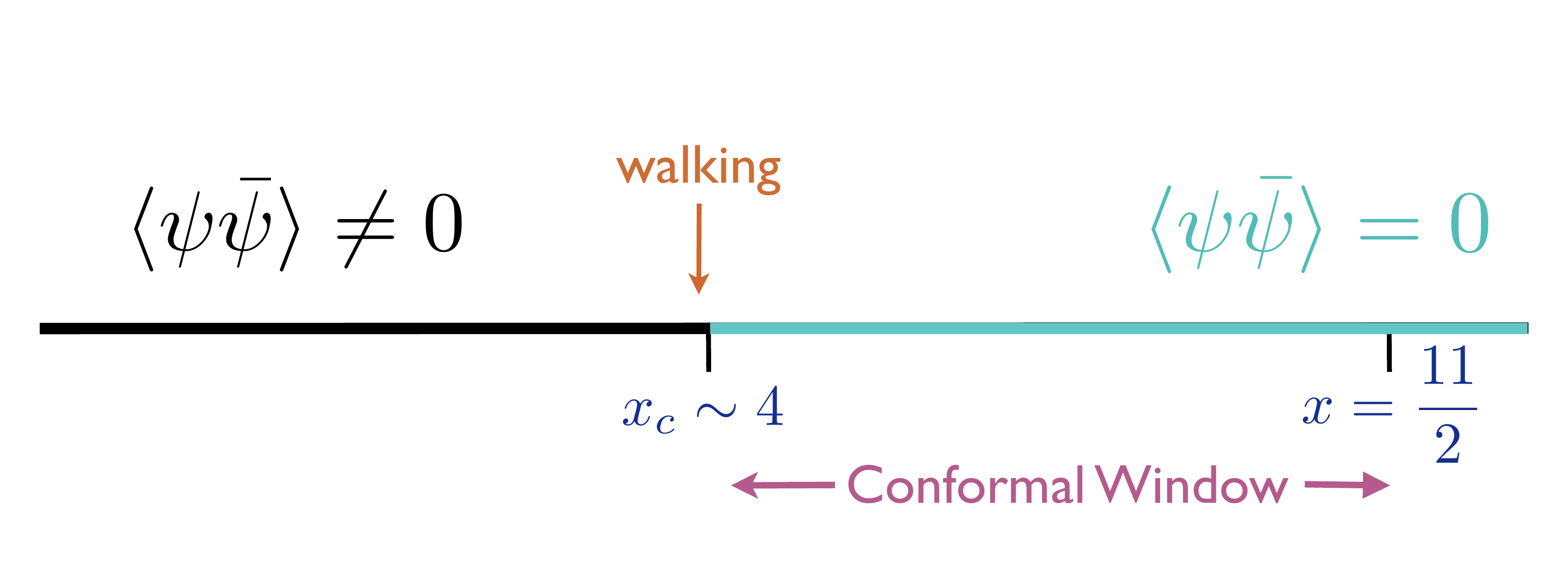}
\caption{\label{fig:vqcdpd1}Zero temperature phase diagrams for a holographic model of QCD in the Veneziano limit.}
\end{figure}

At $x_c$, there is a BKT-type transition, characterized by exponential (Miransky) scaling near the critical point. For example, a given glueball mass $m_n$ will behave like
\begin{equation}
m_n \sim \Lambda \,e^{-\frac{c}{\sqrt{x_c-x}}} \,,
\end{equation}
where $\Lambda$ is a scale with dimensions of energy and $c$ is some dimensionless constant. This scaling behavior is realized in all dimensionful quantities in this regime, including the meson spectrum as computed in \cite{Arean:2012mq}. The regions of interest are summarized in figure \ref{fig:vqcdpd1}.

By extending the analysis to black brane geometries of the type \eno{BHgeom} with a nontrivial horizon function, the VQCD model can be explored at finite temperature as well. Such an investigation results in a phase diagram with two relevant directions related to the temperature $T$ and $x \equiv N_f/N_c$. The general expectation that there exists a low temperature phase with broken chiral symmetry, and a high temperature phase with chiral symmetry restored is indeed born out holographically. 

Again, interesting features of these phase diagrams arise in the approach to $x_c$. As the critical ratio $N_f/N_c\sim 4$ is met from below, the critical temperature signaling the chiral (as well as the confinement/deconfinement) transition falls towards zero. At $x_c$ the transition sharpens into the familiar BKT-type, with the characteristic Miransky scaling noted previously appearing at zero temperature.

\begin{figure}
\centering
\includegraphics[scale=0.26]{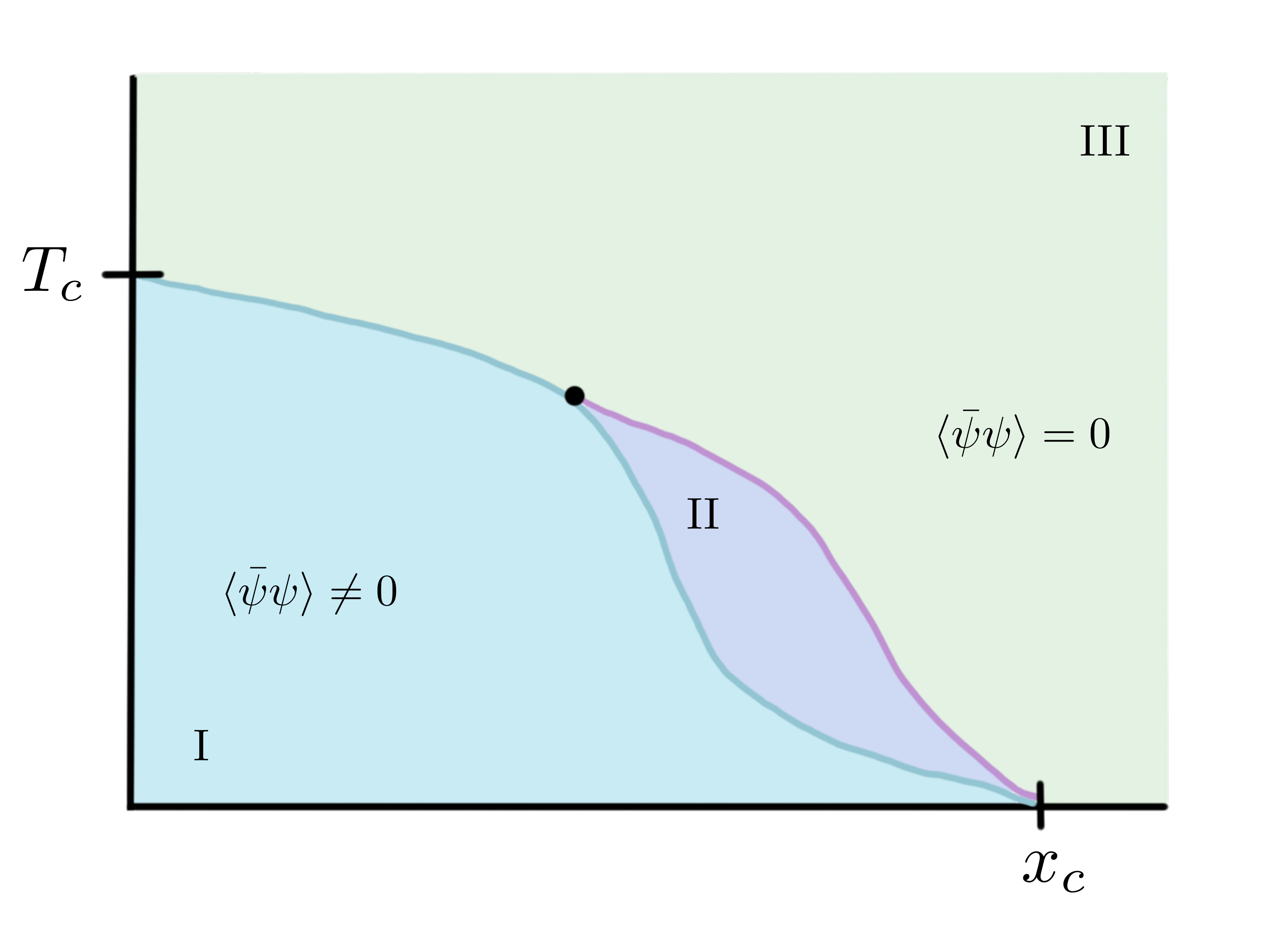}
\caption{\label{fig:VQCDTpd} Cartoon of a ``typical" phase diagram for hot VQCD matter. Regions I and II are phases with broken chiral symmetry, whereas region III has restored chiral symmetry. There is a first order phase transition separating region I from the phases in regions II and III, but a second order transition from region II to region III. The transition between region I and II can be thought of as a ``hadronization" transition, in that it separates a phase with $O(1)$ degrees of freedom (region I) from a phase with $O(N_c^2)$ (region II). A tri-critical point (the large black dot) exists at the point where the lines of transition meet.}
\end{figure}

Importantly, the bottom-up nature of the holographic model does not fully constrain the properties of the dual theory. As a consequence of this freedom, the phase structure of hot VQCD can vary considerably as the potentials of the model are modified. Many examples illustrating this diversity can be found in \cite{Alho:2012mh}. It is worth noting, however, that for many of the models which are qualitatively similar to real world QCD, the phase diagram resembles that of figure \ref{fig:VQCDTpd}. This diagram is characterized by a tri-critical point at the union of two first order lines and a second order line. Generically, there is a phase with broken chiral symmetry at low temperatures, a first order ``hadronization" transition separating regions with broken chiral symmetry but $O(1)$ and $O(N_c^2)$ degrees of freedom, and a second order transition to a region of restored chiral symmetry. In this class of models, the location of the tri-critical point is not robust to changes in the numerical values of the potential parameters, but the overall structure of the phase diagram is.

\section{Nonzero density and the phase diagram of QCD}
\label{PHASE}

Beyond the applications only to QCD matter at finite temperature, holography has also been employed to study QCD at finite density of baryon number as well. In this section, we review the expectations for the QCD phase diagram, and discuss several programs attempting to model it using gauge/gravity duality.

\subsection{Phase diagram of QCD}

\begin{figure}[tb]
\begin{center}
\includegraphics[width=3.5in]{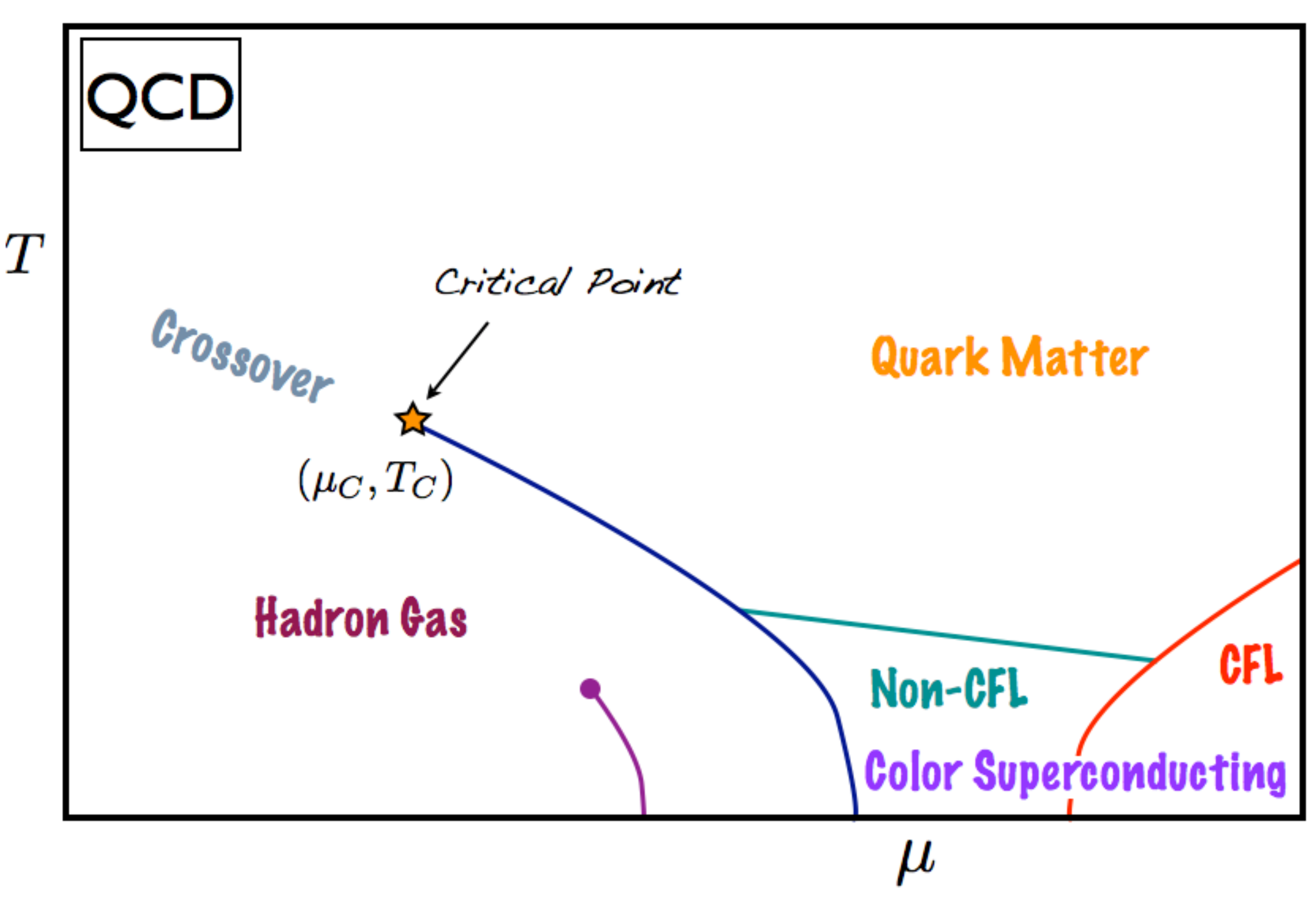}
\begin{minipage}[t]{16.5 cm}
\caption{A cartoon of the QCD phase diagram, showing the crossover and first-order line between the hadron gas and quark matter phases, the critical point, the nuclear matter line and various color superconducting phases. From  \cite{DeWolfe:2010he}. (Color online.) 
\label{Fig:PhaseDiag}}
\end{minipage}
\end{center}
\end{figure}

We briefly review the expectations for the phase diagram of QCD; for more discussion see for example \cite{Kogut:2004su,Stephanov:2007fk, Alford:2007xm}.
A prominent feature of the QCD phase diagram is the transition from chirally broken to chirally unbroken phases. When all quarks are taken massless, chiral symmetry is an exact symmetry of the QCD Lagrangian, and the broken symmetry phase at low $T$ and $\mu$ and the restored symmetry phase at high $T$ and/or $\mu$ are distinct and must be separated by a line of true phase transitions. This line of phase transitions is expected to be first-order near the $\mu$-axis, and can remain first-order as it reaches the $T$-axis in the case of three massless quarks, or can turn into a line of second-order transitions via a tricritical point for two massless quarks.  In the real world, quarks are massive and chiral symmetry is not an exact symmetry of QCD, and the transition near the $T$-axis is known from lattice studies not to be a sharp transition but instead a crossover.  It is widely expected that at sufficiently large chemical potential $\mu$ the first-order line returns; the line then terminates at a critical endpoint at some $(T_c, \mu_c)$.  This is displayed in figure~\ref{Fig:PhaseDiag}.

The critical endpoint is an object of substantial interest and speculation.   It is expected to lie in the universality class of the 3D Ising model, like the standard liquid/gas transition of fluids.
It is anticipated that depending on its location on the phase diagram, future heavy ion experiments such as those at RHIC or FAIR may produce a quark-gluon plasma lying close to the critical point at freeze-out, which could lead to information about its properties (see for example \cite{Aggarwal:2010cw,Staszel:2010zza, Stephanov:1998dy}.)   

Other features of the QCD phase diagram are also visible in figure~\ref{Fig:PhaseDiag}. The nuclear matter transition rises from the $\mu$-axis to end in another critical point. At high chemical potential one encounters various color superconducting phases, some color-flavor locked and some not; agreement on this region of the diagram is not universal and so it is represented schematically.

Theoretical exploration of the phase diagram using lattice gauge theory techniques is constrained by the  sign problem, which makes computations at nonzero $\mu$ very difficult. Lattice methods to avoid the sign problem, such as Taylor expansions around $\mu=0$, are possible
 \cite{Fodor:2001pe, Allton:2002zi, deForcrand:2002ci}, but nonetheless it is natural to pursue other theoretical techniques not affected by the sign problem. The gauge/gravity correspondence is one such method.

As usual, the lack of an exact dual description for QCD means one must formulate a model which hopefully captures the most salient features. Here we describe several attempts to capture the dynamics of the QCD phase diagram.

\subsection{Holographic QCD phase diagram}
\label{PHASEQCD}

One approach to modeling the QCD phase diagram is to try to distill the theory down to its most crucial features, match these to the gravity dual, and use whatever freedom is available to constrain the dynamics. As has been mentioned in other contexts, symmetries are in general the easiest features to match across a gauge/gravity duality. The ``Holographic Phase Diagram" model of \cite{DeWolfe:2010he} goes about this as follows. The essential features of QCD are chosen to be:
\begin{itemize}
{\item It is a non-abelian gauge theory}
{\item The running gauge coupling breaks conformal symmetry, though softly in the ultraviolet}
{\item The presence of quarks generates a conserved $U(1)$ charge, baryon number.}
\end{itemize}
These three features are modeled in the gravity dual with three fields: an asymptotically anti-de Sitter metric with a black hole horizon, a scalar field $\phi$ with a nontrivial profile in the radial direction, and a $U(1)$ gauge field with an electric potential turned on to generate a chemical potential, in the form of an Einstein-Maxwell-scalar theory as previously discussed. Note the chiral condensate does not occur as an explicit degree of freedom; however, it is expected \cite{Son:2004iv} that only one combination of the condensate and the baryon density survives as a hydrodynamic mode, and so it is reasonable to expect that the baryon density can serve as an suitable variable.

Field redefinitions of the metric and scalar can be used to eliminate any functions of the scalar multiplying the Einstein or scalar kinetic terms; once a Chern-Simons term is taken to vanish, the only freedom in this minimal model is in the scalar potential $V(\phi)$ and the gauge kinetic function $f(\phi)$. The philosophy then is to tune these functions to generate as realistic a match to QCD as possible.

Again, one thinks of the running of the scalar as encoding the beta function of the dual gauge theory.
Since the  QCD beta function runs slowly for high and moderate energies, the dual operator is chosen to be almost, but not quite, marginal, with $0 < 4 - \Delta_\phi \ll 1$. The remaining freedom in $V(\phi)$ and $f(\phi)$ can then be fixed by matching to something well-known: QCD lattice data for thermodynamics at zero chemical potential, where the sign problem does not apply. Note that although the gravity dual presumably corresponds to a large-$N_c$ gauge theory, the lattice data used as constraining input is that of ordinary 3-color QCD. Thus the model is attempting to generate a large-$N_c$ limit for QCD with the thermodynamics fixed at what one obtains from three colors.

The equation of state $s(T)$ can be calculated at $\mu=0$ using lattice techniques; the dimensionless quantity $s/T^3$ rises rapidly at the crossover before asymptoting at high temperatures. In \cite{Gubser:2008yx}, it was shown that such behavior can be captured by a potential of the form
\eqn{VChoice}{
  V(\phi) = {-12 \cosh \gamma\phi + b\phi^2 \over L^2} \qquad
    \hbox{with $\gamma = 0.606$ and $b = 2.057$} \,,
 }
 which results in $\Delta_\phi \approx 3.93$. The equation of state is indifferent to the form of the gauge kinetic function $f(\phi)$, but this can then be constrained by matching the lattice results for quark susceptibility $\chi_2$ at $\mu=0$; an effective choice is \cite{DeWolfe:2010he}
\eqn{fChoice}{
  f(\phi) = {{\rm sech} \left[ {6 \over 5} (\phi-2) \right] \over {\rm sech} {12 \over 5}} \,.
 }
It should be emphasized that both functions are chosen with some broad input (the generic presence of exponential functions of supergravity scalars) but are otherwise rather arbitrary, and many other functional forms may possibly produce similar behavior. Thus this choice should be regarded as a proof of principle in generating a holographic QCD phase diagram, with other possible functions still to be explored.

Once the Lagrangian is determined by matching $\mu=0$ data, black hole solutions can be generated at nonzero values of $\mu$ by turning on the gauge field. In practice, these solutions were generated by numerically solving the equations of motion out from the horizon to the boundary. The two initial conditions are the value of the scalar and the value of the electric field at the horizon; the near-boundary form of each solution then determines the thermodynamics $T$, $\mu$, $s$, $\rho$ as previously described. Generic solutions give rise to a source term $\phi_{(4 - \Delta_\phi)}$ \eno{PhiProfile} for the operator dual to the scalar, introducing the scale $\Lambda$ \eno{LambdaScale} that breaks conformal invariance.
A coordinate transformation to make the asymptotic behavior of the scalar field universal across all solutions is equivalent to setting the scale $\Lambda$ \eno{LambdaScale} to unity; thus the quantities $T$ and $\mu$ should properly be thought of as the dimensionless ratios $T/\Lambda$ and $\mu/\Lambda$.

Given that the existence of the crossover was built in by using the zero-density thermodynamics as input, one then wishes to see whether it is indeed the case that this crossover sharpens at nonzero $\mu$ into a line of first-order phase transitions ending at a critical point. By definition, the first-order line is the location in the $T$-$\mu$ plane where multiple thermodynamically stable phases coexist with the same free energy; near the first order line, both phases will continue to exist even though one becomes energetically favored. In addition, a thermodynamically unstable solution exists alongside the stable ones, the three corresponding to the multivalued behavior of the entropy density $s$ and baryon density $\rho$ near the line. In general one expects that all three phases will correspond to distinct black hole configurations, with identical temperature and chemical potential but other thermodynamic quantities distinct. It is easiest to uncover the first-order line by searching for the thermodynamically unstable black holes first.

\begin{figure}[tb]
\begin{center}
\includegraphics[width=3.5in]{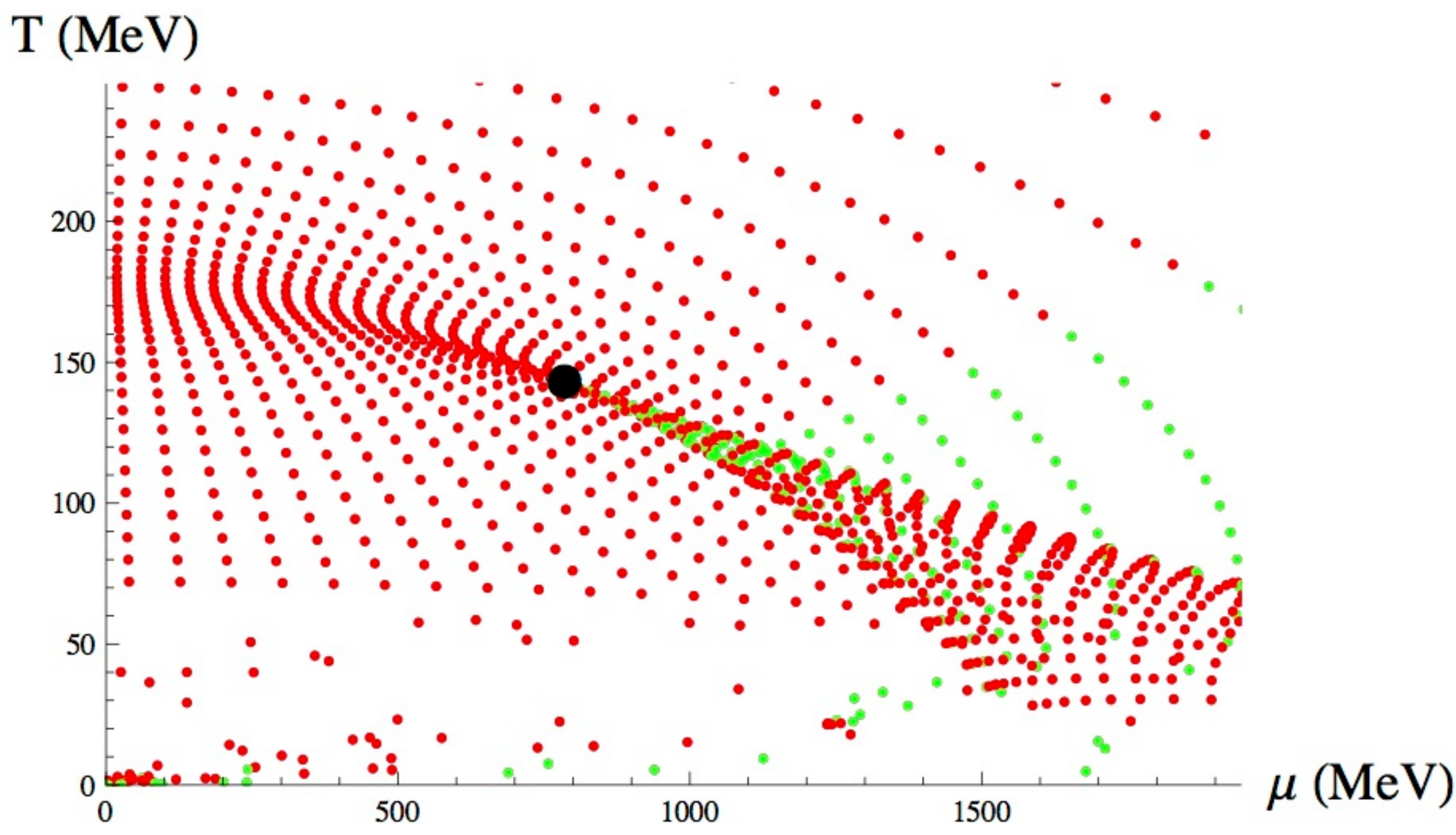}
\begin{minipage}[t]{16.5 cm}
\caption{Locations of numerically-generated black hole solutions in the $T$-$\mu$ plane, with red dots being thermodynamically stable and green dots thermodynamically unstable. The black circle is the location of the critical point. From  \cite{DeWolfe:2010he}. (Color online.) 
\label{Fig:Plane}}
\end{minipage}
\end{center}
\end{figure}

\begin{figure}
\begin{center}
\includegraphics[scale=0.57]{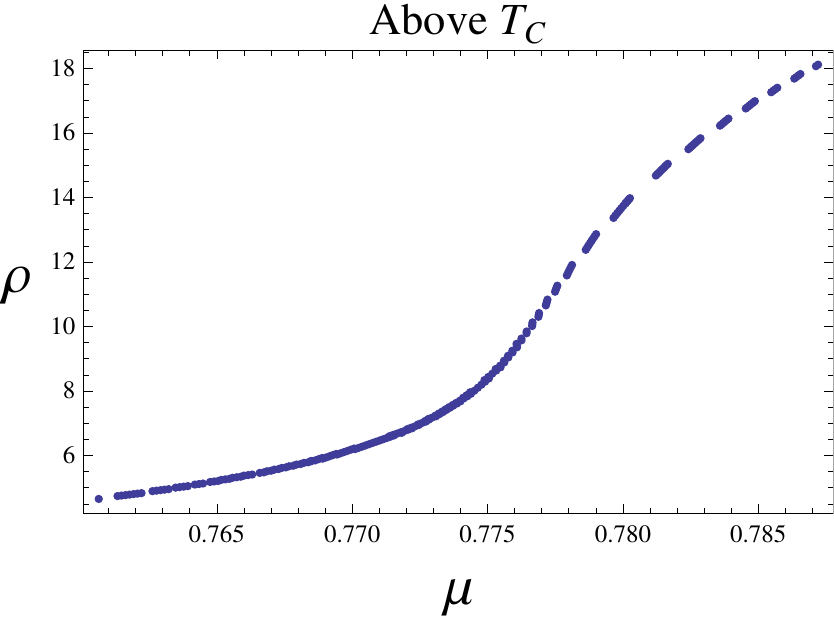}
\includegraphics[scale=0.48]{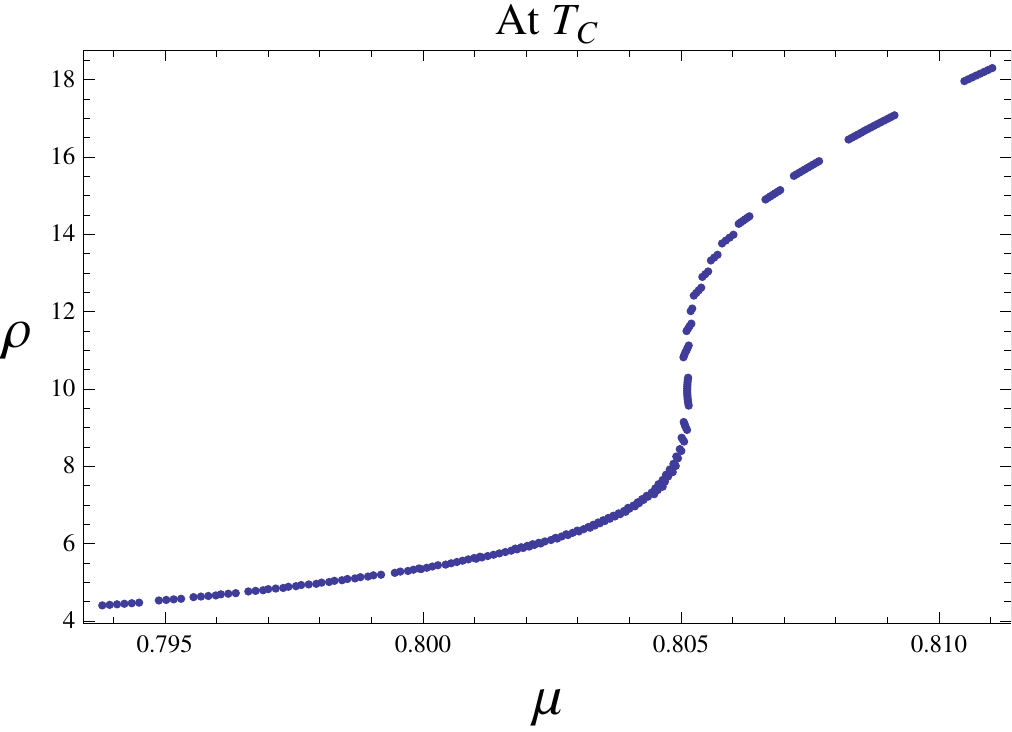}
\includegraphics[scale=0.48]{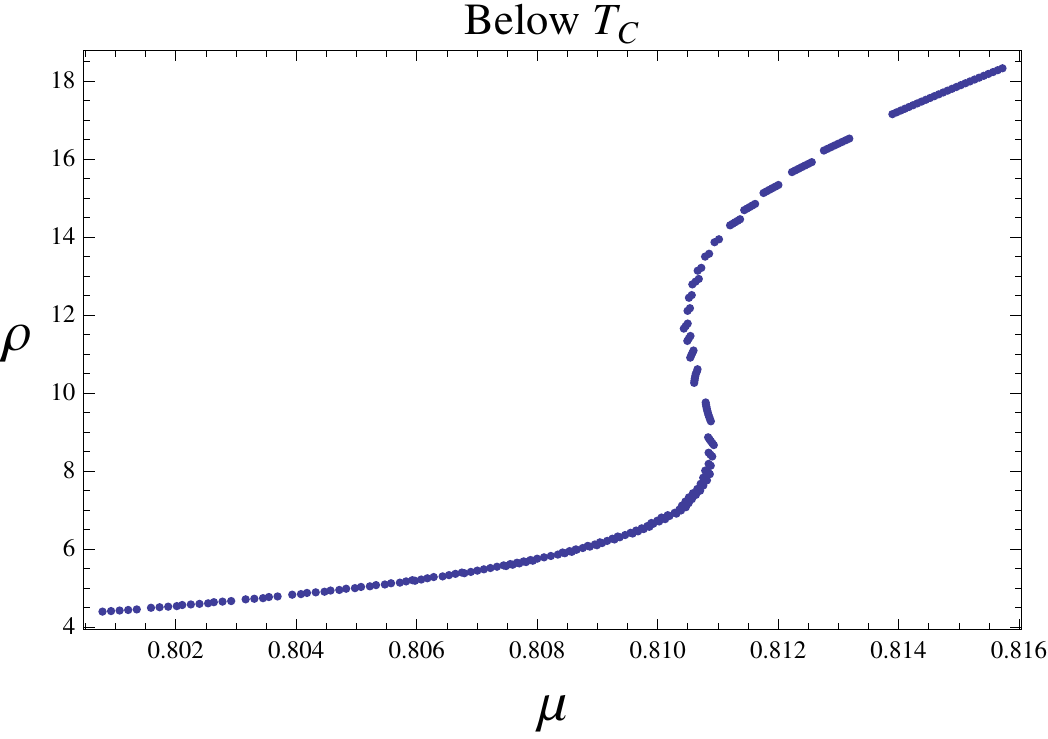}
\caption{The baryon density $\rho$ as a function of chemical potential $\mu$ for several values of $T$ near the critical point.  For $T > T_c$, $\rho(\mu)$ is single-valued (left), while for $T < T_c$ it is multi-valued (right).  At $T= T_c$ the slope is infinite (middle). From  \cite{DeWolfe:2010he}.
\label{fig:rhoMu}}
\end{center}
\end{figure}

A scan of several thousand black hole solutions covering a region of the $T$-$\mu$ plane indeed reveals a narrow strip where unstable and stable solutions coexist; see figure~\ref{Fig:Plane}. Looking at the baryon density $\rho$ (the entropy shows the same behavior) at a fixed temperature indeed shows the multivalued behavior characteristic of a first-order line, shown in figure~\ref{fig:rhoMu}. Varying the fixed temperature, one can determine where the multivaluedness disappears, and thus determine the location of the critical point. Setting the scales to match the input lattice data, the location of the critical point is found to be \cite{DeWolfe:2010he}
 \eqn{CriticalPosition}{
  T_c = 143 \ {\rm MeV} \qquad \mu_c = 783 \ {\rm MeV} \,.
 }
The  slopes of the baryon density and entropy density are the heat capacity $C$ and baryon susceptibility $\chi_2$, and these diverge at the critical point. Second-order phase transitions are classified into universality classes based on the values of a number of critical exponents characterizing these divergences. The primary exponents characterizing the thermodynamics are $\alpha$, $\beta$, $\gamma$ and $\delta$, defined by the power-law behavior of various quantities approaching the critical point from various distinct directions:
\eqn{AlphaDef}{
C_\rho \sim |T - T_c|^{-\alpha} \,, \quad \quad \quad {\rm along \; first \; order \; axis} \,.
}
\eqn{BetaDef}{
\Delta \rho \sim (T_c - T)^\beta \,, \quad \quad \quad {\rm along \; first \; order \; line} \,.
}
\eqn{GammaDef}{
\chi_2 \sim |T - T_c|^{-\gamma} \,, \quad \quad \quad {\rm along \; first \; order \; axis} \,.
}
\eqn{DeltaDef}{
\hskip-.5in \rho - \rho_c \sim |\mu - \mu_c|^{1/\delta} \,, \quad \quad  \quad \;\; {\rm for}\ T = T_c \,.
}
A much larger data set of numerical black holes near the critical point is necessary to estimate these exponents. A set on the order of 120,000 reveals exponents estimated as \cite{DeWolfe:2010he}
\eqn{}{
\alpha = 0 \,, \quad \quad \beta \approx 0.482, \quad \quad \gamma \approx 0.942 \,, \quad \quad
\delta \approx 3.035 \,.
}
These values are consistent with the scaling relations,
\eqn{Scaling}{
\alpha + 2\beta + \gamma  = 2  \,, \quad \quad
\alpha+ \beta(1+\delta) = 2\,,
}
establishing nontrivial consistency with the hypothesis that the thermodynamics of the set of black holes undergoes a genuine second-order phase transition. Moreover, the values for the critical exponents are consistent with mean-field behavior,
\eqn{}{
\alpha =0 \,, \quad \quad \beta = 1/2 \,, \quad \quad \gamma = 1 \,, \quad \quad \delta = 3.
}
Since mean-field behavior results when fluctuations are neglected, a natural hypothesis is that the large-$N_c$ limit, corresponding to neglecting quantum corrections in the gravity dual, has suppressed non-mean field behavior.

\begin{figure}
  \centerline{\includegraphics[width=4in]{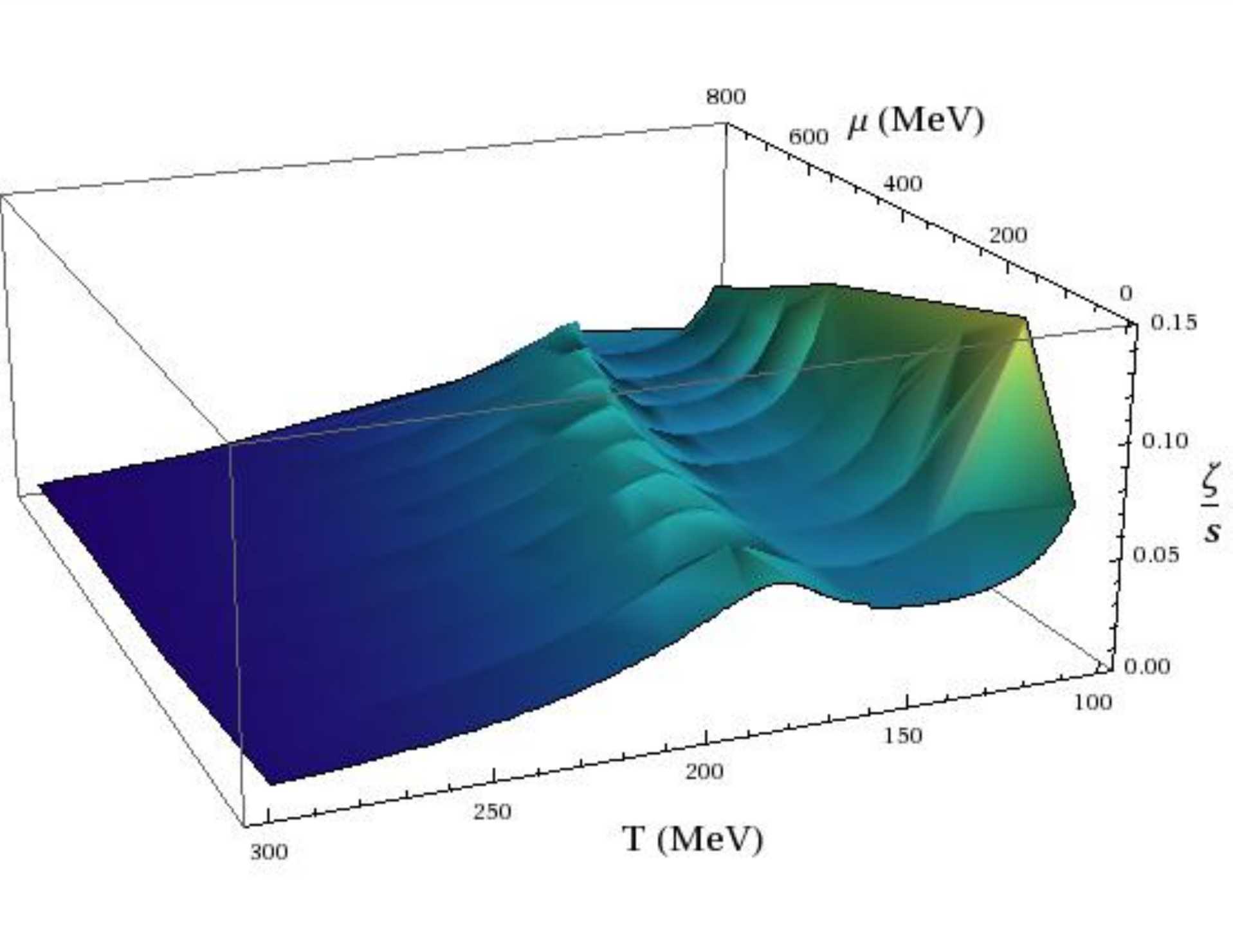}}
    \caption{The bulk viscosity over entropy density over the $T$-$\mu$ plane for the QCD-like black hole solutions. From \cite{DeWolfe:2011ts}.}\label{QCDZetaFig1}
 \end{figure}

One can proceed to study the transport properties of the critical point by studying linearized fluctuations around the nearby black hole solutions \cite{DeWolfe:2011ts}. The shear viscosity, of course, shows the universal behavior of two-derivative gravity theories already discussed. One can additionally study the conductivity $\lambda$ and bulk viscosity $\zeta$ from fluctuations of the gauge field and coupled scalar/graviton system; the latter is plotted normalized to $s$ in figure~\ref{QCDZetaFig1}, while the former displays similar behavior. Visible features include a bump corresponding to the crossover that sharpens into a peak with singular slope at the critical point, and a low-temperature divergence.

Interestingly, while the transport coefficients show divergent slopes at the critical point, they themselves do not diverge, contrary to the expectations \cite{Son:2004iv} that QCD fits into the classification of dynamic critical phenomena \cite{Hohenberg:1977ym} inside so-called model H, which possesses both energy-momentum transport and transport of a conserved charge (determined in \cite{Son:2004iv} to be the hydrodynamic mode combining the baryon density and the chiral condensate). Instead the models seem to fall inside model B, where energy-momentum transport is suppressed. It had been previously hypothesized by Natsuume and Okamura \cite{Natsuume:2010bs} that large-$N_c$ effects could provide exactly such a suppression; these results seem to confirm this hypothesis.

The holographic QCD phase diagram therefore captures a number of features of the expected QCD phase diagram, including a first-order line ending in a critical point with exponents near the values for the 3D Ising model. The differences from expectation --- namely the mean field exponents and the suppression of energy-momentum transport --- all may be explainable as a consequence of large-$N_c$ effects. An open question is then whether real-world QCD is close enough to ``large" $N_c$ to manifest these effects, or whether they should be considered an unrealistic artifact of the large-$N_c$ model.

\subsection{Phase diagram with fundamental matter}
\label{PhaseFund}

Another approach to the phase diagram of QCD is to incorporate fundamental matter and  chiral symmetry breaking explicitly.  ${\cal N}=4$ Super-Yang-Mills theory can be deformed by the addition of $N_f$ ${\cal N}=2$ fundamental hypermultiplets; this is the worldvolume theory of $N_c$ D3-branes near $N_f$ D7-branes. The holographic dual consists of adding a set of $N_f$ probe D7-branes to the $AdS_5 \times S^5$ geometry, and in the limit $N_f \ll N_c$, the backreaction of these branes may be neglected. This theory may be referred to as ``Quenched ${\cal N}=2$ quark matter". One may then 
add a conformal-symmetry breaking effect to emulate the effects of $\Lambda_{\rm QCD}$, and
 investigate the phase diagram of this theory for finite temperature and chemical potential.

In the presence of the probe D7-branes it is natural to split the $AdS_5$ radial coordinate, corresponding to the radius in the six directions perpendicular to the branes, into a radial coordinate $\rho$ for the $\mathbb{R}^4$ spanned by the D7-branes, and a radial coordinate $L$ for the perpendicular $\mathbb{R}^2$. A general embedding of the D7-branes is then determined by the function $L(\rho)$, which for large $\rho$ (near the boundary of $AdS_5$) schematically takes the form
\eqn{}{
L(\rho) \sim m_q + {\langle \bar{q} q \rangle \over \rho^2} \,,
}
where the constant term is the quark mass (the minimum length of a stretched string between the D3- and D7-branes), while the leading non-constant term indicates that the brane must pick an angle in the $\mathbb{R}^2$, and hence breaks the associated $U(1)$ symmetry; this $U(1)$ is associated with chiral symmetry and the coefficient of its breaking gives the chiral condensate $\langle \bar{q} q \rangle$.

A temperature may be turned on by placing a black hole inside $AdS_5$, as usual. There is also a $U(1)$ gauge field living on the D7-brane under which the quarks are charged, which is naturally associated with baryon number. One may thus turn on a chemical potential for baryon number by introducing an electric potential $A_t(\rho)$ for $U(1)_B$. Furthermore, one wishes to emulate QCD by introducing a conformal-symmetry breaking scale to play the role of $\Lambda_{\rm QCD}$; in the literature this has been addressed by turning on a spatial magnetic field $B$ for the baryon number gauge field \cite{Filev:2007gb}. (Note that this breaks four-dimensional Lorentz invariance.)  One then measures $T$ and $\mu$ in units of the scale $B$.

One may then solve for the embedding of the D7-brane given these choices; this corresponds to extremizing the D7-brane action, given by \eno{Brane} with $p=7$.
In general multiple brane configurations may exist, and the thermodynamically preferred one is found by minimizing the free energy.
Note that since the D7-brane is treated as a probe that does not backreact on the $AdS$-Schwarzschild geometry, the adjoint (``glue") dynamics remain those of ${\cal N}=4$ Super-Yang-Mills.

Two distinct kinds of phase transitions emerge, corresponding to distinct types of brane embedding solutions  \cite{Evans:2010iy}. The chiral symmetry breaking transition from $\langle \bar{q} q \rangle = 0$ to $\langle \bar{q} q \rangle \neq 0$ corresponds to whether the ``flat" brane embedding is or is not preferred to one that curves as it moves down the throat. Furthermore, embeddings may end on the black hole or may miss the horizon. When the embedded brane misses the horizon (a ``Minkowski embedding") there is a stable mesonic spectrum associated to D7-brane fluctuations, while if the brane goes into the horizon, mesonic excitations couple to the black hole quasinormal modes and become unstable. The transition between these two situations is referred to as the ``meson melting" transition; for a review of meson dynamics in AdS/CFT see \cite{Erdmenger:2007cm}.

\begin{figure}
  \centerline{\includegraphics[width=3.5in]{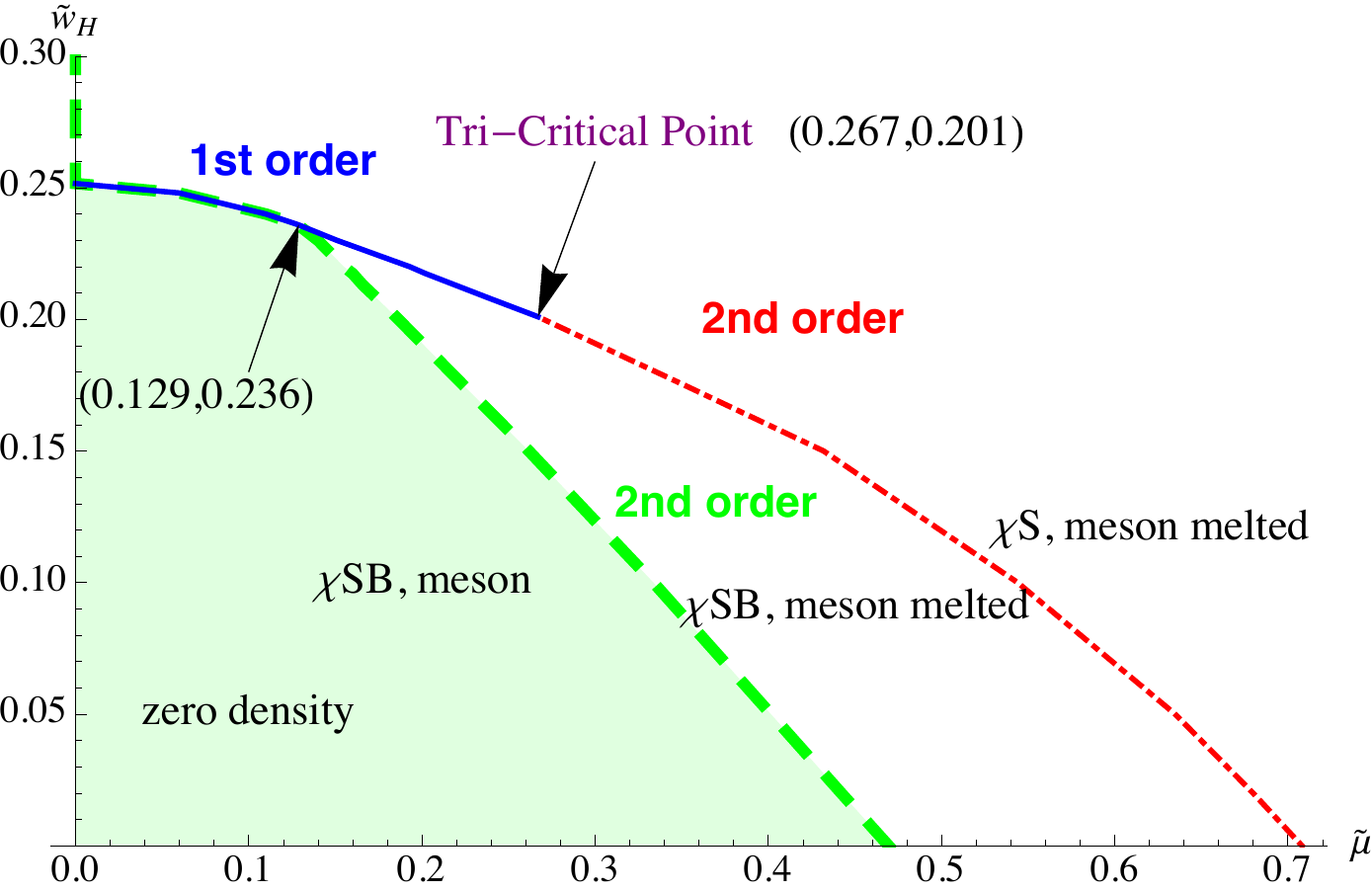}}
    \caption{The holographic phase diagram of ${\cal N}=4$ Super-Yang-Mills with ${\cal N}=2$ hypermultiplets and conformal symmetry broken by a worldvolume magnetic field. From \cite{Evans:2010iy}.}\label{Tvsmufig}
 \end{figure}

The results for the phase diagram for $m_q = 0$ are shown in figure~\ref{Tvsmufig}; the vertical axis is proportional to $T/B$ while the horizontal axis is $\mu/B$. Three phases are apparent. There is a chiral symmetry breaking transition stretching from one axis to the other; unlike the usual expectation for QCD, this is second-order near the $\mu$-axis, while it is first-order near the $T$-axis, with a tricritical point in between. The region with broken chiral symmetry is further divided by a second-order meson melting transition into a region with $\chi$SB and stable mesons near the origin, and a region with $\chi$SB and melted mesons at higher chemical potential.

One may also consider the phase diagram with $m_q \neq 0$; in this case the second-order transition near the $\mu$-axis becomes a crossover.
Thus  quenched ${\cal N}=2$ quark matter displays an interesting phase diagram including a chiral symmetry breaking transition; however, the order of the transitions are contrary to QCD expectations, with in particular the second-order/crossover region near the $\mu$-axis while the first-order transition is near the $T$-axis. Various refinements and generalizations have been attempted, including adding an electric field and studying transport \cite{Evans:2010np, Evans:2011mu} and introducing a probe dilaton profile \cite{Gwak:2011wr, Evans:2011eu}.

To move beyond the quenched approximation, one must take the backreaction of the D7-branes into account. This was studied for nonzero temperature by \cite{Bigazzi:2009bk} and at nonzero temperature and density by \cite{Bigazzi:2011it, Bigazzi:2013jqa}, building on earlier work at zero temperature and density in \cite{Benini:2006hh}. Here the D7-branes are homogeneously smeared and their backreaction included in a perturbative expansion in $\lambda N_f/N_c$. This study was able to see the leading effects of non-conformality in the thermodynamics, as well as predicting an enhancement in the jet quenching parameter.

\subsection{Color superconductivity}

While most efforts to explore the phase diagram of QCD have focused on the chiral symmetry phase transition and the associated critical point, another interesting region is that of large $\mu$, where color superconductivity is expected to occur; for a review see \cite{Alford:2007xm}. An attempt to study this phenomenon from a bottom-up approach was pursued in \cite{Basu:2011yg}, similar in spirit to the study of the critical point discussed in section~\ref{PHASEQCD}. Again, the thermodynamics result from a black brane geometry with a horizon and a gauge field providing the baryon number chemical potential. Again, some additional scale must be introduced to play the role of $\Lambda_{\rm QCD}$; rather than using an explicit scalar field, the authors formulated the gravity dual as a six-dimensional theory living on $AdS_5 \times S^1$, and allowed the radius of the extra dimension to introduce the scale.

Superconductivity is generally associated to the condensation of an operator transforming non-trivially under the gauge group, analogous to the Cooper pair; in the case of color superconductivity the quark bilinear ${\cal O}_{qq} \sim \langle q q \rangle$ typically plays this role. However, on the gravity side of the AdS/CFT correspondence, only gauge-invariant operators are visible, so this quark bilinear has no direct representation on the gravity side. However, the relevant physics should be visible in gauge invariant composites made from the bilinear; consequently the authors of \cite{Basu:2011yg} introduce a scalar on the gravity side $\psi$ dual to the gauge-invariant combination ${\cal O} {\cal O}^\dagger$, taken in this elementary model to have a mass term only.  

\begin{figure}
\begin{center}
\includegraphics[scale=0.35]{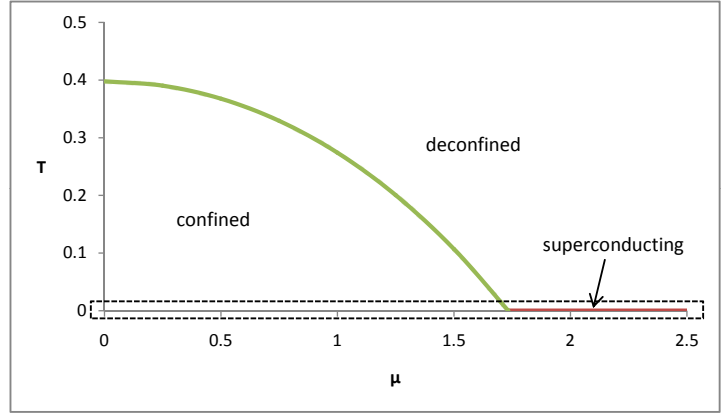}
\includegraphics[scale=0.35]{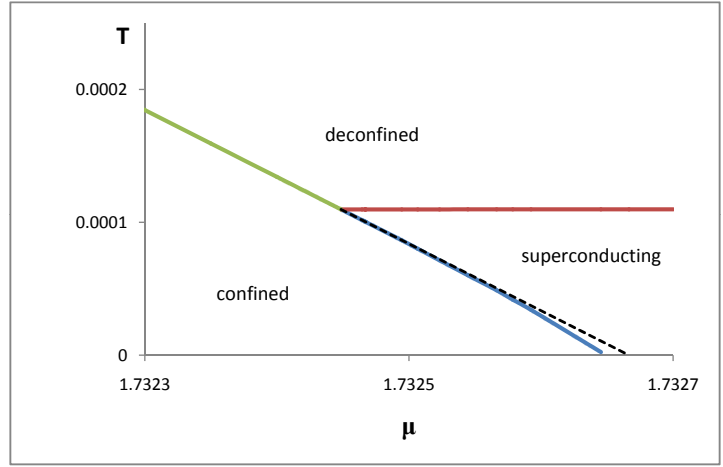}
\caption{ The phase diagram of the superconductivity model with particular values of the scalar mass and extra dimension radius ($m^2 L^2 = -6$ and $R=2/5$). The second figure zooms in on the transition region. From \cite{Basu:2011yg}.
\label{ColorSup}}
\end{center}
\end{figure}

This model was explored throughout the $T$-$\mu$ plane, and the gravitational action, standing for the free energy, was minimized for various configurations throughout the phase diagram. At small values of $\mu$, a transition between confined and deconfined phases was observed, with the scalar vanishing in the preferred configurations. Above a certain critical value of $\mu$, the scalar developed an instability and began to condense, signaling the onset of the superconducting phase (see figure~\ref{ColorSup}). 

While an elementary model, it is interesting to see the basic desired phase structure emerging.
Like the phase diagram study described in section~\ref{PHASEQCD}, but unlike the case in described in section~\ref{PhaseFund}, the fermions are not introduced explicitly via brane dynamics; this can be understood since color superconductivity is expected to arise in the $N_f \sim N_c$ regime, which is beyond the probe limit and branes would be expected to be replaced by their backreacted geometry.

The study of color-flavor locking is also complicated by the fact that it requires $N_f \sim N_c$. In \cite{Chen:2009kx}, $N_f \ll N_c$ explicit D7-branes are introduced, and then of order $N_f$ of the $N_c$ color D3-branes are separated out and their interactions with the D7-branes examined. Color-flavor locking is associated with the dynamical tendency for the D7-branes to absorb the D3-branes and turn them into worldvolume instantons. In the melted meson phase, this dynamical absorption is indeed observed. While these steps have been taken, it is evident that much more could be known about holographic realizations of this corner of the phase diagram.

\section{Future directions}
\label{FUTURE}

We close with some suggestions for future directions of research.

The generalizations of Bjorken flow discussed in section~\ref{BJORKEN} have either transverse structure, or longitudinal structure, but not both.  An obvious next step, then, is to try to combine the $SO(3)$ conformal symmetry with complexified longitudinal boost symmetry to obtain a conserved stress tensor which is hydrodynamical except for edge regions.  It is probably too ambitious to demand that the stress tensor should satisfy positive energy conditions throughout the future wedge of the collision plane; however, it may be possible to impose positivity conditions in some smaller region which nevertheless includes the future of most of the colliding matter.

Numerical studies in heavy ion applications of AdS/CFT have mostly focused on
$2+1$-dimensional problems in the bulk, corresponding to $1+1$-dimensional
problems in the boundary theory.  It is straightforward in principle to extend
the numerical methods described in section~\ref{RELATIVITY} to include more
independent variables.  The attraction of these methods is that they can
accommodate non-hydrodynamic regimes at early times, with realistically rapid
thermalization time scales.  Once one goes beyond $1+1$-dimensional problems in
the boundary theory, one can probe (for example) the development of spatial
inhomogeneities before, during, and after thermalization.  This
is particularly important in light of recent results on very high multiplicity
proton-nucleus collisions, which  exhibit azimuthal correlations reminiscent of flow \cite{CMS:2012qk,Abelev:2012ola,Aad:2012gla,Adare:2013piz}. The degree of
thermalization in these small systems can be quantified using  gauge-gravity duality, which can gracefully interpolate between the initial state and hydrodynamic paradigms that are used to understand the observed correlations \cite{Dusling:2012iga,Dusling:2012cg,Bozek:2012gr}. 

As reviewed in section~\ref{LIGHT}, there are competing efforts to understand
energy loss from high-momentum light particles in AdS/CFT.  Even the scaling of
energy loss with length comes out differently in the different approaches.  It
should be possible to make clearer statements about the differential rate
$dE/dx$ of energy loss in the falling string picture if one can compute the
response $\langle T_{\mu\nu} \rangle$ in the field theory.  Experience with
heavy quarks suggests that one must proceed to length scales well below where
hydrodynamics is applicable in order to fully understand the angular structure
of the emitted radiation. In addition, as the energy of the high-momentum light
particles is increased, the fluctuations become more important, and quantifying
these fluctuations is an essential part of a complete description of energy
loss. However, as described in \Sect{twopnt}, such stochastic phenomena can only be addressed by going beyond
the supergravity approximation \cite{Arnold:2012qg}. 

Charged black holes provide a simple way to study holographic gauge theories at finite density and temperature, as reviewed in section~\ref{PHASE}.  But little work has gone into understanding dynamical properties of the finite density states.  Phenomenologically interesting questions include how chemical potential varies with rapidity and how energy loss of hard probes varies jointly with chemical potential and temperature. There are also a number of questions  arising from the work on the phase diagram outlined in section~\ref{PHASE}: first whether the mean field critical exponents and apparent suppression of energy-momentum transport can be understood as large-$N_c$ effects and whether $1/N_c$ corrections can be understood; secondly how robust the results are to the precise form of the Lagrangian chosen, and whether the holographic model can make an at least approximately quantitative prediction for where in the phase diagram the critical point might lie, or whether the results are too sensitive to assumptions; and thirdly  whether the color superconducting region can be further characterized in a more sophisticated model.

\section*{Acknowledgments}

The work of O.D.\ was supported by the Department of Energy under Grant
No.~DE-FG02-91-ER-40672.
The work of S.S.G.\ was supported in part by the Department of Energy under Grant No.~DE-FG02-91ER40671, and by a Simons Fellowship, award number 230492. 
The work of C.R.\ was supported in part by EU grants  PERG07-GA-2010-268246 and the EU program ``Thales'' ESF/NSRF 2007-2013.
It has also been co-financed by the European Union (European Social Fund, ESF) and
Greek national funds through the
 Operational Program ``Education and Lifelong Learning'' of the National Strategic
 Reference Framework (NSRF) under
 ``Funding of proposals that have received a positive evaluation in the 3rd and 4th Call of ERC Grant Schemes''.
 D.T.\ was supported by the Department  of Energy, DE-FG-02-08ER4154, and as a RIKEN-BNL fellow.

\vfill
\eject

\bibliographystyle{JHEP}
\bibliography{refs}

\end{document}